\documentclass[hidelinks,12pt]{article}
\usepackage[export]{adjustbox} 
\usepackage[super,sort&compress,comma]{natbib}
\usepackage{color}
\usepackage{graphicx}	
\usepackage{amssymb}
\usepackage{enumitem}
\usepackage{bm}
\usepackage{caption}
\usepackage{etoolbox} 
\usepackage{subcaption}
\usepackage{mathtools}
\usepackage{titlesec}
\usepackage{bbm}

\usepackage{colortbl}
\definecolor{Gray}{gray}{0.9}
\usepackage{hyperref}
\makeatletter
\renewcommand*{\@biblabel}[1]{#1.}
\makeatother

\makeatletter
\@addtoreset{equation}{myequation}
\makeatother
\makeatletter
\renewcommand*{\@textcolor}[3]{%
\protect\leavevmode
\begingroup
\color#1{#2}#3%
\endgroup
}
\makeatother

\usepackage{times}
\usepackage[letterpaper, top=2.7cm, bottom=2.1cm, left=2.74cm, right=2.85cm, includehead=false, includefoot=false]{geometry}

\emergencystretch=\maxdimen
\usepackage[labelsep=period]{caption}
\usepackage[labelfont=bf]{caption}
\usepackage{amsmath,accents}
\usepackage[para,online,flushleft]{threeparttable}
\usepackage{multirow}
\usepackage{xltabular}
\usepackage{threeparttablex}
\usepackage{booktabs, caption}
\usepackage{makecell} 

\newenvironment{sciabstract}{%
\begin{quote} \bf}
{\end{quote}}

\usepackage[table]{xcolor}
\AtBeginEnvironment{figure}{\footnotesize}
\AtBeginEnvironment{table}{\footnotesize}
\AtBeginEnvironment{table*}{\footnotesize}
\AtBeginEnvironment{tabularx}{\footnotesize}
\AtBeginEnvironment{xltabular}{\footnotesize}
\AtBeginEnvironment{threeparttable}{\footnotesize}
\AtBeginEnvironment{threeparttablex}{\footnotesize}

\usepackage{xstring}
\hypersetup{bookmarksdepth=2}

\makeatletter
\newcommand*{\JP@tocparagraphline}[2]{%
\@dottedtocline{4}{3.8em}{3.2em}{#1}{#2}
}
\def\JP@lparagraph@hyper\Hy@toclinkstart#1\Hy@toclinkend\JP@stop#2{%
\IfEndWith{#1}{.}%
{\StrGobbleRight{#1}{1}[\JP@tocparagraph]\JP@tocparagraphline{\Hy@toclinkstart{\JP@tocparagraph}\Hy@toclinkend}{#2}}%
{\JP@tocparagraphline{\Hy@toclinkstart{#1}\Hy@toclinkend}{#2}}%
}
\renewcommand*{\l@paragraph}[2]{%
\JP@lparagraph@hyper#1\JP@stop{#2}%
}
\makeatother

\usepackage{xcolor}

\usepackage{tabularx}
\usepackage{array}
\usepackage{textcomp}

\title{\textbf{Atomistic Modeling of Chemical Disorder in Materials: Bridging Conventional Methods and AI-Assisted Approaches}}

\author{
\hspace{0mm}Jiayu Peng,$^{1,*}$ Peichen Zhong$^{2,*}$\\
\\
\hspace{0mm}\normalsize{$^{1}$Department of Materials Design and Innovation,}\\
\hspace{0mm}\normalsize{University at Buffalo, Buffalo, NY 14260, USA;}\\
\hspace{0mm}\normalsize{$^{2}$Department of Materials Science and Engineering,}\\
\hspace{0mm}\normalsize{National University of Singapore, Singapore 117575, Singapore;}\\
\hspace{0mm}{$^{*}$\small Correspondence: jypeng@buffalo.edu (J.P.), zhongpc@nus.edu.sg (P.Z.)}\\
}

\date{}

\begin{document}


\maketitle 

\baselineskip18pt

\clearpage
\phantomsection
\addcontentsline{toc}{section}{Abstract}
\begin{sciabstract}

Chemical disorder, originating from the mixed occupation of crystallographic sites by multiple elements, is widespread in metal alloys, ceramics, and other compositionally complex materials, where short- and long-range orderings can strongly influence their properties. Despite this importance, a central obstacle is the representation gap between experiments and simulations. Experiments often report disorder as partial occupancies and ensemble-averaged behaviors, whereas atomistic simulations and AI workflows usually require fully specified configurations. Tackling this gap requires computational methods that convert averaged disorder descriptions into representative configurational ensembles, while balancing cost, bias, and fidelity. This challenge has become more urgent in AI-driven computational discovery, where ignoring disorder may lead to AI workflows that misrank stability, misjudge novelty, and misdirect experiments with too-idealized representations. This Review systematically highlights how conventional and AI-driven methods bridge this representation gap. We assess the pros and cons of conventional and AI-enabled approaches, e.g., mean-field theories, cluster expansion, quasi-random approximations, Monte Carlo, and emerging schemes powered by the recent development of universal interatomic potentials and generative models. We further highlight how AI can accelerate various computational schemes by reducing the cost of microstate evaluation, configurational exploration, and atomistic-to-thermodynamic closure. We also emphasize how AI can enable disorder-native capabilities, including disorder-aware workflow triage, new ordering-sensitive and alchemical representations, generative models of disordered structures and configurational distributions, and kinetics-aware prediction of processing-dependent disorder. Together, this framework outlines a practical roadmap toward disorder-native AI, which can transform chemical disorder from a representational obstacle into an essential, controllable parameter for realistic AI-accelerated materials discovery.

\end{sciabstract}

\clearpage
\phantomsection
\pdfbookmark[1]{Table of Contents}{toc}
\renewcommand{\contentsname}{Table of Contents}
\tableofcontents

\clearpage
\phantomsection
\addcontentsline{toc}{section}{1. Introduction}
\section*{1. Introduction}

Chemical disorder is widespread in materials and can often play an essential role in their phase stability and functional properties. It appears across a broad range of material classes, including metal alloys, ceramics, and other compositionally complex solids, where multiple elements may share crystallographic lattices in ways that are not perfectly ordered.\cite{Simonov2020} Such disorder can exhibit many forms, such as substitutional disorder in alloys,\cite{Shockley1938} cation disorder in oxides,\cite{Peng2024a} anion disorder in mixed anion compounds, \cite{Young2023} and vacancy disorder in defect-rich materials.\cite{Zhang2023} The landscape of disorder becomes even richer in high-entropy and compositionally complex compounds, where five or more principal elements coexist on one or more sublattices.\cite{Han2024a,Hsu2024,Ding2026} As these variations in local chemical compositions directly reshape bonding environments, they can strongly influence how a material stores charges,\cite{Kang2024} conducts electrons,\cite{Hirai2026} controls diffusion,\cite{Cheng2026} catalyzes reactions,\cite{Jang2026} and deforms under stress\cite{Chen2021a} and radiation.\cite{Zhang2022} For example, chemically disordered rocksalt cathodes\cite{Lun2021} and high-entropy solid electrolytes\cite{Zeng2022} can exhibit unique transport and electrochemical behaviors relevant to energy storage in Li-ion batteries, while high-entropy alloys with different degrees of chemical disorder can possess significant differences in surface reactivity\cite{Peng2021} and durability.\cite{Peng2025} Therefore, chemical disorder is not a niche structural detail, but a crucial and tunable parameter that must be defined and understood clearly before optimizing materials systematically.

\begin{figure}
\phantomsection
\begin{center}
\includegraphics[max size={\textwidth}{\textheight}]{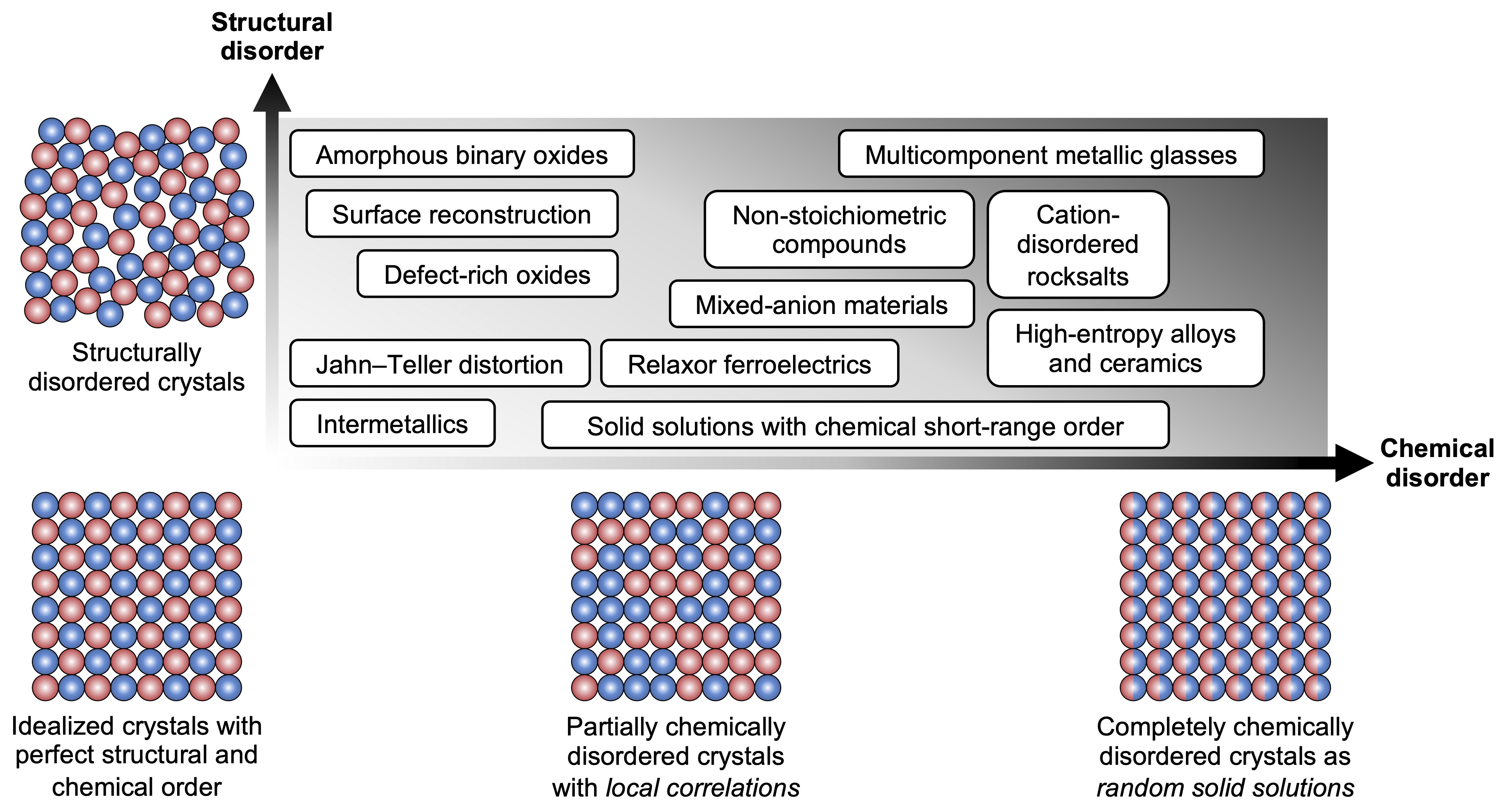}
\caption{
\textbf{Conceptual distinction and coupling between chemical and structural disorder.}
Schematic illustration of disorder in solid materials along two conceptual axes: chemical disorder (horizontal axis), describing the extent to which different species occupy crystallographic sites in an ordered, correlated, or random manner, and structural disorder (vertical axis), showing deviations of atomic positions from an ideal periodic lattice. The lower schematics indicate a continuum of chemical ordering, ranging from a perfectly ordered crystal (left), to a partially disordered crystal with local correlations or chemical short-range order (middle), to a completely chemically disordered crystal as a random solid solution (right). The upper-left schematic shows a structurally disordered material with markedly displaced atomic positions. Red and blue spheres denote different chemical species. Representative classes of solid materials are positioned within this space to illustrate typical combinations of chemical and structural disorder. The placements are qualitative to emphasize that disorder in materials is a continuum, rather than a binary classification.
}
\label{fig:chemical_vs_structural_disorder}
\end{center}
\end{figure}

In solid materials, it is crucial to conceptually differentiate chemical (occupational) disorder from structural (positional) disorder, while recognizing that compositionally complex systems frequently couple the two (Fig. \ref{fig:chemical_vs_structural_disorder}). Broadly speaking, disorder in materials refers to deviations from perfect crystalline order, but these deviations can arise either from how chemical species are distributed across lattice sites or from where atoms are positioned in space.\cite{Hass1984,Sokolovskiy2012,Moniri2023} Specifically, chemical disorder captures the probabilities of having multiple elements occupying equivalent crystallographic sites.\cite{Simonov2020} For instance, metal alloys with the lowest or highest degree of chemical disorder are intermetallics or random solid solutions, respectively.\cite{HumeRothery1969} By contrast, for structural disorder, the dominant deviation lies in atomic positions rather than site identity,\cite{Zacharias2025} as in crystals with substantial lattice distortions, defects, or displaced atoms that disrupt ideal periodicity. As examples, perovskite oxides can exhibit various types of structural disorder through octahedral tilting, Jahn--Teller distortion, and oxygen non-stoichiometry,\cite{Pena2001} while materials with the utmost degree of structural disorder are amorphous solids lacking any crystalline lattices.\cite{Cheng1987} At the same time, these two families of disorder are often linked in practice rather than cleanly separable. In high-entropy alloys and oxides, the random placement of differently sized and bonded atoms can perturb the surrounding lattices and induce local displacements, in which chemical disorder is accompanied by local structural distortions.\cite{Song2017,Barber2025} Considering such distinction and correlation, this Review particularly focuses on disordered materials dominated by chemical disorder, even though readers can look elsewhere\cite{Drabold2009,Biswas2009,Berthier2011,Jones2020,Berthier2023,Liu2025a,Wolf2025,Madanchi2025,Ma2026} for broader perspectives on structural disorder.

Critically, chemical order and disorder are not a simple dichotomy, but instead they define a continuous spectrum that might span fully ordered, partially ordered, and dominantly disordered states (Fig. \ref{fig:chemical_vs_structural_disorder}). Under various conditions, a material can possess long-range chemical order,\cite{Schlegel2025} lose this order only partially,\cite{Wyatt2025} or appear disordered on average while still retaining meaningful local correlations.\cite{Huang2024,Vogl2025} On the one hand, compositionally complex materials with many elements that tend to achieve chemically disordered states may instead form highly ordered intermetallic lattices.\cite{Wang2009,Cui2022,Wang2022,Liu2024a} On the other hand, the absence of long-range order in materials does not necessarily imply completely random mixing, because short-range order can persist through preferred local motifs, clustering tendencies, or avoidance between specific species.\cite{Chen2021b,Chen2023a} This continuity makes chemical disorder fundamentally a question of degree, length scale, and correlation, rather than a binary classification problem. These intermediate states can be the most scientifically crucial, as a modest change in local ordering may alter the distribution of atomic environments, without modulating the average crystal symmetry or electronic structure in an abrupt way.\cite{Pei2026} As a result, materials that appear similarly disordered at the average-structure level may behave differently if their local chemical correlations differ. Framing chemical disorder as a continuum thus offers a robust foundation for understanding how to co-design short- and long-range orderings.

\begin{figure}
\phantomsection
\begin{center}
\includegraphics[max size={\textwidth}{\textheight}]{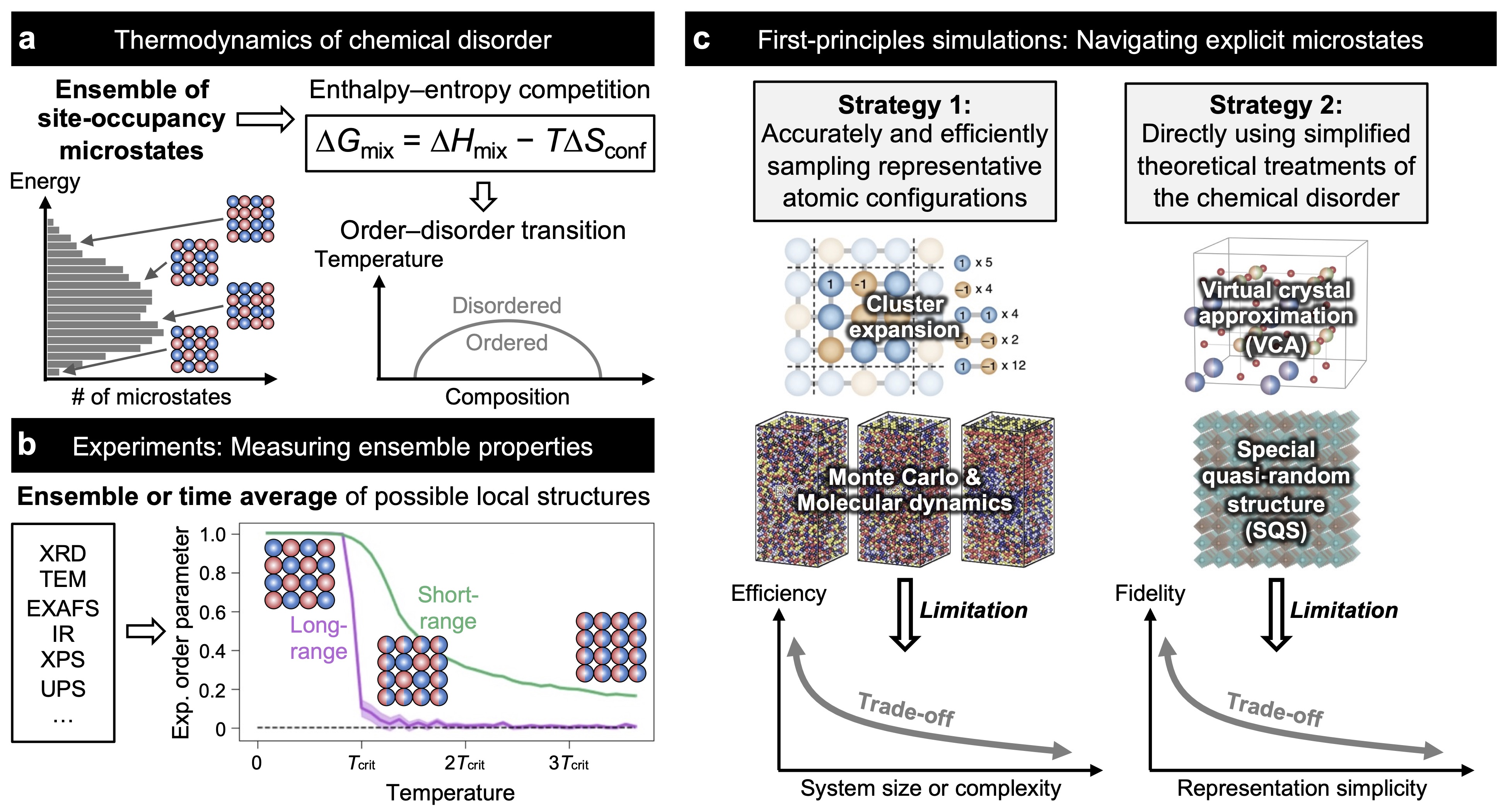}
\caption{
\textbf{Comparing thermodynamic, experimental, and computational perspectives on chemical disorder.}
(a) Chemical disorder is fundamentally a statistical-mechanics problem, stemming from many possible site-occupancy microstates with various energies. Its equilibrium properties are governed by the competition between the enthalpy of mixing and configurational entropy, which gives rise to temperature- and composition-dependent order--disorder transitions.
(b) Experiments examine chemical disorder primarily through ensemble- or time-averaged observables rather than a unique atomically resolved configuration. Key techniques, e.g., X-ray diffraction (XRD), transmission electron microscopy (TEM), extended X-ray absorption fine structure (EXAFS), infrared spectroscopy (IR), X-ray photoelectron spectroscopy (XPS), and ultraviolet photoelectron spectroscopy (UPS), offer information on average structure, local bonding, chemical states, electronic structure, and spatial heterogeneity across disordered materials, yielding averaged order descriptors across many local structures. The schematic indicates that the long-range order descriptor vanishes near a critical temperature, whereas the short-range order descriptor may persist to much higher temperatures, reflecting surviving local correlations in many disordered solids.
(c) This leads to a critical challenge in disorder modeling: although experiments can directly get averaged properties of interest in disordered materials, to compute such properties, atomistic simulations still need to assess numerous explicit microstates over a massive configurational ensemble. Two broad strategies have been used. The first strategy explicitly samples representative configurations via conventional methods (e.g., cluster expansion, Monte Carlo, and molecular dynamics), but it faces a trade-off between efficiency and the need to increase system size or complexity. The second strategy builds upon simplified theoretical treatments of disorder (e.g., the virtual crystal approximation and special quasi-random structures), which can enhance tractability but may sacrifice fidelity to local chemical correlations. Together, these panels show the gap between the characterization and modeling of chemically disordered solids.
Panel (b) adapted with permission from ref.\cite{Ferrari2023}, copyrighted by Springer Nature. Panel (c) adapted from refs.\cite{Santodonato2018,Peng2024a,EkborgTanner2024}, CC BY 4.0.
}
\label{fig:thermo_vs_computational_vs_experimental_gap}
\end{center}
\end{figure}

This continuum of chemical order and disorder can be physically understood by treating it as a statistical-mechanics problem, where the structure is linked to an ensemble of site-occupancy microstates selected by free-energy minimization. In this framework, disorder does not simply imply the lack of any visible ordering, but rather reflects the number of microscopic possibilities that are compatible with the macroscopic constraints of composition, temperature, pressure, and other parameters. Specifically, for chemically disordered solids, the key degrees of freedom are configurational: each microstate corresponds to a particular assignment of chemical species to crystallographic sites, and the resulting material behaviors depend on the statistical distribution of these site-occupancy patterns rather than on any configuration alone (Fig. \ref{fig:thermo_vs_computational_vs_experimental_gap}a). This treatment connects chemical disorder to the configurational entropy.\cite{Brahlek2022,Aamlid2023} For a fully random solid solution (i.e., the ideal mixing limit), the configurational entropy ($\Delta S_{\mathrm{conf}}$) reaches its highest value:
\begin{equation}
\label{eqn:configurational_entropy}
\Delta S_{\mathrm{conf}} = -N k_{\mathrm{B}} \sum_{i} x_i \ln x_i
\end{equation}
where $N$ is the total number of lattice sites, $k_{\mathrm{B}}$ is the Boltzmann constant, and $x_i$ is the fraction of sites occupied by species $i$. While this ideal expression is useful, it assumes that all possible microscopic atomic arrangements have exactly the same energy and are thus equally probable. Real materials, however, are rarely this simple, as various pairs of chemical species can interact differently, making specific local atomic arrangements energetically more favorable than others. A simple way to move beyond ideal mixing is to keep the entropy term but add an energetic cost for unlikely neighbors. An example is the regular solution model,\cite{Bragg1934,Bragg1935,Williams1935} which keeps the entropy of ideal mixing, but with an extra enthalpic penalty or preference for driving random mixing of different elements. In this framework, the Gibbs free energy of mixing ($\Delta G_{\mathrm{mix}}$) is:
\begin{equation}
\label{eqn:free_energy_of_mixing}
\Delta G_{\mathrm{mix}} = \Delta H_{\mathrm{mix}} - T \Delta S_{\mathrm{conf}}
\end{equation}
where $T$ is the temperature, and $\Delta H_{\mathrm{mix}}$ is the enthalpy of mixing. For a binary system with site fractions $x_\mathrm{A}$ and $x_\mathrm{B}$ (i.e., $x_\mathrm{A} + x_\mathrm{B} = 1$), the enthalpic term is often approximated as:
\begin{equation}
\label{eqn:enthalpy_of_mixing}
\Delta H_{\mathrm{mix}} = N \Omega x_\mathrm{A} x_\mathrm{B}
\end{equation}
where $\Omega$ is an effective interaction parameter that describes whether two physically dissimilar neighbors (e.g., A--B) are enthalpically more favorable than similar ones (e.g., A--A and B--B): having $\Omega < 0$ facilitates the mixing of species A and B, while $\Omega > 0$ discourages their mixing. In this form, the enthalpic term represents the energetic preference for particular local chemical motifs, whereas the entropic term reflects the statistical tendency to access many configurations. Order--disorder transitions, therefore, emerge naturally from the compositional and temperature dependence of this entropy--enthalpy competition.\cite{Wyatt2025,Toher2019,Dey2024,Almishal2025,Sivak2025,Dicks2026} As an example, at lower temperatures, the enthalpic contribution dominates and favors ordered states, while at higher temperatures, the growing entropic contribution gives rise to more random, chemically disordered configurations. In mean-field pictures,\cite{Shockley1938} the long-range order parameter can decrease continuously to zero at the critical ordering temperature, but even when long-range order disappears, local energetic biases can still preserve short-range chemical correlations, so that nominally disordered materials often remain far from completely random solid solutions. Altogether, thermodynamics determines the equilibrium tendency toward chemical order or disorder. Furthermore, kinetics might play a key role, as diffusion barriers, finite annealing times, and non-equilibrium operating conditions may hinder the system from reaching its thermodynamic equilibrium, preserve metastable disorder, or freeze in localized chemical correlations during synthesis, processing, and usage.\cite{Schlegel2025,Xing2024,Han2024b,Bacurau2024,Islam2025,Zhong2025a,Chun2025}

Experimentally, chemical disorder is usually not unambiguously resolved atom-by-atom as a specific configuration, but is instead inferred from measurements that capture an ensemble or time average of atomic structures over many local arrangements. This is exactly why no single technique fully resolves chemical disorder on its own, and why complementary measurements are needed across multiple length scales. In diffraction-based crystallography, Bragg scattering primarily encodes the average periodic structure, such as lattice symmetries, lattice parameters, and fractional site occupancies, while superlattice reflections further reveal the presence of long-range chemical order.\cite{Schlegel2025} When deviations from the average periodic structure become important, diffuse scattering and pair distribution function analysis can be especially valuable, because they retain information about occupational and displacive correlations that are not captured by Bragg peaks alone, including short-range order.\cite{Szymanski2023,Deng2025a} Complementary spectroscopies can further reveal different aspects of element-specific coordination environments.\cite{Joress2023,Morris2026} Extended X-ray absorption fine structure can reveal local environments, symmetry breaking, and preferred nearest-neighbor chemistries; vibrational spectroscopies probe local bonding and vibrational environments; and X-ray and ultraviolet photoelectron spectroscopies further provide surface-sensitive information on elemental chemical states and local electronic structure, respectively. Moreover, imaging and chemical mapping methods can provide projected local motifs, three-dimensional neighborhood statistics, and direct evidence of compositional heterogeneity.\cite{Lun2021,Moniri2023,Vogl2025,He2024} From these experiments, ensemble-averaged order descriptors can be derived (Fig. \ref{fig:thermo_vs_computational_vs_experimental_gap}b). To characterize long-range order, a widely used quantitative descriptor is the sublattice order parameter\cite{Shockley1938} ($\eta$):
\begin{equation}
\label{eqn:long_range_order_parameter}
\eta = c_{\mathrm{X}}^\mathrm{I} - c_{\mathrm{X}}^\mathrm{II}
\end{equation}
for a binary alloy on two sublattices, where $c_{\mathrm{X}}^\mathrm{I}$ and $c_{\mathrm{X}}^\mathrm{II}$ are the fractions of species $\mathrm{X}$ occupying sublattices $\mathrm{I}$ and $\mathrm{II}$, respectively. Having $\eta = 0$ indicates no long-range order, whereas a larger $|\eta|$ indicates stronger long-range chemical ordering. In addition, for short-range order, a widely leveraged shell-resolved order descriptor is the Warren--Cowley parameter\cite{Cowley1950,Norman1951,Ferrari2023} ($\zeta_{ij}^m$):
\begin{equation}
\label{eqn:short_range_order_parameter}
\zeta_{ij}^m = 1 - \frac{P_{ij}^m}{x_j}
\end{equation}
where $P_{ij}^m$ is the probability that an atom of species $i$ has a species $j$ atom in its $m$th coordination shell, and $x_j$ further denotes the overall concentration of species $j$ in the material. By definition, $\zeta_{ij}^m = 0$ represents fully random mixing, whereas $\zeta_{ij}^m < 0$ highlights an enhanced preference for having $i$--$j$ neighbors, and $\zeta_{ij}^m > 0$ indicates avoidance between species $i$ and $j$, which has often been associated with local clustering or atomic segregation. Nonetheless, these experimentally derived order parameters remain statistical averages of an underlying configurational ensemble rather than an atomically resolved and unambiguously identified reconstruction of the full three-dimensional atomistic disorder. This limitation is critical, since average occupancies, shell-wise order parameters, and pair-correlation-based descriptors still cannot uniquely elucidate the exact arrangement of atoms in any given region.\cite{Ding2026,Aamlid2023,Raabe2023} In other words, multiple distinct configurations can share identical pair statistics, and therefore, pair information alone is typically insufficient to fully reconstruct a configuration.\cite{Jiao2010} Moreover, these experimental techniques suffer from high cost and low throughput, which makes them useful for analyzing selected samples but also limits their power in large-scale materials screening. In addition, experimental signatures traditionally regarded as evidence of short-range chemical order, e.g., diffuse features in electron diffraction of random alloys,\cite{Coury2023} can be challenging to authenticate unambiguously, as they may alternatively arise from structural disorder, surface effects, or other symmetry-breaking artifacts.\cite{Walsh2023,Walsh2024} Overall, these pressing limitations necessitate the use and development of methods beyond experimental characterization to rationalize, engineer, and optimize chemical disorder in materials.

Computational approaches have shown great potential in addressing the intricate complexity of chemical disorder, but a crucial challenge in modeling such disorder is the mismatch between experiments and simulations, where experimental characterization typically captures ensemble-averaged behaviors, whereas atomistic modeling is generally performed on explicit microstates (Fig. \ref{fig:thermo_vs_computational_vs_experimental_gap}c). As the properties of interest are usually determined by averages over many accessible atomic arrangements, any realistic prediction of finite-temperature behaviors (or any meaningful comparison with experiments) must account for the configurational ensemble that is consistent with the chemical compositions and thermodynamic constraints of disordered solids, rather than relying on just a single deterministic structure. In practice, this challenge can be approached in two broad ways:\cite{Ferrari2023,Raabe2023} one may explicitly sample many configurations and average their predicted properties, or one may directly employ effective theoretical treatments that represent underlying disorder statistically without enumerating every arrangement in full. Both strategies are highly valuable, but they both become progressively formidable with increasing numbers of elements, sublattices, defects, and local chemical correlations. For example, explicit sampling can quickly encounter a combinatorial explosion in the number of possible configurations, which can push first-principles calculations beyond practical limits.\cite{Toher2019} Moreover, developing effective statistical descriptions can decrease the computational burden, but they might lose physical fidelity if the local chemical correlations that matter the most are oversimplified or neglected.\cite{Chen2023a} Consequently, simulations of chemical disorder are still severely constrained by the persistent trade-off among thermodynamic fidelity, computational cost, configurational bias, and scalability.

\begin{figure}
\phantomsection
\begin{center}
\includegraphics[max size={\textwidth}{\textheight}]{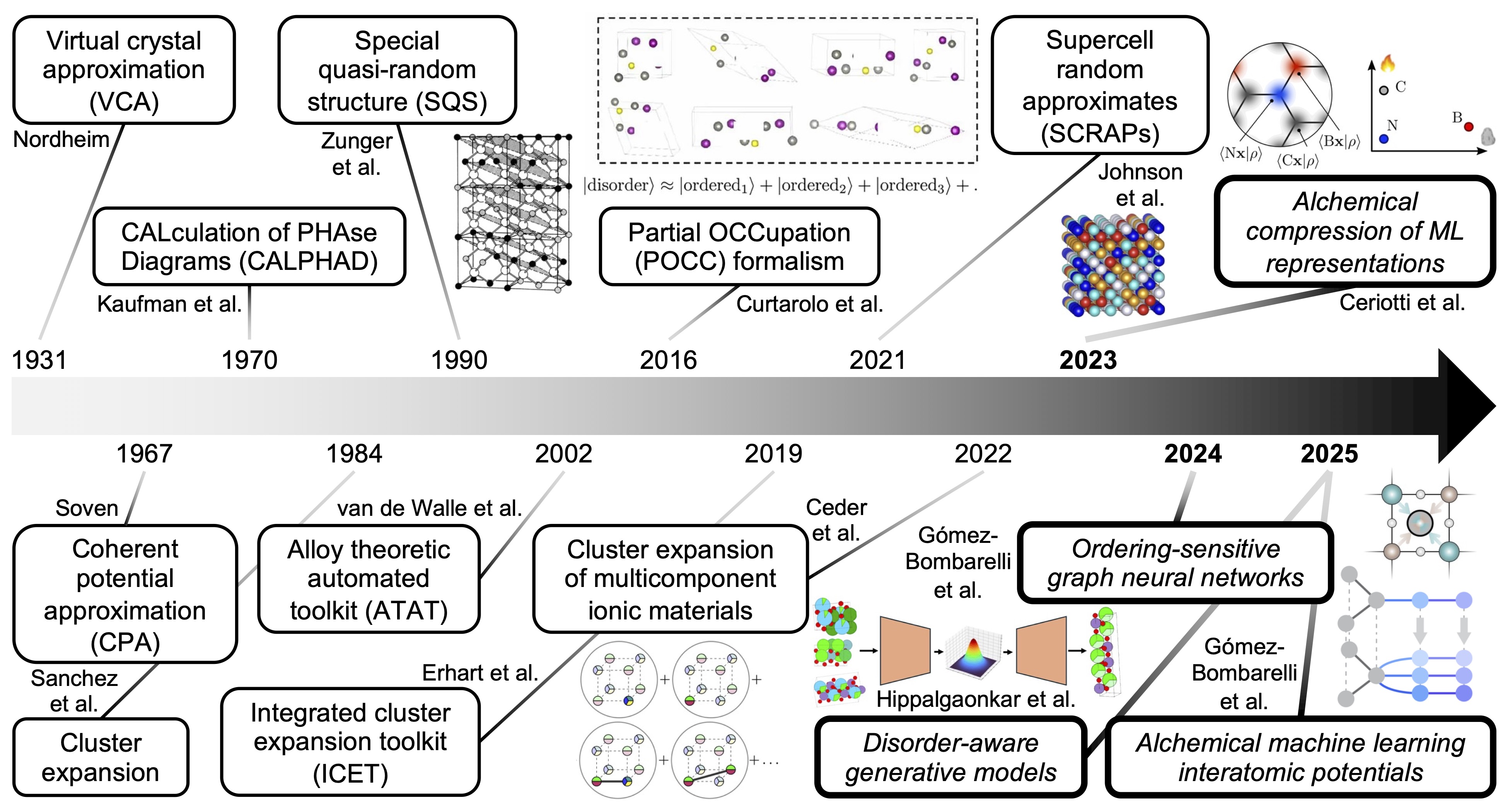}
\caption{
\textbf{Selected milestones in the development of computational methods for modeling chemically disordered materials.}
Chronological schematic illustrating essential methodological advances in modeling chemical disorder over nearly a century. Early computational approaches center on replacing explicit configurational complexity with effective averaged descriptions, which provide tractable mean-field or thermodynamic treatments of substitutional disorder without enumerating all atomic arrangements.\cite{Nordheim1931,Soven1967,Kaufman1970} A complementary line of development introduces more explicit configurational representations via cluster expansion (including its various implementations and extensions to more complex systems), thereby facilitating statistically grounded descriptions of configurational energetics and finite-temperature ordering.\cite{Sanchez1984,vandeWalle2002a,Angqvist2019,BarrosoLuque2022} Moreover, structure-centered approximations are developed to represent chemical disorder with carefully designed quasi-random supercells that aim to mimic selected ensemble-averaged correlation functions, while being computationally practical for first-principles modeling.\cite{Zunger1990,Wei1990,Yang2016,Singh2021} Beyond conventional methods, recent milestones highlight the growing integration of ML into disorder modeling, which promotes configurational exploration, strengthens transferability across diverse complex chemical compositions, and preserves sensitivity to various coordination environments.\cite{Lopanitsyna2023,Peng2024b,Petersen2025,Nam2025} Together, this timeline shows the historical evolution from simplified average-atom and thermodynamic formalisms toward increasingly explicit, scalable, and data-driven treatments of disorder, showing the ongoing convergence of conventional statistical-mechanics methods and modern AI tools. These milestones presented here are intended as representative landmarks, rather than an exhaustive history of the entire research direction.
Figure adapted from refs.\cite{BarrosoLuque2022,Peng2024b,Petersen2025,Nam2025}, CC BY 4.0, and with permission from refs.\cite{Wei1990,Divilov2025,Singh2021,Lopanitsyna2023}, copyrighted by the American Physical Society and Springer Nature, respectively.
}
\label{fig:computational_method_development_timeline}
\end{center}
\end{figure}

The rapid development of machine learning (ML) and artificial intelligence (AI) can narrow the gap between experiments and simulations and ease the accuracy--cost trade-off in disorder modeling by boosting configurational sampling, energy minimization, and structural discoveries without omitting the disorder physics that exist in real materials (Fig. \ref{fig:computational_method_development_timeline}). State-of-the-art data-driven tools, e.g., universal interatomic potentials\cite{Yuan2026} and generative models,\cite{Metni2026} can greatly expand the speed and scale of exploration, making broader sampling and faster screening more feasible than with direct first-principles simulations alone.\cite{Peng2022a} However, recent critiques of AI-accelerated computational materials discovery have also made clear that ML algorithms that cannot account for the various possibilities of chemical disorder can introduce major issues in high-throughput materials screening. High-throughput ML pipelines built around ordered, small-cell, effectively zero-temperature representations might over-predict highly ordered structures that are unlikely to persist at finite temperature, especially when configurational entropy would lead to disorder in real materials. In a recent example,\cite{Cheetham2024} apparently ``new'' ordered predictions have been shown to correspond more realistically to already discovered disordered phases. More fundamentally, if disorder is ignored altogether, AI models may produce both false positives and false negatives by missing stabilizing or destabilizing effects that arise directly from local chemical randomness and short-range correlations.\cite{Ferrari2023,Raabe2023} Therefore, ML and AI are not simply faster replacements for physics-based modeling, but must themselves become physically sound and explicitly disorder-aware to reliably promote the understanding and optimization of chemical disorder in materials. Several dedicated reviews, perspectives, and tutorials have addressed complementary aspects of this broader landscape, including configurational thermodynamics and cluster expansion,\cite{EkborgTanner2024,vanderVen2018,Xie2022} disorder and short-range order in high-entropy and compositionally complex materials,\cite{Pei2026,Ferrari2023,Ghazisaeidi2026,Wang2023a} and physics-driven AI for alloy and compositionally complex materials design.\cite{Raabe2023,Hart2021,Liu2023} However, these previous works generally focus on particular methodological families or material classes, without a strong focus on a unified treatment of how chemical disorder is represented, sampled, and connected to emerging AI-assisted approaches across different materials systems.

In this Review, we underscore the evolution and convergence of first-principles simulation methods and modern ML approaches as complementary computational tools to enable accurate, efficient, and physically robust atomistic modeling of chemically disordered materials (Fig. \ref{fig:computational_method_development_timeline}). Building on the combinatorial nature of configurational ensembles in disordered materials, we frame disorder modeling as a problem of how to represent, sample, and average over chemically complex microstates to ensure sufficient thermodynamic fidelity and minimal selection bias. We systematically compare the major strengths and key limitations of representative computational approaches, such as effective-medium and mean-field theories, cluster expansion, quasi-random structural approximations, ensemble-based averages, stochastic sampling, molecular dynamics, and macroscopic thermodynamic models---alongside rapidly emerging ML-driven schemes that might transform traditional simulation workflows using machine learning interatomic potentials (MLIPs), ordering-aware graph neural networks (GNNs), and generative AI. We emphasize not only the key benefits and drawbacks of individual approaches in throughput, accuracy, bias, and transferability but also show how different tools can be combined, where conventional methods supply interpretable thermodynamic formalisms, while ML further expands sampling scale and screening scope without discarding essential disorder physics. Finally, by examining remaining challenges in disorder modeling, we outline a forward-looking roadmap for developing physics-driven, disorder-predictive AI that integrates seamlessly with established statistical-mechanics formalisms and complex experimental observables, with the goal of promoting robust, unbiased elucidation and enabling accurate, efficient optimization of chemically disordered materials.

\begin{table}[hbt!]
\phantomsection
\centering
\begin{threeparttable}
\caption{
\textbf{Comparison of representative conventional simulation methods for modeling chemical disorder.}
}
\label{table:comparing_conventional_methods}
\begin{tabular}{llccccc}
\Xhline{3\arrayrulewidth}
Method category & Disorder representation & Sampling & Fidelity & Efficiency & Scalability \\
\Xhline{3\arrayrulewidth}
\begin{tabular}{@{}l@{}}Effective-medium and \\ mean-field theories \\ (e.g., VCA and CPA)\end{tabular} & \begin{tabular}{@{}l@{}}Direct partial occupancy \\ via averaged medium\end{tabular} & No & Very low & High & High \\
\hline
\begin{tabular}{@{}l@{}}Cluster-expansion-based \\ on-lattice statistical \\ mechanics (possibly \\ with Monte Carlo)\end{tabular} & \begin{tabular}{@{}l@{}}Explicit on-lattice \\ microstates via fitted \\ Hamiltonian\end{tabular} & Extensive & High & Medium & Medium \\
\hline
\begin{tabular}{@{}l@{}}Quasi-random structural \\ approximations (e.g., SQS)\end{tabular} & \begin{tabular}{@{}l@{}}One or a few supercells \\ matching desired local \\ chemical correlations\end{tabular} & No\tnote{1} & Medium & High & Medium \\
\hline
\begin{tabular}{@{}l@{}}Ensemble-based microstate \\ averages (e.g., POCC)\end{tabular} & \begin{tabular}{@{}l@{}}Many explicit microstates \\ with statistical averaging\end{tabular} & Limited & High & Low & Low \\
\hline
\begin{tabular}{@{}l@{}}Stochastic sampling and \\ molecular dynamics\end{tabular} & \begin{tabular}{@{}l@{}}Explicit site occupations, \\ usually coupled with \\ atomic relaxations\end{tabular} & Extensive & Very high & Very low\tnote{2} & Low \\
\hline
\begin{tabular}{@{}l@{}}Macroscopic thermodynamic \\ models (e.g., CALPHAD)\end{tabular} & \begin{tabular}{@{}l@{}}Direct partial occupancy \\ via analytical, empirical \\ free-energy models\end{tabular} & No & Low & High & High \\
\Xhline{3\arrayrulewidth}
\end{tabular}
\begin{tablenotes}
\footnotesize
\item[1] Excluding the negligible, one-time sampling during SQS construction. \\
\item[2] Might be accelerated substantially through parallel computing with GPU implementations, but remains expensive at the direct first-principles level.
\end{tablenotes}
\end{threeparttable}
\end{table}

\phantomsection
\addcontentsline{toc}{section}{2. Traditional computational methods for disorder modeling}
\section*{2. Traditional computational methods for disorder modeling}

Traditional methods for modeling chemical disorder in solid materials can be understood most clearly with a crucial question: how can experimentally averaged, partially occupied disordered states be translated into a form that first-principles simulations can actually compute? Different methods address this question in fundamentally dissimilar ways (Table \ref{table:comparing_conventional_methods}). Mean-field methods replace disordered crystals with an effective average medium. On the contrary, the most explicit approaches represent disorder through individual atomic configurations and recover observables by configurational averaging. Between these two limits lie a variety of modeling strategies, each providing a distinct balance among physical realism, computational cost, and generality. Hence, the key distinction among diverse methods is not simply whether disorder is included, but rather how such disorder is represented, whether explicit sampling is required, and how experimentally meaningful properties are recovered from the underlying configurational landscape.

\phantomsection
\addcontentsline{toc}{subsection}{2.1 Effective-medium and mean-field theories}
\subsection*{2.1 Effective-medium and mean-field theories}

Effective-medium and mean-field theories\cite{Nordheim1931,Soven1967} offer the simplest way to connect experimentally reported partial occupancies with electronic-structure calculations. In experiments, a disordered crystal is typically described with an average representation, where a lattice site can be ``partially species $i$ and partially species $j$.'' By contrast, standard first-principles simulations require fully specified crystal structures, where every crystallographic site is assigned to a definite atom. This mismatch is essential for density functional theory (DFT), a first-principles electronic-structure computational framework in which the ground-state properties of a system are determined from its electron density.\cite{Hohenberg1964,Kohn1965} In practical modeling of solid materials, DFT is applied to a specified arrangement of nuclei, rather than directly to the configurational ensemble average over various possible crystallographic site occupations. Consequently, conventional DFT for periodic solids cannot directly represent chemical disorder as partial site occupancy, unless one either samples many explicit configurations or uses an approximate averaged description of disorder.\cite{Ferrari2023,Raabe2023}

Diverse effective-medium and mean-field methods have been developed precisely to bypass this limitation by replacing the many possible local atomic environments of a disordered crystal with a single averaged medium that can be directly treated within a periodic electronic-structure framework (Fig. \ref{fig:conventional_method_comparison}a). The central idea behind them is to trade atomistic details for computational efficiency. Instead of enumerating or sampling the exponentially large set of configurations, one creates a translationally invariant ``effective crystal structure'' whose properties approximate the configurational average of the truly disordered system. This strategy can be particularly natural for substitutional disorder on a well-defined parent lattice, where the goal is to estimate average trends in phase stability,\cite{Ramer2000} electronic structure,\cite{Eckhardt2014} or other key properties\cite{Bellaiche2000} as the composition changes. This approximation has been especially influential in modeling metallic alloys, where the underlying disorder mainly broadens electronic states and renormalizes average interactions, while the crystalline lattice structures of these materials remain relatively well-defined.

\paragraph{Virtual crystal approximation (VCA).}
VCA is the simplest realization of this averaging idea and is best viewed as a DFT-compatible approximation to partial site occupancy.\cite{Nordheim1931} Specifically, in VCA, a disordered site shared by two species $i$ and $j$ is replaced by a ``fictitious atom'' whose effective ionic potential ($V_{\mathrm{VCA}}(\mathbf{r})$) is the composition-weighted average of the constituents:
\begin{equation}
\label{eqn:virtual_crystal_approximation}
V_{\mathrm{VCA}}(\mathbf{r}) = x_i V_i(\mathbf{r}) + x_j V_j(\mathbf{r})
\end{equation}
where $V_i(\mathbf{r})$ and $V_j(\mathbf{r})$ indicate the site-centered ionic potentials of species $i$ and $j$ at position $\mathbf{r}$, respectively, while $x_i$ and $x_j$ are their corresponding concentrations. Through this treatment, the crystal remains perfectly periodic, so the problem can be treated using essentially the same standard first-principles machinery as an ordered solid. 
In practice, VCA is often implemented in periodic DFT calculations with averaged pseudopotentials\cite{Ramer2000} or closely related mixed-species constructions,\cite{Bellaiche2000} making it widely used to quickly estimate composition-dependent trends.

VCA is useful for reducing chemical disorder to a single periodic DFT calculation, but such a simplification is also its central weakness, as it removes the local fluctuations that often matter the most in disordered materials. Since no supercell construction or configurational sampling is required, VCA keeps the computational cost close to that of an ordered DFT calculation and is thus well suited for rapid scans of alloy compositions, approximate band filling, average lattice parameters, or broad energetic trends.\cite{Ramer2000,Eckhardt2014,Bellaiche2000} This high efficiency has been particularly beneficial when the constituent elements in alloys are chemically similar, the lattice remains close to ideal, and the properties of interest depend primarily on smooth average behaviors.\cite{Winkler2002,Iniguez2003} For instance, VCA has recently been incorporated into a broader computational framework for assessing Li-ion transport barriers and dynamics in mixed-anion solid-state electrolytes.\cite{Payne2025} Nevertheless, as every site experiences the same averaged potential, VCA neglects site-to-site fluctuations, alloy scattering, local charge redistribution, unequal bond lengths, and short-range chemical order.\cite{Bellaiche2000} Therefore, it cannot capture disorder-induced band broadening, local lattice relaxations caused by size mismatch, or environment-specific bonding effects. As a result, VCA becomes markedly unreliable when disorder strongly couples to strain, localized electronic states, electronegativity differences, or bonding heterogeneity. However, the loss of species-resolved information does not necessarily give rise to a large error in every averaged observable. In Pb(Zr\textsubscript{0.5}Ti\textsubscript{0.5})O\textsubscript{3}, VCA reproduces the averaged B-site Born effective charge closely (e.g., $6.47$ vs. $6.40$ in an explicit ordered supercell), even though the separate Zr and Ti ``ghost''-atom contributions are $9.62$ and $3.32$, compared with $6.43$ and $6.37$ for Zr and Ti in the explicit supercell.\cite{Bellaiche2000} This example has illustrated both the strength and limitation of VCA: substantial species-resolved differences can cancel in an averaged response. Accordingly, VCA is most computationally reliable when the target is an averaged property and local-environment variations are secondary, whereas explicit configurational treatments become important when local environments control the observable.

\paragraph{Coherent potential approximation (CPA).}
CPA goes beyond VCA's simple representation by replacing a disordered material not with an averaged potential, but using an effective medium that reproduces the average scattering of electrons from its random lattice.\cite{Soven1967} Thus, CPA offers a more natural electronic-structure description of lattice disorder, especially when such disorder broadens bands and influences transport. Formally, CPA approximates the exact configurational average of the one-electron Green's function of a randomly substituted alloy:\cite{Yonezawa1973}
\begin{equation}
\label{eqn:coherent_potential_approximation}
\left\langle \mathcal{G}(z;\boldsymbol{\sigma}) \right\rangle_{\boldsymbol{\sigma}}
\approx \overline{\mathcal{G}}(z) =
\left[z - \hat{\mathcal H}_{0} - \Sigma_{\mathrm{CPA}}(z)\right]^{-1}
\end{equation}
where $\mathcal{G}(z;\boldsymbol{\sigma})$ denotes the one-electron Green's function for a particular atomic configuration $\boldsymbol{\sigma}$, and $\left\langle \mathcal{G}(z;\boldsymbol{\sigma}) \right\rangle_{\boldsymbol{\sigma}}$ further represents the configurational average over all possible occupations. Moreover, $\overline{\mathcal{G}}(z)$ denotes the Green's function of the coherent effective medium, $z$ is the complex energy with an infinitesimal positive imaginary part, $\mathcal{H}_{0}$ is the periodic part of the Hamiltonian defined on the parent lattice, and $\Sigma_{\mathrm{CPA}}(z)$ denotes the energy-dependent coherent potential (or effective self-energy) that represents the average scattering of this random alloy. Although this formalism might appear slightly abstract, its physical meaning is straightforward: CPA seeks a self-consistent medium in which replacing an average site by a real atomic species can produce, on average, no additional scattering. Hence, standard CPA is a single-site, mean-field theory, in which each site is embedded in an averaged environment. In first-principles calculations, CPA is most commonly combined with DFT through Green's-function-based implementations, such as Korringa--Kohn--Rostoker CPA, where the substitutional disorder in mixed alloys is represented naturally through site-resolved scattering matrices on their fixed parent lattices.\cite{Soven1970,Shiba1971,Johnson1986,Johnson1990,Singh2015}

CPA is powerful as it captures the average electronic consequences of substitutional disorder much more realistically than VCA while still avoiding explicit configurational sampling, but its single-site, effective-medium nature still limits the local physics it can describe. Compared with VCA, by retaining alloy scattering and disorder-induced spectral broadening, CPA gives a more realistic description of densities of states, magnetic moments, and transport properties, making it especially useful for random metallic alloys where electrons remain reasonably itinerant and the main role of disorder is to smear or renormalize the otherwise Bloch-like band structures.\cite{Gyorffy1972} As an example, CPA has been found to generate simulation results in reasonable agreement with available experimental data for the optical gap in disordered transition metal dichalcogenides.\cite{AlcazarRuano2025} Nevertheless, regarding limitations, because standard CPA still replaces the true configurational landscape with an averaged medium, it does not explicitly capture short-range order, clustering, pair or cluster correlations, local motif-dependent electrostatics, or local lattice relaxations such as unequal bond lengths and strain fields around mismatched atoms.\cite{Soven1967} These omissions become vital when local chemistry strongly affects the material behaviors, such as in some high-entropy alloys, ionic compounds, and covalent semiconductors with significant mismatch in atomic size or electronegativity.\cite{Yonezawa1973} Cluster and nonlocal extensions of CPA formalisms may recover part of this missing physics,\cite{Rowlands2008} but the standard single-site form remains most reliable for substitutional disorder on an approximately undistorted parent lattice, e.g., in simple alloys.

Collectively, VCA and CPA occupy the most approximate but also the most computationally efficient end of the disorder modeling spectrum. Both methods can bypass explicit sampling by replacing a disordered ensemble with an averaged medium, and both can be coupled with DFT to access composition-dependent properties at relatively low cost. VCA is the simpler and more broadly accessible approximation within periodic DFT workflows, while CPA is a more realistic choice when disorder scattering and spectral broadening are essential. However, both methods inherit the limitation of mean-field thinking: they capture only the average disordered state, not the distribution of local environments. This drawback has motivated the next family of methods, which keep the occupation variables explicit and treat configurational correlations directly.

\phantomsection
\addcontentsline{toc}{subsection}{2.2 Cluster-expansion-based on-lattice statistical mechanics}
\subsection*{2.2 Cluster-expansion-based on-lattice statistical mechanics}

\begin{figure}
\phantomsection
\begin{center}
\includegraphics[max size={\textwidth}{\textheight}]{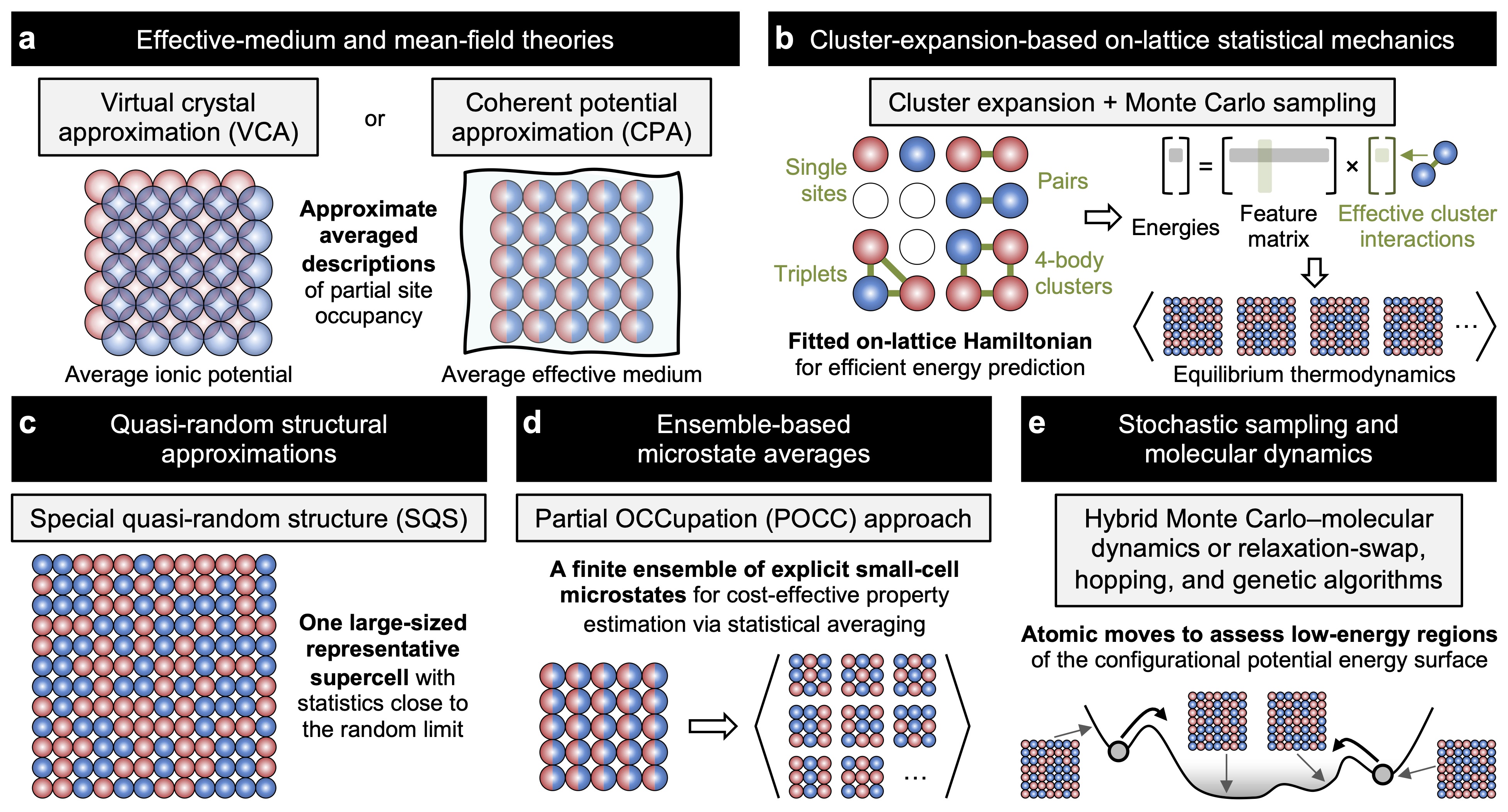}
\caption{
\textbf{Conceptual differences among various atomistic simulation methods for modeling chemical disorder.}
(a) Effective-medium and mean-field theories (e.g., VCA and CPA) replace explicit disorder with a single averaged periodic medium that can be handled within a periodic electronic-structure simulation framework, trading atomistic details for computational efficiency. In VCA, constituent atoms are merged into an average ionic potential, whereas in CPA, disorder is represented through an effective medium that reproduces configurationally averaged scattering behaviors. These methods intentionally smooth out the local configurational fluctuations that exist in real disorder.
(b) Cluster-expansion-based statistical mechanics retains configurational physics more explicitly by expressing the energy as a sum of effective atomistic interactions across single sites, pairs, triplets, and higher-order clusters. Once fitted, this lattice Hamiltonian can enable sampling (e.g., by Monte Carlo) to examine equilibrium thermodynamics, finite-temperature orderings, and short-range correlations across vast configurational ensembles.
(c) Quasi-random structural approximations (e.g., SQS) represent disorder with one or a few representative supercells whose selected correlation functions approximate those of a fully random solid solution, allowing direct first-principles simulations without full ensemble sampling.
(d) Ensemble-based microstate-average methods, illustrated by POCC, describe a disordered crystal as a finite statistical mixture of explicit ordered microstates, usually generated within a restricted cell size and fixed parent lattice. Observable properties are then obtained by averaging over this manageable set of symmetry-distinct configurations, providing an intermediate picture between single-structure approximations and extensive thermodynamic sampling.
(e) Stochastic sampling and molecular dynamics probe the low-energy region of the configurational PES directly through atomic moves or dynamical trajectories. These approaches are valuable when disorder is accompanied by strong local relaxation, off-lattice distortions, or many nearly degenerate minima, such that the relevant disordered structures are better regarded as a collection of low-energy basins than as a single averaged crystal or an exhaustively enumerated ensemble. Red and blue spheres denote different chemical species.
}
\label{fig:conventional_method_comparison}
\end{center}
\end{figure}

Cluster-expansion-driven methods approach chemical disorder in a fundamentally different way from the effective-medium and mean-field strategies. Rather than replacing a disordered crystal with an average representation, they treat this solid as an ensemble of explicit lattice-occupancy microstates and predict key properties by statistical averaging over the ensemble (Fig. \ref{fig:conventional_method_comparison}b). This distinction matters, as local ordering, clustering, and order--disorder transitions are controlled by how many configurations are populated and how their energetics differ, not by a single averaged crystal. Hence, cluster expansion keeps the configurational physics that mean-field descriptions intentionally smooth out, while offering a practical route to finite-temperature prediction. Here, we summarize the key formalism, practical considerations, and limitations of cluster expansion in the context of chemical disorder, whereas dedicated reviews and tutorials on cluster expansion offer more detailed technical discussions of its construction, validation, and sampling.\cite{EkborgTanner2024,Xie2022,vanderVen2018}

\paragraph{Theoretical foundations and formalism.}
The technical starting point of cluster expansion is to describe chemical disorder via a configurational ensemble on a parent lattice. In the canonical ensemble, the probability ($P_s$) of a microstate $s$ with energy $E_s$ at temperature $T$ is:\cite{Chandler1987}
\begin{equation}
\label{eqn:cluster_expansion_canonical_probability}
P_s = \frac{\exp(-\beta E_s)}{Z}
\end{equation}
where $\beta = 1/(k_{\mathrm{B}}T)$, $k_{\mathrm{B}}$ is the Boltzmann constant, and $Z$ is the partition function:
\begin{equation}
\label{eqn:cluster_expansion_canonical_partition_function}
Z = \sum_s \exp(-\beta E_s)
\end{equation}
For chemically disordered materials, one can coarse-grain the problem by grouping together all vibrational and electronic states relevant to an occupational configuration $\boldsymbol{\sigma}$. This treatment can lead to a configuration-dependent free energy $F(\boldsymbol{\sigma})$ and the equivalent expression:\cite{vandeWalle2002b}
\begin{equation}
\label{eqn:cluster_expansion_coarse_grained_partition_function}
Z = \sum_{\boldsymbol{\sigma}} \exp[-\beta F(\boldsymbol{\sigma})]
\end{equation}
where the sum runs over all allowed occupational configurations. In this form, $F(\boldsymbol{\sigma})$ works as an effective free energy for the slow configurational degrees of freedom after the faster vibrational and electronic degrees of freedom have been coarse-grained out. In many practical applications, especially when the focus is on chemical ordering instead of vibrational thermodynamics, $F(\boldsymbol{\sigma})$ can be approximated by a lattice Hamiltonian built from the relaxed ground-state energy of each configuration.\cite{Ceder1993} Although earlier studies have shown that this approximation might shift order--disorder transition temperatures quantitatively,\cite{Garbulsky1994} it typically preserves the overall topology of the computed phase diagrams.\cite{Garbulsky1996,vanderVen2000} The magnitude might nevertheless be substantial in specific model systems. For instance, for a phase-separating face-centered-cubic alloy model, including configuration-dependent lattice vibrations reduces the transition temperature by a factor of two vs. neglecting these contributions.\cite{Garbulsky1994} This coarse-graining approximation is what converts the complex many-body problem of chemical disorder into a lattice statistical-mechanics problem.

Cluster expansion provides the effective Hamiltonian that makes this configurational picture practical.\cite{Sanchez1984,vandeWalle2002a,Angqvist2019,BarrosoLuque2022,Laks1992,Zarkevich2004,Sanchez2010,Puchala2023,BarrosoLuque2024} Given a selected parent lattice, the configurational energy can be written as a sum over orbit-averaged cluster correlation functions, which allows both metallic alloys\cite{Sanchez1984,Sanchez2019} and lattice-based ionic systems\cite{Tepesch1995,Seko2009} to be effectively treated using the same general framework. For a multicomponent system, the configurational energy ($E(\boldsymbol{\sigma})$) is:\cite{EkborgTanner2024}
\begin{equation}
\label{eqn:cluster_expansion_hamiltonian}
E(\boldsymbol{\sigma}) = \sum_{\omega} m_{\omega} J_{\omega} \Pi_{\omega}(\boldsymbol{\sigma})
\end{equation}
where $\omega$ denotes an orbit of symmetrically equivalent clusters, $m_{\omega}$ shows its multiplicity, $J_{\omega}$ is the effective cluster interaction (ECI), and $\Pi_{\omega}(\boldsymbol{\sigma})$ is the orbit-averaged correlation function:
\begin{equation}
\label{eqn:cluster_expansion_orbit_correlation}
\Pi_{\omega}(\boldsymbol{\sigma}) =
\left\langle \Phi_{\boldsymbol{\alpha}\in\omega}(\boldsymbol{\sigma}) \right\rangle_{\omega}
\end{equation}
Here, $\Phi_{\boldsymbol{\alpha}}(\boldsymbol{\sigma})$ is the corresponding basis function for a particular decorated cluster $\boldsymbol{\alpha}$:
\begin{equation}
\label{eqn:cluster_expansion_basis_function}
\Phi_{\boldsymbol{\alpha}}(\boldsymbol{\sigma}) = \prod_i \phi_{\alpha_i}(\sigma_i)
\end{equation}
where $\sigma_i$ specifies which species occupies site $i$, $\phi_{\alpha_i}(\sigma_i)$ is a site basis function, and $\left\langle \Phi_{\boldsymbol{\alpha}\in\omega}(\boldsymbol{\sigma}) \right\rangle_{\omega}$ shows the average of $\Phi_{\boldsymbol{\alpha}}(\boldsymbol{\sigma})$ over all clusters in orbit $\omega$. In a nutshell, the correlation functions describe how atoms are arranged across the lattices, while the ECIs quantify how strongly those patterns contribute to the total energy of the chemically disordered systems.

Following the mathematical framework in Eqs. \eqref{eqn:cluster_expansion_hamiltonian}--\eqref{eqn:cluster_expansion_basis_function}, a cluster expansion can be defined by how the interaction basis is chosen and fitted. In binary systems, the cluster expansion often reduces to an Ising-like representation with occupation variables analogous to $\{-1,1\}$, whereas multicomponent and multi-sublattice systems require a larger set of orthogonal basis functions on each site, such as sinusoidal or indicator bases.\cite{vandeWalle2009,Zhang2016,BarrosoLuque2021,Rigamonti2024} Although the total number of possible configurations is enormous, the number of interaction terms that matter physically is often much smaller when the energetics are dominated by relatively local clusters, in particular if the fitting results respect the natural hierarchy from point terms to pairs, triplets, and higher-order atomic interactions.\cite{Mueller2009,Nelson2013,Zhong2022} In principle, for a fixed parent lattice and a complete basis, cluster expansion is formally exact for representing on-lattice configurational energetics, but in practice it must be truncated and regularized to remain computationally and statistically tractable.\cite{Xie2022} Collectively, this formalism of cluster expansion highlights that its accurate and efficient construction is both a physics problem and a regression problem: one must choose a parent lattice, define a tractable cluster space, build representative first-principles training structures, and eventually, fit a sparse, yet predictive Hamiltonian. Just as importantly, ensuring low fitting errors alone is not enough. A cluster expansion may predict training energies well but still yield incorrect thermodynamics if the low-energy motifs that dominate phase behaviors are not captured correctly, which is why ground-state-preserving fitting strategies can be crucial in configurational thermodynamics.\cite{Huang2017}

\paragraph{Practical use and considerations.}
Cluster expansion is most helpful when it is coupled with statistical sampling, as this combination converts a fitted lattice Hamiltonian into a vital tool for finite-temperature thermodynamics. Specifically, once the ECIs are known, the energy change associated with a trial configurational move can be evaluated very cheaply, making Monte Carlo sampling feasible over an extremely large number of atomic configurations. In canonical Monte Carlo, one proposes composition-conserving moves (such as swapping the species on two sites), computes the resulting energy change, and accepts the move with a probability\cite{Lun2021,Ouyang2019,Ji2019} (see Eq. \eqref{eqn:metropolis_algorithm} for details). In (semi-)grand-canonical sampling, the composition is allowed to vary under an imposed chemical potential, so the acceptance criterion also depends on the applied chemical potential and the corresponding change in particle number or species identity.\cite{vandeWalle2002c,Wang2020,Cheng2023,Xie2023} This cluster expansion--Monte Carlo workflow has therefore become a standard route for predicting phase diagrams, order--disorder transformations, equilibrium long- and short-range orderings, and ion intercalation thermodynamics in diverse disordered solids.\cite{Wolverton1998,Chen2023b,Guo2023} While Monte Carlo has been the most common method, the same fitted Hamiltonians may be analyzed with cluster-variation-type approaches when an analytical free-energy treatment of short-range order is preferred.\cite{Fu2024}

The strength of cluster expansion is that it retains explicit configurational correlations while remaining much cheaper than direct first-principles sampling. As the sampled states are explicit microstates, rather than a single averaged medium, this method captures short-range order, local motif populations, ordering and clustering transitions, and composition-dependent connectivity trends that are outside the natural scope of effective-medium and mean-field theories.\cite{Lun2021,Ji2019,Roy2026,Liu2026a} As an example, a recent cluster expansion--Monte Carlo study on entropy-stabilized oxides has shown that these chemically disordered materials can still retain measurable short-range order under synthesis-relevant conditions, even though increasing the chemical complexity suppresses the order--disorder transition and moves the structures closer to the ideal random limit\cite{Aamlid2024} (Fig. \ref{fig:conventional_method_demonstration}a). Equally importantly, the fitted Hamiltonian remains strongly physically interpretable: pair terms often reflect nearest-neighbor preferences or aversions, while higher-order terms capture cooperative atomic motifs that cannot be directly reduced to pairwise interactions. For instance, recent cluster-decomposition analyses\cite{BarrosoLuque2024} have made model interpretability more quantitative by linking the fitted Hamiltonian to chemically meaningful local contributions. Collectively, these characteristics make cluster expansion especially useful for substitutional alloys, finite clusters, nanoparticles, intercalation compounds, and other crystalline solid solutions that can be mapped cleanly onto well-defined parent lattices with only moderate local relaxation.\cite{Sanchez1984,Sanchez2019,Tepesch1995,Seko2009}

Ionic or chemically complex systems typically require additional care, as a short-range real-space cluster expansion may not capture long-range interactions efficiently enough on its own.\cite{BarrosoLuque2022} As an example, in multicomponent oxides,\cite{Ouyang2019,Aamlid2024} halide frameworks,\cite{Zhong2024} and multi-redox battery electrode materials,\cite{Lun2021,Ji2019,Huang2023} long-range Coulomb interactions, charge-balance constraints, and coupled cation--anion disorder can still remain essential even when the short-range chemistry is already represented reasonably well in cluster expansion. To address this challenge, a common extension is to augment the cluster expansion with an explicit electrostatic term:\cite{BarrosoLuque2022}
\begin{equation}
\label{eqn:cluster_expansion_electrostatic_correction}
E(\boldsymbol{\sigma}) =
\sum_{\omega} m_{\omega} J_{\omega} \Pi_{\omega}(\boldsymbol{\sigma}) + \frac{E_0}{\varepsilon_r}
\end{equation}
where $E_0$ denotes the unscreened Coulomb energy from an Ewald summation based on nominal ionic charges, and $\varepsilon_r$ is the effective dielectric constant. This electrostatically corrected cluster expansion has been especially useful for accurately modeling cation--anion short-range order,\cite{Zhong2020} heterovalent chemistry,\cite{Yang2022} and coupled vacancy ordering\cite{Zhong2023a} in complex disordered materials. Beyond improving energetic accuracy, this term also helps suppress highly charge-imbalanced and otherwise unphysical configurations during statistical sampling, keeping the configurational space under cluster-expansion-driven investigation more chemically meaningful. For example, in disordered rocksalt battery cathode materials, cluster-expansion-driven atomistic simulations have indicated that cation short-range order can strongly influence Li transport kinetics\cite{Ji2019} (Fig. \ref{fig:conventional_method_demonstration}b). Additionally, charge-decorated, sparse lattice models combined with semi-grand-canonical cluster expansion--Monte Carlo can further infer Li-vacancy orderings, recover Li intercalation voltage profiles, and disentangle complex transition-metal and oxygen redox processes.\cite{Zhong2023a}

Broadly speaking, the core weaknesses of cluster expansion follow directly from its lattice-centered construction. Since cluster expansion is fundamentally based on on-lattice models, it is less reliable when disorder induces large off-lattice displacements, major coordination changes, defect formation, or strong local strain.\cite{EkborgTanner2024,Xie2022} In addition, its complexity also grows rapidly with the number of components and candidate clusters, which makes basis selection, regularization, and training-set design increasingly crucial.\cite{BarrosoLuque2021,Zhong2023a,Muller2025,Wang2026a} The related fitting error can also increase with configurational dimensionality. In a three-sublattice Li--Mn--O--F ionic cluster expansion, for example, the out-of-sample root-mean-square error converges near $35$ meV/atom, compared with reported energy errors around $8$ meV/atom for lower-dimensional ternary rocksalt cluster expansions.\cite{Yang2022} Although such values depend on the chemistry, dataset, and validation protocol, they illustrate why high-dimensional configurational spaces might be more difficult to represent accurately with a tractable fitted Hamiltonian. Moreover, recent advances have introduced low-rank or tensor-factorized representations of atomic interactions that greatly compress interaction representations and strengthen the tractability of these approaches in chemically complex, high-dimensional configurational spaces.\cite{Drautz2019,Darby2023,Vorotnikov2025} In addition, long-range electrostatic interactions and elastic effects might converge slowly in a conventional real-space cluster expansion and require mixed-space treatments or auxiliary strain variables.\cite{Blum2004,Wang2023b} In certain ionic materials, electronic degrees of freedom might not be fully separable from configurational disorder, which can limit the accuracy of a purely occupational Hamiltonian.\cite{Zhou2006} These issues do not remove the value of cluster expansion, but they do define the range of systems for which it is most reliable.

Overall, cluster-expansion-based statistical mechanics provides a rigorous bridge between first-principles energetics and finite-temperature configurational thermodynamics. Its dominant advantage over effective-medium and mean-field theories is that it keeps explicit configurations and their correlations, which allows it to describe ordering physics with much greater detail and interpretability. Its central cost is that one must first build and validate a reliable, accurate lattice Hamiltonian, and this requirement becomes even more demanding for disordered solids with strong relaxation, long-range interactions, or high chemical complexity. These central strengths and weaknesses explain why cluster expansion remains one of the most powerful computational frameworks for chemical disorder, but also why one may sometimes leverage simpler modeling strategies when full configurational sampling is unnecessary or impractical.

\phantomsection
\addcontentsline{toc}{subsection}{2.3 Quasi-random structural approximations}
\subsection*{2.3 Quasi-random structural approximations}

Quasi-random structural approximations take a different route from the two methods discussed above. Effective-medium and mean-field strategies, such as VCA and CPA, replace a disordered crystal with an averaged medium, whereas cluster-expansion-based statistical mechanics builds an effective Hamiltonian and samples many configurations to recover equilibrium behaviors. In contrast, quasi-random structural approximations replace the full configurational ensemble with one (or a few) carefully chosen periodic supercells, whose local statistics mimic the disordered state (Fig. \ref{fig:conventional_method_comparison}c). In this sense, they are representative-structure approaches: they retain an explicit atomistic lattice in first-principles simulations, but avoid full thermodynamic sampling.

\paragraph{Special quasi-random structure (SQS) method.}
The best-known approximation is the SQS approach. The key idea of SQS is to construct a supercell whose low-order correlation functions match those of an infinite random solution as closely as possible at a fixed composition. Notably, the original SQS demonstrations have shown that even small supercells can capture the first few physically important correlations of random alloys far better than a naive random occupation of the same cell.\cite{Zunger1990,Wei1990} Once these supercells are created, they can be directly used in standard DFT simulation workflows as explicit structural models of the disordered phase.

The foundations of SQS are relevant to the cluster-correlation language in cluster expansion, while the goal is different. SQS uses the orbit-averaged correlation functions $\Pi_{\omega}(\boldsymbol{\sigma})$ and cluster basis functions $\Phi_{\boldsymbol{\alpha}}(\boldsymbol{\sigma})$ in Eqs. \eqref{eqn:cluster_expansion_orbit_correlation} and \eqref{eqn:cluster_expansion_basis_function}, but they are leveraged to identify supercells whose correlations best reproduce those of the target disordered state. In a perfectly random alloy, the desired value of this correlation is characterized by the ensemble average ($\overline{\Pi}_{\omega}^{\,\mathrm{rand}}$):
\begin{equation}
\label{eqn:sqs_correlation_function_random_average}
\overline{\Pi}_{\omega}^{\,\mathrm{rand}}
=
\sum_{\boldsymbol{\sigma}} P_{\mathrm{rand}}(\boldsymbol{\sigma}) \Pi_{\omega}(\boldsymbol{\sigma})
\end{equation}
where $P_{\mathrm{rand}}(\boldsymbol{\sigma})$ is the probability of the configuration $\boldsymbol{\sigma}$ in this random ensemble. Specifically, for a binary alloy with occupation variables $\sigma_i \in \{-1,+1\}$, this target becomes:
\begin{equation}
\label{eqn:sqs_correlation_function_binary_target}
\overline{\Pi}_{\omega}^{\,\mathrm{rand}} = (2x-1)^{|\omega|}
\end{equation}
where $x$ denotes the concentration of the $+1$ species, and $|\omega|$ represents the number of sites in the cluster. At equimolar composition ($x = 0.5$), a perfectly random binary alloy exhibits zero average correlations for all single-site clusters, pairs, triplets, and higher-order clusters.\cite{Zunger1990,Wei1990}

Viewed this way, the SQS construction becomes a finite-cell optimization problem. Among all configurations compatible with a chosen composition and supercell size, the algorithm seeks the structure whose correlation vector is closest to the target vector over a selected set of cluster orbits $\mathcal{O}$. A common objective function ($\mathcal{E}_{\mathrm{SQS}}(\boldsymbol{\sigma})$) for this optimization task is:
\begin{equation}
\label{eqn:sqs_objective_function}
\mathcal{E}_{\mathrm{SQS}}(\boldsymbol{\sigma})
=
\sum_{\omega \in \mathcal{O}} w_{\omega}
\left[\Pi_{\omega}(\boldsymbol{\sigma}) - \overline{\Pi}_{\omega}^{\,\mathrm{target}} \right]^2
\end{equation}
where $w_{\omega}$ is the weight assigned to orbit $\omega$, and $\overline{\Pi}_{\omega}^{\,\mathrm{target}}$ shows the target correlation for the orbit. This weighted squared-deviation form in Eq. \eqref{eqn:sqs_objective_function} offers a convenient generic representation of the underlying optimization task to capture the correlation-matching spirit of SQS,\cite{Zunger1990,Wei1990} although many recent implementations have introduced alternative objective functions, including criteria that focus on either prioritizing exact matching of the maximum number of target correlations \cite{vandeWalle2013} or explicitly balancing correlation accuracy against DFT cost.\cite{Kadzielawa2026} Thus, while SQS borrows the correlation language of cluster expansion, it does not fit interaction coefficients or construct an effective Hamiltonian. Instead, the correlations are used only to build representative structures. Hence, SQS is not random in a literal sense, but rather the periodic configuration whose selected low-order statistics best approximate the target disordered ensemble within a finite supercell.

The main appeal of SQS is that it converts disorder into one or a few explicit supercells that can be used directly in first-principles DFT simulations. As SQS supercells can be relaxed, local lattice distortions and charge redistribution can emerge naturally,\cite{Jiang2004,Shin2006} making SQS much more realistic than averaged-medium approaches for various structural and electronic problems. This practical advantage explains why SQS has been applied successfully across a massive range of materials systems. The correlation matching also allows SQS to outperform an arbitrary random supercell of the same size, as it reduces finite-size artifacts that may otherwise arise from small periodic cells. Classical work has shown the usefulness of SQS for metallic alloys\cite{Zunger1990,Wei1990,Jiang2004,Shin2006} and ceramic solid solutions.\cite{Jiang2016a} More recent studies on high-entropy alloys continue to treat SQS as one of the most practical DFT-ready structural models, particularly when the key objective is to capture local-environment effects, rather than exact finite-temperature thermodynamics.\cite{Gao2016,Gehringer2023,Lian2025} Moreover, the broad use of SQS has been supported by steady advances in structure generation. Modern implementations typically search in the discrete configurational space using stochastic optimization, systematic enumeration, or hybrids of the two. A practical consequence is that, as the number of components and relevant pair or multi-site correlations increases, accurate SQS models require larger supercells, which raises DFT cost. A widely used implementation is the \texttt{mcsqs} package\cite{vandeWalle2013} within the alloy theoretic automated toolkit (ATAT),\cite{vandeWalle2002a} which can search for supercells that exactly match as many target correlations as possible, while optimizing the shape and atomic occupations of these supercells using a well-established multi-sublattice framework. Representative modern toolchains also include newer ecosystems, such as the integrated cluster expansion toolkit (ICET),\cite{Angqvist2019} as well as reproducible workflow layers (e.g., SimplySQS).\cite{Lebeda2026} With tools such as ATAT and ICET, SQS has become a practical and broadly used approximation for assessing the phase stability and functional properties of chemically disordered solids.\cite{Liu2025b,Wang2025a}

At the same time, the same simplification that offers SQS its practical power also defines its limitations. As one or a few supercells are used to replace an entire ensemble, conventional SQS does not by itself provide configurational entropy, temperature-dependent short-range order, or the distribution of fluctuations across many microstates.\cite{Schuler2024,Li2024a} SQS is therefore best viewed as a practical structural approximation to a chosen disordered state, rather than as a complete theory of disorder. Even when the target disordered state is fully random, finite-size convergence can remain nontrivial in heterovalent systems. For MgAl\textsubscript{2}O\textsubscript{4} and ZnSnP\textsubscript{2}, linear extrapolation to the perfectly disordered limit gives excess energies of $0.15$ and $0.21$ eV/cation, respectively, about $0.02$ and $0.05$ eV/cation above the corresponding values obtained from the largest ($6\times6\times6$) SQS cells.\cite{Seko2015} Thus, SQS works best when the physically relevant disordered state is close to the ideal random limit. The reliability of SQS decreases significantly when this ideal random state is no longer the physically relevant target, especially in systems where long-range electrostatics or strong local chemical preferences drive substantial short-range order.\cite{Seko2015} For instance, related difficulties have been increasingly observed in cation-disordered rocksalt materials, where local ordering, electrostatics, and relaxation effects are strongly coupled.\cite{Szymanski2023} Such limitations motivate future quasi-random approximations to target non-random local correlations more explicitly.

\paragraph{Relevant quasi-random approximations.}
The SQS idea can be extended naturally when the physically relevant disordered state is not perfectly random. Instead of matching the correlation functions of an ideal random alloy, one can construct a representative supercell that matches the correlation functions of a short-range-ordered state. This task can be achieved using the special quasi-ordered structure (SQoS) approach, where the short-range-order correlations are obtained from Monte Carlo simulations on a DFT-based cluster expansion model, and the SQoS method further constructs DFT-manageable supercells whose correlations approximate these ensemble-averaged short-range-order states.\cite{Liu2016} More generally, this example shows that cluster expansion and Monte Carlo can be used to determine the equilibrium correlation pattern, whereas a short-range-order-targeted representative structure (such as SQoS) provides one explicit configuration that approximates it for detailed first-principles calculations. Furthermore, a related but distinct framework is the supercell random approximate (SCRAP), introduced for accelerated modeling of high-entropy alloys.\cite{Singh2021} Similar to SQS, SCRAP replaces the full configurational ensemble by an optimized explicit supercell. However, this new method is formulated around target one-site and pair probabilities and uses a hybrid cuckoo-search strategy to explore large compositional and ordering spaces efficiently (Fig. \ref{fig:conventional_method_demonstration}c). By design, SCRAP can target non-random short-range order as well as the fully random limit, so it is best regarded as a broader relative of SQS, rather than simply another implementation. Broadly speaking, these methods tackle the key weakness of conventional SQS. Their limitation, however, is that the target correlations must come from elsewhere, e.g., an experiment or another thermodynamic model, so they do not actually remove the need for statistical mechanics when the goal is to determine the equilibrium state itself.

Taken together, these quasi-random approximations occupy an intermediate regime between averaged-medium descriptions and full statistical-mechanics sampling. Traditional SQS is most effective when the relevant disordered states are close to the ideal random limit, while beyond-random variants such as SQoS and SCRAP extend the same representative-supercell concept to systems with significant short-range order. Their main advantage is that they retain explicit local environments in a DFT-manageable form, whereas their main limitation is that they approximate a chosen disorder state rather than derive the full equilibrium ensemble. In addition, as they are ultimately static approximations, finite-temperature properties still require additional treatment, e.g., via phonon calculations, molecular dynamics, or Monte Carlo sampling built on top of the representative structures. For these reasons, these static quasi-random structural approximations are generally most valuable as baseline models of random disorder, representative structures for first-principles calculations, or inputs to more complete finite-temperature frameworks.

\phantomsection
\addcontentsline{toc}{subsection}{2.4 Ensemble-based microstate averages}
\subsection*{2.4 Ensemble-based microstate averages}

Ensemble-based microstate-average methods describe a chemically disordered crystal as a finite statistical mixture of various explicit ordered configurations, which makes them fundamentally distinct from the three strategies discussed above. In effective-medium and mean-field methods, disorder is replaced by an averaged environment, so specific local arrangements are not modeled explicitly. In cluster expansion, the core object is a fitted effective Hamiltonian, which is further sampled across a large configurational space. In SQS, disorder is approximated by one or a few specially designed supercells that reproduce selected correlation functions of random alloys. In contrast, ensemble-based methods begin by choosing a finite set of explicit microstates that all satisfy the target composition and parent lattice, and then recover observables by averaging over this set. The central approximation is thus that, instead of treating the full (and usually massive) set of possible local atomic arrangements, these ensemble-based approaches average over only a computationally manageable subset of representative configurations (Fig. \ref{fig:conventional_method_comparison}d). In this context, they can be viewed as a middle ground between SQS and cluster expansion: they retain explicit atomistic configurations like SQS, but estimate properties from statistical averages over multiple configurations, as in cluster-expansion-based statistical mechanics.

\paragraph{General framework.}
The formalism is straightforward. Let $s_i$ denote microstate $i$, meaning one fully occupied ordered configuration consistent with the target composition. Moreover, let $g_i$ be the configurational degeneracy of such a microstate. $F_i(T)$ is its free energy at temperature $T$. Then, the probability of microstate $i$ within the chosen ensemble ($P_i(T)$) is:
\begin{equation}
\label{eqn:finite_ensemble_probability}
P_i(T) = \frac{g_i \exp[-\beta F_i(T)]}{\sum_j g_j \exp[-\beta F_j(T)]}
\end{equation}
where $\beta = 1/(k_{\mathrm{B}}T)$, $k_{\mathrm{B}}$ is the Boltzmann constant, and the denominator is the partition function of the finite ensemble. Any observable $A$ can then be estimated as:
\begin{equation}
\label{eqn:finite_ensemble_average}
\langle A \rangle_T = \sum_i P_i(T) A_i
\end{equation}
where $A_i$ represents the value of this observable for microstate $i$. In the simplest case, $F_i(T)$ is approximated by the relaxed DFT energy (or enthalpy) of this microstate, although vibrational, magnetic, or electronic contributions can be further added when necessary. An early conceptual foundation for this category of methods rests on formulating symmetry-adapted configurational modeling in a reduced space of independent site-occupancy configurations, while keeping their degeneracy explicit.\cite{GrauCrespo2007,Hart2008} In practice, the central task becomes how to construct the finite set of microstates in a way that is both computationally feasible and physically representative. Hence, the main approximation in these ensemble-based microstate-average methods does not lie in the statistical-mechanics formalism itself, but in replacing the immense number of possible atomic arrangements with a manageable finite ensemble of representative microstates.

\begin{figure}
\phantomsection
\begin{center}
\includegraphics[max size={\textwidth}{\textheight}]{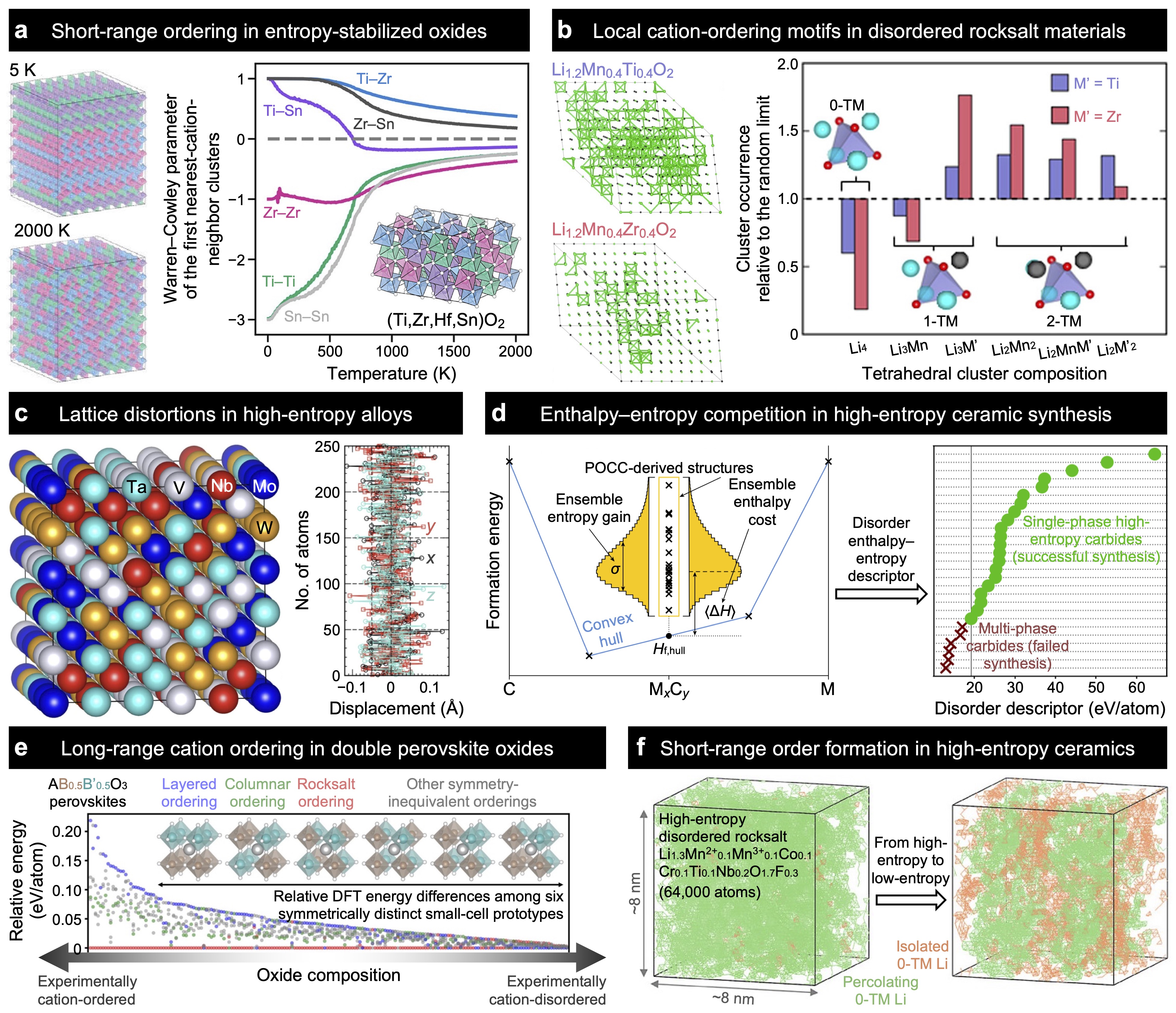}
\caption{
\textbf{Representative applications of conventional first-principles methods for modeling chemical disorder.}
(a) Cluster expansion--Monte Carlo can reveal persistent chemical short-range ordering in entropy-stabilized oxides at elevated temperatures.\cite{Aamlid2024}
(b) In disordered rocksalts, cluster-expansion-based Monte Carlo simulations quantify local motifs that can affect Li transport, where 0-transition-metal-neighbored (0-TM) tetrahedral channels for facile Li ion migration are highlighted by green bonds.\cite{Ji2019}
(c) A SCRAP-optimized 250-atom supercell for an equimolar TaNbMoWV random solid solution shows no short-range order in the first three nearest-neighbor shells, exhibiting local heterogeneity in lattice distortions.\cite{Singh2021}
(d) A finite set of POCC-derived ordered crystal structures can give rise to a distribution of formation energies around the convex hull, from which a disorder enthalpy--entropy descriptor can be constructed to rationalize and screen whether multicomponent ceramic syntheses are likely to lead to single-phase high-entropy products or instead decompose into multi-phase mixtures.\cite{Divilov2024}
(e) A compact finite ensemble of six B-site ordering prototypes is sufficient to predict experimental long-range cation order vs. disorder in A\textsubscript{2}BB'O\textsubscript{6} double perovskite oxides across a massive chemical space.\cite{Peng2024a}
(f) Stochastic relaxation-swap sampling drives initially random high-entropy ceramic structures toward lower-energy short-range-ordered states, leading to the emergence of critical isolated 0-TM channels that can significantly influence long-range Li ion transport.\cite{Liao2025}
Panel (a) adapted with permission from ref.\cite{Aamlid2024}, copyrighted by the American Chemical Society. Panels (b), (d), and (e) adapted from ref.\cite{Ji2019}, ref.\cite{Divilov2024}, and ref.\cite{Peng2024a}, respectively, CC BY 4.0. Panel (c) adapted with permission from ref.\cite{Singh2021}, copyrighted by Springer Nature. Panel (f) adapted with permission from ref.\cite{Liao2025}, copyrighted by John Wiley and Sons.
}
\label{fig:conventional_method_demonstration}
\end{center}
\end{figure}

\paragraph{Partial OCCupation (POCC) formalism.}
The best-known realization of the finite-ensemble microstate-average concept is the POCC approach\cite{Yang2016} used in AFLOW high-throughput atomistic modeling workflows.\cite{Divilov2025,Oses2023} With this method, a partially occupied structure is mapped to a set of ordered derivative supercells that reproduce the target composition as closely as possible within a chosen finite cell size, following the broader logic of combinatorial supercell construction for atomic substitutions.\cite{Yang2016,Okhotnikov2016} Symmetry is further used to identify the distinct microstates, and only symmetry-inequivalent configurations are retained explicitly, with multiplicities restored via the degeneracy factors $g_i$. After the energies and other properties of these microstates are computed, ensemble averages are obtained from their statistical weights. Representative applications have shown that POCC can be applied not only to describe partial occupancy itself, but also to derive disorder-aware descriptors for driving the discovery of chemically complex solids (Fig. \ref{fig:conventional_method_demonstration}d). For instance, POCC configurational energy distributions from ordered microstates have been used to uncover synthesizable single-phase high-entropy carbides\cite{Sarker2018} and later extended into disordered enthalpy--entropy descriptors to facilitate broader high-entropy ceramics discovery.\cite{Divilov2024}

This explicit-supercell strategy is powerful because it retains real-space local environments and enables direct averaging of diverse properties, but it is also costly because many microstates must be computed explicitly. Every member of the ensemble is a standard crystal structure with full site occupancies. Thus, standard first-principles workflows can be applied directly without modifying the underlying electronic-structure method. As each microstate can also be computed independently, these workflows are naturally parallel. This makes the method especially useful when one wants to evaluate ensemble-averaged properties beyond total energies, such as lattice dynamics,\cite{Esters2021} optical response,\cite{Calzolari2022} and thermal expansion behaviors.\cite{Esters2023} Notably, a useful way to view its relation to SQS is that an SQS can be regarded as one member of a larger configurational ensemble. SQS is designed to mimic the random limit with one carefully chosen structure, while ensemble averaging allows the relative importance of different atomic configurations to change with temperature via their Boltzmann weights. This finite-temperature flexibility is particularly essential in chemically complex ceramics, where disorder can strongly affect phase stability and vibrational properties.\cite{Toher2022} Nonetheless, the trade-off is that the number of distinct configurations that are expected to be computed grows rapidly with cell size and chemical complexity. Hence, the POCC approach remains constrained by finite supercell size and computational cost.

\paragraph{Relevant finite-ensemble approaches.}
Recent studies have further broadened the category of finite-ensemble microstate-average approaches beyond canonical POCC, while maintaining the same central idea: a chemically disordered crystal can be approximated as a weighted collection of explicit microstates. Some methods replace exhaustive enumeration with random or targeted sampling, which can make microstate averaging more practical for chemically complex systems while preserving this ensemble-averaging logic.\cite{Jiang2016b,Sorkin2021,Kristoffersen2022,Novick2023,Kuner2024} Ensemble-based averaging has also been leveraged to reproduce local-environment-sensitive spectroscopic observables, where signals of nuclear magnetic resonance (NMR) are modeled by adding the contributions from many ordered microstates.\cite{Moran2019} Moreover, a particularly succinct example has been shown for multicomponent double perovskite oxides (Fig. \ref{fig:conventional_method_demonstration}e), where the distribution of the DFT-derived energies of only six symmetrically distinct small-cell cation-ordered prototypes is sufficient to accurately predict the experimental likelihood of long-range B-site cation disorder across a massive compositional space.\cite{Peng2024a} Taken together, these examples highlight the vast diversity of finite-ensemble strategies that complement POCC for different materials classes, observables, and computational scales.

Overall, these microstate-average approaches occupy a valuable middle ground in chemical disorder modeling. They are more physically explicit than mean-field methods as they examine local environments directly, more flexible than SQS as they average over multiple configurations rather than just one, and typically simpler than cluster expansion as they avoid fitting an effective Hamiltonian. Their main strength is generality: in principle, any property that can be computed for an ordered microstate can be averaged across the ensemble afterward. Nonetheless, they are limited by combinatorial growth: as the number of components, partially occupied sublattices, or relevant correlation lengths increases, the explicit ensemble needs to be truncated or sampled approximately, and the final result depends on whether the retained microstates still capture the key local physics of chemical disorder. Thus, these methods are often best viewed as controlled finite-ensemble approximations, rather than exact descriptions of the thermodynamic limit.

\phantomsection
\addcontentsline{toc}{subsection}{2.5 Stochastic sampling and molecular dynamics}
\subsection*{2.5 Stochastic sampling and molecular dynamics}

Stochastic sampling and molecular-dynamics-based methods become particularly helpful when chemical disorder cannot be described adequately by any single one of the atomistic simulation frameworks discussed earlier (Table \ref{table:comparing_conventional_methods}). Notably, in cluster expansion, Monte Carlo mainly acts as a statistical-mechanics solver for a pre-fitted lattice-model Hamiltonian. Additionally, quasi-random structural approximations (e.g., SQS) work by compressing complex chemical disorder into one or a few representative supercells, while ensemble-based methods (e.g., POCC) instead average over the full set of symmetry-distinct microstates within a limited cell size. By contrast, the present category of methods explores the low-energy region of the configurational potential energy surface (PES) more directly through stochastic moves or dynamical trajectories, without requiring either a pre-fitted lattice model, representative supercell configurations, or exhaustive enumeration of the entire small-cell ensemble. This perspective is especially useful in modeling chemically disordered materials showing large local structural relaxation, off-lattice distortions, or many nearly degenerate configurations, where the statistically important atomic structures are often best viewed as a subset of low-energy basins on a rough PES landscape (Fig. \ref{fig:conventional_method_comparison}e). For this reason, purely random configurational sampling is usually inefficient for chemically disordered materials. The number of possible occupational arrangements grows combinatorially, while the statistically essential low-energy configurations occupy only a small fraction of the overall PES, particularly when local relaxation is strong. Random structures are thus usually more helpful as initial seeds or reference configurations than as the primary means of sampling disorder.

\paragraph{Direct Monte Carlo sampling.}
Direct sampling using Monte Carlo is a natural starting point, as it generates a Markov chain whose long-time distribution approaches the target ensemble. In the standard Metropolis scheme, a trial configuration is accepted with probability ($P_{\mathrm{acc}}$):
\begin{equation}
\label{eqn:metropolis_algorithm}
P_{\mathrm{acc}} = \min\left[1,\exp(-\beta \Delta E)\right]
\end{equation}
where $\beta = 1/(k_{\mathrm{B}}T)$, $k_{\mathrm{B}}$ is the Boltzmann constant, $T$ is the temperature, and $\Delta E = E_{\mathrm{new}} - E_{\mathrm{old}}$ is the energy change between the trial and current states. For chemically disordered solids, this trial move may involve a species swap, vacancy motion, a change in site occupancy, or a coupled occupational--displacive update.\cite{Santodonato2018,Shi2024,Roy2024,Tang2025} For example, DFT-based Metropolis Monte Carlo has been combined with full structural relaxation after each swap to reveal pronounced short-range order in SiSn alloys.\cite{Jin2021} Moreover, a Monte Carlo framework with selective atom swapping and repeated relaxation has been used to predict the order-to-disorder transitions in multicomponent MXenes.\cite{Oyeniran2026} The main benefit of this approach is that it targets equilibrium statistics without first reducing the problem to a fitted effective Hamiltonian. However, this method can be inefficient when each atomic move must be evaluated on a rugged, continuously relaxing PES.

\paragraph{Enhanced Monte Carlo strategies.}
Various enhanced Monte Carlo methods can improve on plain Metropolis sampling by helping the system cross barriers more efficiently. As an example, simulated annealing gradually lowers the temperature parameter in Monte Carlo simulations so that the search first explores broadly and then concentrates on lower-energy regions.\cite{Kirkpatrick1983} Replica-exchange Monte Carlo (i.e., parallel tempering) runs multiple replicas at different temperatures and swaps them occasionally, which allows low-temperature trajectories to benefit from barrier crossing at higher temperatures.\cite{Hukushima1996,Earl2005,Kasamatsu2019} These approaches remain close to Boltzmann sampling, but they differ in cost: simulated annealing is simple and widely used, whereas replica exchange is better suited to rugged PES landscapes at the price of running multiple replicas.

\paragraph{Hybrid Monte Carlo--molecular dynamics methods.}
Molecular dynamics can complement Monte Carlo by evolving atoms continuously on their PES landscape and thus capturing thermal motion, anharmonicity, and local structural relaxation naturally. This approach is essential when the energy depends not only on what species occupy which sites, but equally importantly on how the surrounding lattice responds. Nevertheless, direct molecular dynamics is usually inefficient for equilibrating substitutional disorder because changes in site occupancy are rare on accessible simulation time scales.\cite{Zhu2026a} Fortunately, hybrid Monte Carlo--molecular dynamics methods tackle this mismatch by coupling discrete chemical updates with continuous relaxation, e.g., by adding short molecular dynamics segments between Monte Carlo moves.\cite{Neyts2012} This coupling is physically well motivated as occupational variables are typically slow and activated, while local positional relaxation is much faster, enabling Monte Carlo to update the chemical states of atomic motifs, whereas molecular dynamics can re-equilibrate the surrounding structural basin. For instance, a hybrid Monte Carlo--molecular dynamics approach has been developed to examine how boron and carbon interstitials drive distinct short-range-ordered motifs in Inconel-type superalloys.\cite{Dolezal2025} Additionally, combined Monte Carlo and molecular dynamics simulations of CoCuFeNiPd and CoCuFeNiTi have further traced atomic segregation, short-range ordering, and cluster formation in high-entropy alloys to chemical-affinity disparity and exclusivity among the constituents.\cite{Chen2021c} In the most rigorous formulation, a trajectory of molecular dynamics can be used as a proposal move and coupled with a Metropolis correction to recover the target ensemble exactly, whereas many materials-oriented implementations instead apply short segments of molecular dynamics more pragmatically to relax atomic positions between occupational updates using Monte Carlo. These schemes are well-suited to disordered materials in which the energetic cost of a chemical rearrangement depends strongly on local distortion, including systems with large size mismatch, heterovalent substitution, or soft structural modes. However, because each occupational update must still be coupled to explicit dynamical relaxation, hybrid Monte Carlo--molecular dynamics schemes can remain computationally expensive and may still struggle to navigate highly rugged PES landscapes efficiently, motivating the more aggressive search strategies discussed next.

\paragraph{Relaxation-swap methods.}
Relaxation-swap approaches can push the logic of hybrid Monte Carlo--molecular dynamics further through replacing finite-temperature segments of molecular dynamics simulations with direct structural relaxation of each altered configuration.\cite{Xu2025a} Starting from an occupational state $\boldsymbol{\sigma}_{n-1}^{*}$ and relaxed coordinates $\mathbf{R}_{n-1}^{*}$, such approaches propose a new occupational state $\boldsymbol{\sigma}_{n}^{\prime}$ and obtain the corresponding relaxed structure ($\mathbf{R}_{n}^{\prime}$) from:
\begin{equation}
\label{eqn:relaxation_swap_proposal}
\mathbf{R}_{n}^{\prime} = \operatorname*{argmin}_{\mathbf{R}} U(\boldsymbol{\sigma}_{n}^{\prime},\mathbf{R})
\end{equation}
where $U(\boldsymbol{\sigma},\mathbf{R})$ represents the potential energy as a function of occupational state $\boldsymbol{\sigma}$ and atomic coordinates $\mathbf{R}$. The move is then evaluated using the relaxed energy difference ($\Delta E_{n}$):
\begin{equation}
\label{eqn:relaxation_swap_evaluation}
\Delta E_{n} = U(\boldsymbol{\sigma}_{n}^{\prime},\mathbf{R}_{n}^{\prime}) - U(\boldsymbol{\sigma}_{n-1}^{*},\mathbf{R}_{n-1}^{*})
\end{equation}
This move can be further accepted or rejected by designing a criterion based on $\Delta E_n$. Notably, this strategy can be particularly beneficial when local relaxation contributes substantially to the energetics of chemical disorder. By relaxing altered configurations before comparing energies, these methods can avoid the low acceptance rates that can arise if strained, unrelaxed structures are evaluated directly. Crucially, a practical modern example is the short-range order swapping method, which uses descriptor-guided atom swaps, feature selection, and Bayesian optimization to construct atomistic structural models efficiently from initially random crystal lattices, thereby providing a strongly practical heuristic alternative to costly explicit thermodynamic sampling\cite{Liao2025} (Fig. \ref{fig:conventional_method_demonstration}f). Nevertheless, the thermodynamic meaning of these relaxation-swap strategies should be stated carefully: they sample relaxed inherent structures, rather than exploring the full finite-temperature vibrational ensemble. Therefore, the central value of such relaxation-swap methods is in identifying statistically important low-energy local structural motifs efficiently.

\paragraph{Basin-hopping and minima-hopping methods.}
These two types of hopping methods should be better regarded as landscape-exploration tools than as exact ensemble samplers. Specifically, basin-hopping perturbs the lattice structure and then immediately relaxes it, converting the PES into a network of local minima and jumps between basins.\cite{Wales1997} As an example, a basin-hopping scheme for multi-cation perovskites has been established to identify low-energy cation-ordering patterns more efficiently than cluster-expansion-based modeling.\cite{Zhang2024a} Moreover, minima-hopping combines short escape trajectories from molecular dynamics with local relaxation to search for new low-energy minima efficiently.\cite{Goedecker2004, Talapatra2015} These methods are especially helpful when positional relaxation is as important as site occupancy, as they do not require a fixed-lattice representation. However, because these methods prioritize efficient discovery of low-energy minima rather than rigorous thermodynamic weighting, they do not directly result in equilibrium ensemble averages and can also be too sensitive to the design of the perturbation or escape moves.

\paragraph{Genetic and evolutionary algorithms.}
Genetic algorithms and evolutionary methods address the same low-energy search problem from a population perspective by evolving many candidate configurations simultaneously. Candidate structures are selected according to a fitness measure and combined via crossover and mutation to produce new populations.\cite{Le2016} For instance, a genetic algorithm has been used to predict stable chemical orderings in bimetallic nanoparticles across a broad size, shape, and compositional space of FePt nanoparticles.\cite{Dean2020} The primary advantage of these approaches is that they can assess multiple competing low-energy basins at once, which is helpful for chemically disordered materials with many nearly degenerate occupational patterns. Symmetry-adapted crossover schemes and more recent hybrids that incorporate Metropolis-like acceptance rules have further improved their efficiency and statistical relevance.\cite{Mohn2009,Anand2023} However, these algorithms can be computationally demanding because many candidate structures must be evaluated simultaneously. In addition, the convergence of evolutionary approaches may depend sensitively on their representation, fitness function, and crossover or mutation operators.

\paragraph{Kinetic and force-bias Monte Carlo.}
The sampling and search strategies above mainly target equilibrium ensembles or low-energy configurations, whereas chemical disorder can also evolve kinetically during processing. Kinetic Monte Carlo explores the configurational space through discrete activated events, e.g., vacancy-mediated site exchanges, using transition rates to access diffusion and ordering over time scales beyond conventional molecular dynamics.\cite{vanderVen2018,Voter2007} As an example, kinetic Monte Carlo has been used to model the formation kinetics of chemical short-range order in multi-principal element alloys and build time--temperature--ordering relations.\cite{Shen2021} Chemical ordering can in turn substantially influence diffusion and couple the evolving disorder state to kinetic transport.\cite{Xu2023,Li2024} The predictive accuracy of kinetic Monte Carlo depends on the completeness of the underlying atomistic event space and the accuracy of the associated barriers and rate prefactors. Time-stamped force-bias Monte Carlo instead uses force-biased continuous atomic displacements to connect the Monte Carlo evolution with an effective time scale, offering another route to boost activated relaxation and solid-state diffusion without a predefined discrete event catalog.\cite{Mees2012} Nevertheless, the effective time in this method does not necessarily correspond exactly to real physical time. Therefore, the predicted kinetic time scales should be interpreted cautiously and, where possible, benchmarked against molecular dynamics or experiment.\cite{Bal2014}

Overall, these atomistic modeling approaches span a continuum from equilibrium sampling to PES exploration and to kinetic evolution. Direct and replica-exchange Monte Carlo, as well as hybrid Monte Carlo--molecular dynamics methods, remain closest to statistical mechanics in the strict sense, whereas relaxation-swap, hopping, and genetic algorithms are most effective as tools for locating the low-energy configurations that dominate disordered material behaviors. In practice, their usefulness depends strongly on access to fast, accurate energy prediction models, which is why they have become substantially more powerful when coupled with modern MLIPs and relevant surrogate Hamiltonians.\cite{Sivak2025,Zhong2025a,Guan2025,Fang2025} Their practical reach has been expanded further by large-scale parallel computing and implementations with graphics processing units (GPUs). For example, the recently proposed SMC-X method exploits generalized checkerboard updates on accelerator hardware to push Monte Carlo on chemically complex metallic alloys to billion-atom scales.\cite{Liu2025c,Liu2025d} Moreover, PyHEA combines GPU-accelerated Monte Carlo with global and local search strategies to boost short-range-order simulations in high-entropy alloys.\cite{Niu2025} Looking further ahead, although quantum computing remains at an early stage, recent work suggests that quantum annealing may offer a complementary route for sampling thermodynamically relevant low-energy atomic configurations within chemically disordered materials.\cite{Camino2025}

\phantomsection
\addcontentsline{toc}{subsection}{2.6 Macroscopic thermodynamic models}
\subsection*{2.6 Macroscopic thermodynamic models}

In distinction to atomistic simulation approaches, macroscopic thermodynamic models describe chemical disorder at the macroscopic phase level rather than with explicit atomic configurations. Instead of enumerating countless microstates and averaging over them, these methods represent each disordered phase directly by its free energy (or other important thermodynamic properties) as a function of temperature, pressure, and composition. Therefore, they are best leveraged as a phase-level layer on top of atomistic disorder modeling, integrating thermodynamic information from atomistic modeling and experiments to predict phase boundaries, chemical potentials, and synthesis windows, rather than resolving the exact atomic motifs within a disordered structure.

\paragraph{Simplified thermodynamic screening descriptors.}
The simplest macroscopic approach is to leverage the same enthalpy--entropy competition introduced earlier in Eqs. \eqref{eqn:configurational_entropy}--\eqref{eqn:enthalpy_of_mixing}, but apply it mainly as a screening tool rather than as a full phase-equilibria model. In this picture, disorder is favored when the entropic stabilization is large enough to offset the enthalpic penalty of mixing. Typical implementations fall into two related forms. Some retain an explicit but simplified free-energy model by coupling ideal configurational entropy with approximate mixing enthalpies to screen random solid solutions against competing ordered phases. As an example, DFT-derived binary interaction parameters have been combined with the regular solution model to unveil new compositions likely to stabilize single-phase high-entropy alloys across the periodic table.\cite{Chen2023c} In addition, other approaches move one step further toward large-scale screening by replacing one or both thermodynamic terms with more empirical but still physically motivated descriptors for order--disorder tendency or synthesizability.\cite{Wyatt2025,Dey2024,Almishal2025} Related descriptor ideas can also be extracted from explicit microstate ensembles, providing a useful bridge to the ensemble-based approaches discussed earlier, e.g., for driving computational discovery of high-entropy ceramics.\cite{Sarker2018,Divilov2024}

These simplified methods are attractive because their speed, interpretability, and ease of use make them powerful tools for screening broad disorder trends across large chemical spaces, but these simplifications are also their central weaknesses because they limit thermodynamic rigor. By relying on a small number of physically meaningful quantities, they can quickly distinguish general tendencies toward random solid solutions, ordered compounds, or phase separation and are therefore especially helpful for high-throughput studies of high-entropy alloys and ceramics without requiring a full thermodynamic database.\cite{Wyatt2025,Dey2024,Almishal2025,Chen2023c} Nonetheless, since they are not full free-energy models, they often rely on ideal (or near-ideal) entropy expressions, coarse enthalpy estimates, or empirical decision boundaries. Furthermore, they usually do not enforce full phase coexistence, mass balance across multiple competing phases, or detailed sublattice chemistries. Thus, they are fast screening tools, rather than full substitutes for phase-equilibria modeling.

\paragraph{CALculation of PHAse Diagrams (CALPHAD).}
CALPHAD\cite{Kaufman1970} has long been known as an established macroscopic framework for modeling chemical disorder. In CALPHAD, each phase is assigned a parameterized Gibbs free-energy model whose form describes the relevant physics of that phase, such as substitution on one or more sublattices, vacancies, magnetic ordering, and non-stoichiometry. At fixed temperature $T$, pressure $p$, and overall composition, equilibrium is obtained by minimizing the total Gibbs free energy of the multicomponent system ($G_{\mathrm{sys}}$):
\begin{equation}
\label{eqn:calphad}
G_{\mathrm{sys}} = \sum_{\phi} n_{\phi} G^{\phi}(T,p,\{y_i^{\phi}\})
\end{equation}
where $\phi$ labels the phases present, $n_{\phi}$ denotes the amount of phase $\phi$, $G^{\phi}$ shows the molar Gibbs free energy of phase $\phi$, and $y_i^{\phi}$ is the composition or site-fraction variable in this phase model. For substitutional solutions, CALPHAD follows exactly the same free-energy logic as Eq. \eqref{eqn:free_energy_of_mixing}, but it further extends it into a flexible phase-modeling framework.\cite{Lukas2007} In practice, the free energy of each phase is described as a sum of reference-state terms, configurational entropy terms, and non-ideal interaction terms, often using Redlich--Kister expressions for substitutional solutions and compound-energy or sublattice formalisms for ordered, ionic, or non-stoichiometric phases. The model parameters are further examined against diverse inputs, including phase boundaries, activities, calorimetry, order--disorder transition temperatures, and selected first-principles data. During database assessment, these parameters are optimized such that a single set of free-energy functions is able to reproduce the heterogeneous inputs as consistently as possible. At the same time, this assessment preserves the topology of binary, ternary, and higher-order phase relations. For instance, a CALPHAD sublattice description of the Al--Ni system has been created to model its A1-to-L1\textsubscript{2} order--disorder transition via an entirely phase-level, non-atomistic framework.\cite{Ansara1997} Thus, CALPHAD differs formally from atomistic modeling methods, such as cluster expansion, SQS, and Monte Carlo sampling. CALPHAD is a top-down description that starts from phase-level free-energy functions and determines equilibrium by thermodynamic minimization, while atomistic methods are bottom-up descriptions that start from explicit configurations and recover thermodynamics of chemical disorder by statistical averaging over microscopic states.

The crucial strength of conventional CALPHAD is that it converts diverse experimental and computational inputs into a thermodynamically self-consistent free-energy framework that can be applied efficiently, but this coarse-grained treatment is also its key limitation, as it sacrifices direct microscopic details.\cite{Lukas2007} Once a database has been built, CALPHAD can rapidly compute phase diagrams and fractions, chemical potentials, partitioning trends, and their thermodynamic driving forces across broad compositional and temperature spaces, which is why it has become a dominant tool for developing multicomponent alloys and ceramics with well-defined lattices. CALPHAD can also incorporate magnetic, vibrational, and electronic free-energy contributions via semi-empirical but thermodynamically consistent expressions, which are often cumbersome to assess in traditional atomistic workflows, e.g., cluster-expansion-based statistical mechanics. Its flexibility is further showcased by compound-energy and sublattice formalisms, which allow partial ordering, vacancies, non-stoichiometry, and substitutions to be described within a unified Gibbs free-energy framework. However, because conventional CALPHAD represents disorder mainly at the phase and mean-field levels, it cannot explicitly resolve local atomic arrangements, short-range order, or site-specific electronic structure. Its interaction parameters should thus be interpreted as effective thermodynamic descriptors, rather than direct microscopic interactions. As a result, the predictive reliability of pure CALPHAD methods depends heavily on the quality and transferability of the underlying assessment, particularly in chemically complex systems or poorly characterized compositional spaces. In addition, database development is labor-intensive as the phase models must be selected carefully, while assessments across binaries, ternaries, and higher-order systems must remain mutually consistent. Overall, these strengths and weaknesses are exactly why CALPHAD remains indispensable for efficient phase-level prediction, yet also why it needs to be extended to further recover the missing local configurational physics.

Recent developments have extended CALPHAD in two critical directions. Specifically, one direction boosts CALPHAD by incorporating more realistic configurational statistics, allowing short-range ordering to be treated more explicitly than in conventional mean-field descriptions. For example, the cluster variation method has recently been combined with CALPHAD and the Fowler--Yang--Li transform to enable a more practical thermodynamic framework with intrinsic chemical short-range order.\cite{Fu2024} The other direction couples CALPHAD with atomistic modeling of chemical disorder, where atomistic simulations offer mixing enthalpies, ordering tendencies, and finite-temperature corrections, while CALPHAD further translates these inputs into phase-diagram-ready thermodynamic functions. Modern tools, such as ATAT,\cite{vandeWalle2002a} offer a practical route for mapping first-principles calculations of atomistic energies onto CALPHAD thermodynamic models in high-throughput workflows.\cite{vandeWalle2017} In addition, for high-entropy alloys, cluster expansion has been coupled with a mean-field statistical-mechanics model and DFT energies to infer solid-solution transition temperatures,\cite{Lederer2018} highlighting how atomistically assessed disorder energetics can be transformed into CALPHAD-driven hierarchical screening of multicomponent materials. Taken together, these developments further establish CALPHAD as an integration layer between microscopic energetics and engineering-scale thermodynamic prediction---an essential role that is becoming even more important, as MLIPs and other AI tools are increasingly being leveraged as fast upstream engines for thermodynamic assessments of chemical disorder.\cite{Shen2025,Zhu2025a,Zhu2025b,Kunselman2025}

\phantomsection
\addcontentsline{toc}{subsection}{2.7 Practical guidance for method selection}
\subsection*{2.7 Practical guidance for method selection}

Taken together, these methods discussed above should be viewed as complementary approaches for different disorder regimes and scientific targets. Hence, method selection should begin with the quantity to be predicted and the thermodynamic regime of interest, e.g., whether the goal is a static property of a prescribed disordered state, an ensemble-averaged property or distribution, equilibrium phase behavior, time-dependent disorder evolution, or phase-level thermodynamics. The next consideration is which degrees of freedom should remain active, ranging from lattice occupations only to further coupled structural relaxation, defects, charge states, and magnetism, together with the spatial and temporal scales required for convergence. Such requirements affect the appropriate balance among disorder simulation approaches. Together, these considerations provide a practical basis for method selection by clarifying the appropriate use, conditions that motivate moving beyond each method, and opportunities for ML acceleration (Table \ref{table:practical_method_selection}).

Practical system size is best interpreted based on method category and the correlation length of the target disorder, rather than as a universal atom-count ceiling. Effective-medium methods do not require explicit disorder supercells, whereas SQS methods, finite-ensemble approaches, and direct first-principles sampling are constrained by the size and number of explicit cells that must be evaluated. Moreover, for cluster expansion, once a reliable Hamiltonian has been fitted, it can sample much larger on-lattice cells at low incremental cost, while CALPHAD operates at the phase level, where atom count is not the relevant measure of scale. The required explicit cell size can nevertheless become substantial. As an example, in CrCoNi, short-range order extends over approximately $25$--$30$ \AA, and about 2,000 atoms are required for size convergence, whereas typical DFT-sized cells produce relative errors of up to $40\%$ in Warren--Cowley parameters,\cite{Cao2025} illustrating why practical scalability of various disorder modeling methods should ultimately be judged against the relevant disorder correlation length, rather than by atom count alone.

\begin{table}[hbt!]
\phantomsection
\centering

\begin{threeparttable}

\caption{
\textbf{Practical guide for selecting computational approaches for modeling chemical disorder.}
}
\label{table:practical_method_selection}

\footnotesize
\setlength{\tabcolsep}{3pt}
\renewcommand{\arraystretch}{1.05}

\begin{tabular}{@{}
>{\raggedright\arraybackslash}m{0.18\textwidth}
>{\raggedright\arraybackslash}m{0.26\textwidth}
>{\raggedright\arraybackslash}m{0.26\textwidth}
>{\raggedright\arraybackslash}m{0.25\textwidth}
@{}}

\Xhline{3\arrayrulewidth}

Method category &
Use when &
Move beyond when &
Potential role of ML \\

\Xhline{3\arrayrulewidth}

Effective-medium and mean-field theories (e.g., VCA and CPA)
&
Average properties of near-random substitutional solids are sufficient and local environments are secondary.
&
Local relaxation, short-range order, charge redistribution, or local clustering becomes highly important.
&
Learn corrections from explicit configurations or predict deviations from mean-field behavior.
\\

\hline

Cluster-expansion-based on-lattice statistical mechanics (possibly with Monte Carlo)
&
Equilibrium ordering, configurational free energies, transition temperatures, or phase boundaries are needed on a parent lattice.
&
Off-lattice relaxation or coupled defect, charge, or spin degrees of freedom strongly affect the relevant disorder state space.
&
Identify informative reference configurations, accelerate generation of configurational energetics, and accelerate sampling.
\\

\hline

Quasi-random structural approximations (e.g., SQS)
&
One or a few representative supercells are sufficient to evaluate properties of a random or prescribed short-range-ordered state.
&
Representative-cell results are not converged, or ensemble averaging and temperature-dependent disorder must be resolved.
&
Accelerate relaxation and property evaluation, enabling larger cells, multiple realizations, and convergence testing.
\\

\hline

Ensemble-based microstate averages (e.g., POCC)
&
A finite set of microstates can capture the required ensemble average or relevant property distribution.
&
The required ensemble becomes too large, or rare states and long-range correlations are not captured.
&
Select representative microstates, accelerate energy evaluation, and guide ensemble refinement.\\

\hline

Stochastic sampling and molecular dynamics
&
Finite-temperature disorder, short-range order, or coupled occupational--structural fluctuations require detailed explicit sampling.
&
Direct first-principles sampling is too costly, or finite-size, slow-mixing, or time-scale limitations prevent efficient convergence.
&
Accelerate energy and force evaluations with MLIPs, enabling larger cells, longer trajectories, and more extensive sampling.
\\

\hline

Macroscopic thermodynamic models (e.g., CALPHAD)
&
Experimentally anchored phase equilibria or synthesis windows across broad composition--temperature spaces are the target.
&
Local environments must be resolved atomistically, rather than represented only through phase-level thermodynamics.
&
Integrate atomistic and experimental data, fit free-energy models, and boost uncertainty-aware phase-diagram prediction.
\\

\Xhline{3\arrayrulewidth}

\end{tabular}

\end{threeparttable}

\end{table}

This decision framework (Table \ref{table:practical_method_selection}) also emphasizes when greater computational complexity or ML acceleration is actually warranted. A simpler representation might be sufficient when the target observable is insensitive to configurational fluctuations, while explicit ensemble sampling becomes crucial when short-range order, configurational entropy, finite-temperature transitions, or kinetic history determine the quantity of interest. Moreover, direct first-principles evaluation remains appropriate when the required number and size of configurations are tractable, whereas disorder-aware AI approaches and ML models become increasingly essential when the relevant configurational space, correlation length, or simulation time exceeds the practical reach of first-principles simulations. Notably, this AI acceleration is valuable only when the ML models can truly capture and preserve the disorder-sensitive physics required by the ultimate scientific goal. Such practical considerations naturally motivate the AI-assisted strategies discussed next.

\phantomsection
\addcontentsline{toc}{section}{3. Modern AI-assisted approaches for disorder modeling}
\section*{3. Modern AI-assisted approaches for disorder modeling}

Modern AI-assisted approaches for modeling chemically disordered materials are generally best regarded not as replacements for classical statistical mechanics, but as a new, complementary set of data-driven tools for boosting the scale, fidelity, efficiency, and transferability with which disordered configurational ensembles are represented, sampled, and interpreted. The task is the same: chemical disorder is determined by a massive set of symmetry-inequivalent occupational microstates whose relative populations depend collectively on energy, entropy, temperature, and processing history (Fig. \ref{fig:thermo_vs_computational_vs_experimental_gap}). Conventional methods offer the physical foundation for tackling this problem, but they are still constrained by the combinatorial growth of the configurational space, the high cost of relaxing and evaluating various possible structures, and the persistent challenge of recovering ensemble-dependent quantities from limited first-principles data. To address these bottlenecks, ML and AI serve two complementary roles. First, when coupled with conventional modeling workflows, ML and AI can accelerate these established workflows by replacing costly energy, force, relaxation, or property evaluations with faster surrogate models that can facilitate configurational exploration, molecular dynamics, and ensemble averaging at much larger length and time scales. Second, beyond acceleration, recent advances have introduced disorder-native capabilities that have been difficult to realize using conventional methods alone, e.g., ordering-sensitive ML representations, disorder-propensity prediction, differentiable chemical identities, and generative models for configurational thermodynamics. Thus, the central distinction among modern AI-driven approaches is not simply whether they use ML, but what physical objects they learn, where they enter the disorder simulation workflows, and whether they primarily boost an existing approximation or provide a brand-new way to represent and model chemical disorder.

\phantomsection
\addcontentsline{toc}{subsection}{3.1 ML as an accelerator for existing modeling formalisms}
\subsection*{3.1 ML as an accelerator for existing modeling formalisms}

\begin{figure}
\phantomsection
\begin{center}
\includegraphics[max size={\textwidth}{\textheight}]{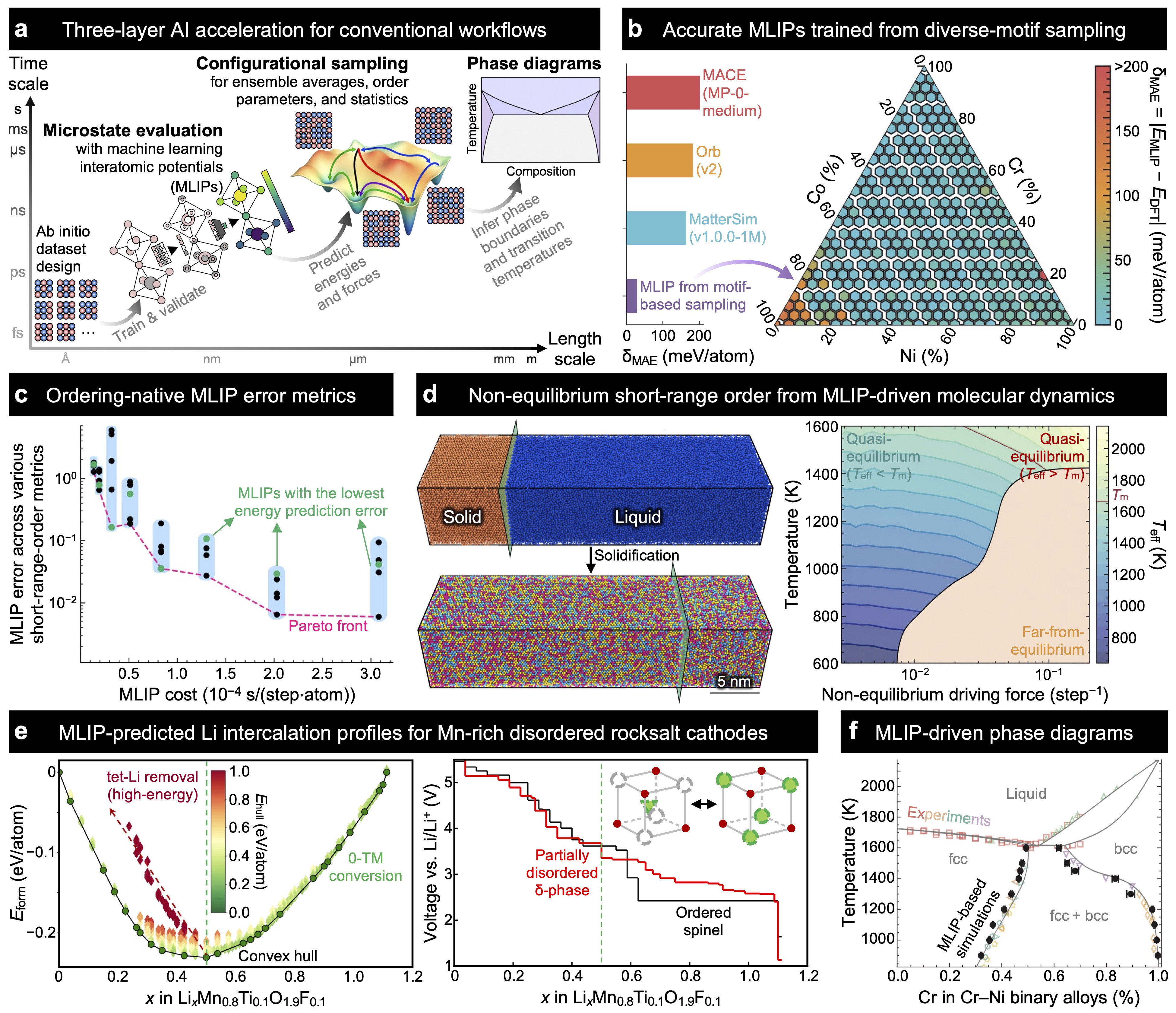}
\caption{
\textbf{ML as an accelerator for established atomistic and thermodynamic formalisms for disorder modeling.}
(a) ML can accelerate all three key layers of conventional disorder modeling workflows, from microstate evaluation to configurational sampling and phase-diagram prediction.
(b) Diverse-motif sampling reduces both the magnitude and compositional variability of energy errors relative to broadly trained universal MLIPs.\cite{Sheriff2026}
(c) Low energy errors alone do not always identify the MLIPs that best predict short-range-order-sensitive properties.\cite{Cao2025}
(d) High-fidelity MLIP-driven molecular dynamics enables large-scale simulations of non-equilibrium processing, highlighting how solidification might give rise to short-range-order states beyond equilibrium annealing pathways.\cite{Islam2025}
(e) Fine-tuned universal MLIP simulations help resolve disorder-enabled Li intercalation pathways in Mn-rich disordered rocksalt cathodes, linking Li--vacancy configurations and phase transformation to voltage profiles and accessible capacity.\cite{Zhong2025a}
(f) MLIP-driven sampling converts atomistic energetics into phase diagrams with phase boundaries comparable to experiments and CALPHAD.\cite{Sheriff2026}
Panels (b), (c), (d), and (f) adapted from ref.\cite{Sheriff2026}, ref.\cite{Cao2025}, and ref.\cite{Islam2025}, respectively, CC BY 4.0. Panel (e) adapted with permission from ref.\cite{Zhong2025a}, copyrighted by the American Physical Society.
}
\label{fig:ml_as_an_accelerator}
\end{center}
\end{figure}

ML acceleration of established disorder modeling formalisms is easiest to understand by asking which bottleneck in traditional modeling workflows the learned ML model is designed to tackle. In this category of data-driven methods, ML does not replace the statistical-mechanics backbone of disorder modeling, but instead lowers the cost of applying established simulation frameworks to chemically complex configurational ensembles. This acceleration can occur at three strongly connected levels (Fig. \ref{fig:ml_as_an_accelerator}a). First, at the microstate level, the learned models can make individual configurations cheaper to evaluate by predicting energies, forces, stresses, structural relaxations, and configuration-level properties. Second, at the ensemble level, they make the exploration and averaging of different microstates more tractable by coupling surrogate evaluations with Monte Carlo, molecular dynamics, active learning, configurational navigation, or ensemble averaging. Third, at the macroscopic level, they help translate atomistic or ensemble-level information into phase-level thermodynamic descriptions, where uncertainty-aware simulations are increasingly needed for complex multicomponent systems. This hierarchy is helpful because it separates the physical object being learned from the modeling framework being accelerated: in all three cases, the configurational ensemble remains the central part of disorder modeling, while ML decreases the cost and increases the scale and data efficiency with which this ensemble is evaluated.

\paragraph{Accelerating microstate evaluations.}
The first and most mature acceleration layer targets the repeated, expensive assessment of various site-occupancy microstates. In SQS, finite-ensemble microstate averaging, hybrid Monte Carlo--molecular dynamics, or finite-temperature sampling, one of the most costly procedures is usually the prediction of relaxed structures, energies, forces, stresses, and configuration-level properties for numerous different atomic arrangements. MLIPs directly address such a bottleneck by learning a surrogate PES, thereby replacing repeated first-principles calculations with much more efficient energy and force evaluations.\cite{Friederich2021,Unke2021,Ko2023,Kalita2025} Descriptor-based\cite{Liu2024b,Ibrahim2026} and graph-based ML models\cite{Chen2021d,Deshmukh2024,Fang2024,Zhang2025,Wang2025b,Fang2026} can also be used to infer scalar quantities directly when forces are not needed. In this sense, ML does not alter the underlying statistical-mechanics problem. Instead, it makes each explicit configuration sufficiently low-cost to evaluate, allowing larger supercells, deeper sampling, and broader chemical spaces to be computationally feasible.

MLIPs have evolved from descriptor-based, system-specific force fields toward increasingly expressive and transferable atomistic models. Early MLIPs represent atomic environments with handcrafted symmetry functions, bispectrum or smooth-overlap descriptors, or relevant atomic representations, to learn the mapping from the descriptors to atomic energy contributions.\cite{Behler2007,Bartok2010,Bartok2013} These models have pioneered the idea that DFT-quality energies and forces can be approximated without explicitly solving the electronic-structure problem, but their prediction performance and model transferability depend heavily on the chosen descriptors and training domain.\cite{Friederich2021,Unke2021} More recent MLIPs increasingly use GNN-based architectures,\cite{Xie2018,Chen2019,Damewood2023,Corso2024,Lunger2024} where atoms or crystallographic sites are treated as graph nodes and local neighbor relationships are treated as graph edges. Such GNN-driven MLIPs learn many-body environment representations directly from data, reducing the need for manual feature engineering and promoting transferability across diverse chemistries and structural motifs.\cite{Schutt2018,Gasteiger2021,Park2021} Equivariant message-passing GNN-powered MLIPs further encode rotational and translational symmetries into the architecture,\cite{Smidt2021,Batzner2022,Batatia2022,Duval2023,Batatia2025a} which can be highly valuable for learning complex forces and tensorial responses in chemically disordered solids.\cite{Wen2024,Yan2025}

This message-passing view provides a compact way to see how modern graph-based MLIPs evaluate a fixed disordered microstate. For a site-occupancy microstate $\sigma$ with atomic positions $\{\mathbf{q}_i\}$, each atom $i$ has a hidden feature vector $\mathbf{u}_i^{(\ell)}$ at the GNN layer $\ell$, initialized from its discrete chemical identity $\sigma_i$ and updated by aggregating information from neighboring atoms:\cite{Damewood2023}
\begin{equation}
\label{eqn:conventional_gnn_1}
\boldsymbol{\mu}_i^{(\ell)}
=
\sum_{j\in\mathcal{N}(i)}
\mathcal{M}_{\theta}^{(\ell)}
\!\left(
\mathbf{u}_i^{(\ell)},
\mathbf{u}_j^{(\ell)},
\mathbf{b}_{ij}
\right)
\end{equation}
\begin{equation}
\label{eqn:conventional_gnn_2}
\mathbf{u}_i^{(\ell+1)}
=
\mathcal{U}_{\theta}^{(\ell)}
\!\left(
\mathbf{u}_i^{(\ell)},
\boldsymbol{\mu}_i^{(\ell)}
\right)
\end{equation}
\begin{equation}
\label{eqn:conventional_gnn_3}
\widehat{E}_{\theta}(\sigma,\{\mathbf{q}_i\})
=
\sum_i
\rho_{\theta}\!\left(\mathbf{u}_i^{(L)}\right)
\end{equation}
\begin{equation}
\label{eqn:conventional_gnn_4}
\widehat{\mathbf{f}}_{i,\theta}
=
-\frac{\partial \widehat{E}_{\theta}}{\partial \mathbf{q}_i}
\end{equation}
Here, $\boldsymbol{\mu}_i^{(\ell)}$ is the aggregated message received by atom $i$ at layer $\ell$, $\mathcal{N}(i)$ is its neighbor set, and $\mathbf{b}_{ij}$ is an edge embedding denoting the local geometry of the $i$--$j$ atomic pair, such as interatomic distance, direction, or other symmetry-adapted geometric features. The functions $\mathcal{M}_{\theta}^{(\ell)}$ and $\mathcal{U}_{\theta}^{(\ell)}$ are learned message and update functions, respectively; $\rho_{\theta}$ is the learned atomic-energy readout; $L$ is the final message-passing layer; and $\theta$ denotes trainable GNN parameters. The hats indicate ML-derived quantities: $\widehat{E}_{\theta}$ is the predicted total potential energy of the specified microstate, and $\widehat{\mathbf{f}}_{i,\theta}$ is the corresponding force acting on atom $i$, obtained by differentiating the predicted energy with respect to its atomic position. The schematic form in Eqs. \eqref{eqn:conventional_gnn_1}--\eqref{eqn:conventional_gnn_4} is not meant to suggest that all MLIPs are graph-based. Rather, it shows the architectural logic of many state-of-the-art MLIPs used for molecular and materials simulations. In disorder modeling, a vital point is that the occupational state is predefined before the MLIP is evaluated: each lattice site $i$ is assigned a definite species $\sigma_i$, and the model further predicts the energy and forces of this specific atomic arrangement. Hence, conventional GNN-driven MLIPs can accelerate the evaluation of explicit configurations, but they do not by themselves determine which microstates should be generated, sampled, and thermodynamically weighted. These MLIPs are best understood as fast evaluators of explicit atomic configurations, whose broader value in disorder modeling eventually depends on how they are coupled to downstream sampling, search, and ensemble-averaging workflows.

Disordered alloys, multicomponent oxides, and high-entropy materials can offer demanding testbeds for microstate-level ML acceleration, as useful MLIP surrogates need to be transferable across chemical spaces while remaining sensitive to local motifs. Large first-principles datasets spanning ordered and disordered structures (including high-entropy alloys) have shown that the prediction errors depend strongly on whether the training data include sufficient configurational diversity and relaxation information.\cite{Wang2025b,Li2024b} Training-set design methods that can sample diverse local motifs have resulted in highly accurate MLIPs for predicting stacking-fault energies, short-range order, heat capacities, and phase diagrams of both simple and high-entropy alloys\cite{Sheriff2026} (Fig. \ref{fig:ml_as_an_accelerator}b). Once such desired local-environment coverage is achieved, MLIPs can transform otherwise prohibitive first-principles calculations into tractable disorder modeling or screening workflows. For instance, MLIPs have enabled large-scale simulations of GeSn alloys, revealing short-range ordering and local structural features that would be difficult to access using direct DFT alone.\cite{Chen2024a} In high-entropy oxides, MLIP-driven computational workflows have been used to map complex compositional spaces, rationalize experimentally observed single-phase oxide compositions via bond-length and mixing-enthalpy descriptors, and further prioritize synthesizable candidates in broader search spaces with MLIP-relaxed structures and energy-variance-based descriptors.\cite{Sivak2025,Dicks2026} Beyond general PES evaluation, this MLIP-based microstate-level acceleration similarly applies to other specialized tasks that repeatedly appear in disorder modeling workflows, e.g., structural relaxation, defect energetics, and atomic segregation. For example, chemistry-driven relaxation models based on chemical subgraph matching can reduce or bypass expensive DFT relaxations in low-energy configurational exploration of disordered crystal structures.\cite{An2025} Collectively, these studies show that useful microstate-level surrogates are not merely generic energy models: their training data, descriptors, and validation targets must be aligned with the local chemical motifs, relaxation modes, and relative energy scales that control realistic disorder physics in materials.

The design of training data and model validation criteria strongly affects whether microstate-level MLIP surrogates are reliable for disorder modeling, rather than being merely accurate on disorder-insensitive test sets.\cite{Choyal2024,Riebesell2025} Efficient small-cell sampling has been recently employed for multi-principal element alloys to show that these representative small-cell structures can provide a high-quality training dataset for modeling phase transitions, local chemical ordering, and other thermodynamic properties in several multicomponent systems.\cite{Liu2025e} Moreover, a systematic study of MLIP-driven short-range-order modeling in CrCoNi alloys has shown that conventional held-out energy errors do not necessarily correlate with the ability to capture local-ordering-sensitive quantities, including stacking-fault energies and phase stability, motivating training-data design principles targeted directly at short-range-order fidelity\cite{Cao2025} (Fig. \ref{fig:ml_as_an_accelerator}c). These cases emphasize a key requirement for disorder-focused MLIPs: a robust validation protocol must assess whether these models learn the relative energetics and ensemble statistics of competing orderings.\cite{Peng2024a,Peng2024b}

Recent advances in universal MLIPs may further expand the reach of microstate evaluation, but their use in disorder modeling still requires careful validation against local ordering physics. Here, universal MLIPs refer to broadly pretrained PES surrogates trained across large chemical and structural datasets\cite{Deng2023,Merchant2023,Schmidt2024,Kaplan2025,Mazitov2025a,Levine2026,Malosso2026,BarrosoLuque2026,Peng2026} to reproduce energies and forces from a specific level of electronic-structure theory.\cite{Yuan2026,Choi2025} These general-purpose MLIPs have made zero-shot or few-shot structural relaxation, formation energy ranking, and large-scale materials screening efficient across broad chemical spaces.\cite{Deng2023,Merchant2023,Chen2022,Zhang2024b,Clausen2024,Batatia2025b,Mazitov2025b,Wood2025} This capability is particularly attractive for chemically disordered materials, as many of them retain well-defined parent lattices and constituent elements, making universal MLIPs helpful first-pass accelerators for estimating relaxation trends, thermodynamic driving forces, and synthetic accessibility.\cite{Zhong2025a,Biswas2026,Ullberg2026,Jakob2026a} However, broad chemical coverage does not automatically guarantee disorder-aware accuracy for universal MLIPs. As an example, accurate local structure relaxation from universal MLIPs might not imply reliability for thermodynamic ranking across chemically distinct microstates, where relative energy differences of only a few meV/atom can strongly affect the Boltzmann weights. Notably, the first generation of universal MLIPs (such as M3GNet,\cite{Chen2022} CHGNet,\cite{Deng2023} and MACE-MP-0)\cite{Batatia2025b} has a systematic PES softening effect, leading to compressed Ca--Mg--O mixing energies that underestimate the miscibility-gap temperature.\cite{Deng2025b} Pretraining on non-equilibrium datasets\cite{Mazitov2025a,BarrosoLuque2026} can largely resolve this softening and make universal MLIPs more useful in disorder modeling.\cite{Creed2026} Moreover, local strain, charge redistribution, high-energy transition states, and near-degenerate microstates may still place the local environments or relative energy scales outside the training data distribution. For instance, universal MLIPs can misrank atomic configurations in mixed-valence systems when electronic entropy and charge ordering are not properly represented.\cite{Petersen2026} Therefore, universal MLIPs should be regarded as powerful accelerators and physically informative priors, instead of automatically reliable substitutes for disorder-specific model benchmarking, fine-tuning, and validation.

Overall, microstate-level ML accelerators address the first layer of the modeling bottleneck. They make explicit configurations much cheaper to relax, evaluate, and rank, but real disordered states are determined by statistical populations of various competing microstates, rather than by any single configuration alone. Hence, the broader scientific value of these accelerators depends on the next workflow layer: coupling fast surrogate evaluations to configurational sampling and ensemble-averaging protocols to correctly recover the statistical nature of chemical disorder.

\paragraph{Accelerating ensemble workflows.}
Once individual microstates become efficient to evaluate, the next acceleration layer turns fast ML evaluations into physically meaningful configurational statistics. Chemical disorder is rarely determined by a single low-energy structure alone. Short-range order, configurational entropy, diffusion-relevant motifs, and functional responses emerge from populations of accessible configurations.\cite{Ferrari2023,Raabe2023,Chen2026} ML is most helpful at this layer when it is embedded into well-established sampling workflows, e.g., Monte Carlo, molecular dynamics, finite-ensemble averaging, and active-learning-driven configurational search. In these methods, ML provides fast prediction of energies, forces, properties, and uncertainty estimates, while the configurational ensemble is still sampled, weighted, and interpreted using statistical mechanics, consistent with the same ensemble averaging formalism introduced in Eqs. \eqref{eqn:finite_ensemble_probability} and \eqref{eqn:finite_ensemble_average}.

ML-accelerated Monte Carlo illustrates this strategy by preserving the statistical-mechanics sampler while replacing costly energetic or property evaluations with learned surrogates. Recent attention-based GNN models coupled with Monte Carlo have been used to infer configurational entropy and order--disorder transition temperatures, with a critical observation that the variance of energy prediction errors across various configurations can matter more for disorder properties than ordinary average test error.\cite{Fang2024} Similarly, lattice GNN surrogates have enabled Monte Carlo simulations of complex Lu--H--N solid solutions, where interstitial disorder and multicomponent occupation make direct first-principles sampling impractical.\cite{Guan2025} Beyond energetics, equivariant GNNs integrated with Monte Carlo have been applied to predict ensemble-averaged functional properties, e.g., electrical and optical conductivities in termination-disordered MXenes.\cite{Fang2025} Such examples show how ML can extend ensemble-based strategies from energy averaging to diverse essential observables, while keeping the final prediction as a statistically robust quantity.

Active learning and configurational search algorithms address a complementary bottleneck: determining which atomistic configurations to evaluate, revisit, or use for retraining. In ternary Pd--Pt--Sn alloys, uncertainty-aware GNN surrogates have been employed to explore structures and predict formation energetics with far fewer DFT calculations than exhaustive high-fidelity workflows.\cite{Deshmukh2024} Graph-theory-based Monte Carlo tree search has also been introduced to identify representative atomic configurations in disordered solids as the supercell size and compositional complexity increase.\cite{He2026} Similarly, ML-accelerated evolutionary Monte Carlo has been coupled with MLIPs to explore phase behavior and segregation of compositionally complex materials in reactive environments.\cite{Han2026} Relevant ML-powered Monte Carlo frameworks have been proposed to uncover order--disorder transitions in complex alloys over time scales inaccessible to previous molecular dynamics simulations.\cite{Zhou2026} These methods are best regarded as accelerators of explicit configurational space navigation, rather than directly generating the ensemble distribution itself.

MLIP-driven sampling and dynamics extend ML acceleration from equilibrium statistics to short-range-order evolution across realistic length scales, time scales, and processing pathways. Large-scale ML-enabled atomistic simulations have helped quantify chemical short-range order in metallic alloys,\cite{Sheriff2023} while MLIP-powered molecular dynamics have shown that solidification, deformation, and thermomechanical processing can generate non-equilibrium short-range-order states distinct from equilibrium predictions\cite{Islam2025,Islam2026} (Fig. \ref{fig:ml_as_an_accelerator}d). Similar methods have been expanded to disorder-to-partial-order transformation in Mn-rich disordered rocksalt cathode materials for Li-ion batteries\cite{Zhong2025a} (Fig. \ref{fig:ml_as_an_accelerator}e), short-range ordering in covalently bonded high-entropy carbides,\cite{Wei2026} and configurational entropy in compositionally complex ferrite spinels.\cite{Jakob2026a} Notably, for the Mn-rich disordered-rocksalt example, a chemistry-specific fine-tuned CHGNet model has reported test errors of $2$ meV/atom in energy and $68$ meV/\AA\ in forces, which has been subsequently used for MLIP-driven molecular dynamics simulations over $2$ ns.\cite{Zhong2025a} Altogether, these examples show how MLIPs can move disorder simulations beyond individual equilibrium configurations toward finite-temperature, history-dependent, and statistically converged descriptions of disorder.

Collectively, these examples show that ensemble-level ML acceleration succeeds only when faster evaluation is converted into statistically faithful disorder observables. Therefore, the core validation target needs to shift from isolated energy or force errors to the fidelity of the sampled ensemble itself, e.g., its underlying microstate rankings, order parameters, and kinetic pathways. The next step is to propagate ensemble averages into macroscopic thermodynamic descriptions, where ML assists with thermodynamic closure, uncertainty quantification, and data integration.

\paragraph{Accelerating thermodynamic closure.}
The third acceleration layer begins where microstate evaluation and ensemble sampling finish: transforming atomistic or configurational information into macroscopic thermodynamic descriptions. Many practical questions relevant to chemically disordered solids ultimately require a deep understanding of phase boundaries, miscibility gaps, order--disorder transition temperatures, and synthesis windows, rather than merely microscopic atomic configurations or sampled local motifs. CALPHAD and related thermodynamic models are indispensable for such a purpose, as they provide phase-level free-energy descriptions across various compositions and temperatures. In this context, ML is most helpful as a data-integration and uncertainty-quantification tool, parameterizing thermodynamic models from first-principles or experimental data, unveiling composition--temperature regions where additional calculations are most informative, and propagating atomistic uncertainty into phase-level predictions.\cite{Shen2025}

Recent operational workflows demonstrate that this thermodynamic closure can be achieved via three concrete routes: automated thermodynamic data generation, experimental-data-guided parameter refinement, and phase-diagram-level benchmarking of ML surrogates. For instance, MLIP-driven CALPHAD workflows can now generate SQS energies, vibrational contributions, and liquid-phase molecular-dynamics-derived data at substantially lowered computational cost and further use the resulting phase diagrams as application-level tests of MLIP accuracy.\cite{Zhu2025a,Zhu2025b} A related Pt--W case study has shown that MLIP-derived thermodynamic data can be combined with experimental phase-equilibria information to refine CALPHAD parameters, with a modest adjustment to physically grounded Gibbs-energy descriptions substantially boosting agreement with experimentally measured phase boundaries.\cite{Kunselman2025} Complementary MLIP studies across alloy compositional spaces have demonstrated that phase diagrams provide stringent thermodynamic validation targets beyond static energy errors, showing that carefully trained MLIPs can lead to experimentally consistent phase boundaries across diverse alloy systems\cite{Sheriff2026} (Fig. \ref{fig:ml_as_an_accelerator}f). Together, these studies help define a practical AI-driven closure route: ML first accelerates the generation, benchmarking, and tuning of thermodynamic inputs, while CALPHAD and relevant phase-level formalisms preserve the interpretable free-energy structure needed for equilibrium prediction.

Atomistic-to-thermodynamic integration becomes particularly valuable in high-dimensional compositional spaces where brute-force phase-diagram construction is difficult. As an example, GNN surrogates can predict configuration- and pressure-dependent equation-of-state quantities that are subsequently transformed into Gibbs energies and phase diagrams for a multicomponent interstitial solid solution.\cite{Guan2025} This workflow is highly useful, as it does not stop at configurational evaluation. Instead, it uses ML-accelerated sampling to produce thermodynamic quantities that can be interpreted at the phase level. A complementary approach has been developed for multi-principal element alloys, where the thermodynamic stability of complex alloys can be estimated from lower-dimensional subsystem energetics with meV-scale accuracy, while retaining a strong connection to cluster-expansion theory and experimentally reported synthesizability trends.\cite{Wang2026a} Collectively, these examples show that ML-assisted atomistic models can supply disorder-aware inputs for thermodynamic assessments, while phase-level constraints can subsequently suggest which compositions, configurations, and temperature ranges deserve further evaluation.

Uncertainty quantification is the key requirement that distinguishes thermodynamic closure from standard high-throughput atomistic prediction. Critical phase boundaries, miscibility gaps, and order--disorder transition temperatures can shift substantially when competing free energies differ by merely a few meV/atom. Hence, single-value ML predictions can be insufficient unless uncertainties are propagated into the thermodynamic model. In chemically disordered materials, this uncertainty can arise from various sources, e.g., first-principles errors, ML surrogate errors, finite-size configurational sampling, incomplete theoretical treatments of vibrational, magnetic, and electronic entropies, and experimental errors.\cite{Lukas2007} To evaluate these uncertainties, Bayesian thermodynamic assessments, Gaussian-process corrections, ensemble-driven models, gradient-informed optimization, and active-learning loops can all help identify composition--temperature regions where predictions remain underconstrained.\cite{Shen2025} Without these analyses, ML-accelerated thermodynamic closure may appear informative while remaining physically undervalidated.

At this closure layer, agentic workflows can be helpful as orchestration tools that coordinate physically constrained simulation steps. Large-language-model-driven computational AI agents may help coordinate first-principles calculations, MLIP relaxations, configurational sampling, thermodynamic parameter fitting, and experimental data assimilation in closed loops.\cite{AspuruGuzik2025,Vriza2026,Liu2026b,Kumar2026} The scientific value of these agentic tools, however, will depend on whether the underlying physical modeling, uncertainty estimates, and thermodynamic constraints remain explicit, auditable, and reproducible.\cite{Shen2025,Xin2025,Xin2026a} A realistic near-term goal is not black-box, autonomous phase-diagram discovery, but robust atomistic-to-thermodynamic pipelines in which ML accelerators offer fast and uncertainty-calibrated inputs, while thermodynamic formalisms further enforce consistency across phases, compositions, and temperatures to facilitate reliable phase-level prediction.

Taken together, these three acceleration layers show that ML can greatly expand the practical reach of established disorder formalisms without changing their physical foundations. It makes microstate evaluation less expensive, ensemble exploration broader, and thermodynamic closure more scalable, while the physical description and predictive interpretation of chemical disorder remain deeply grounded in classical statistical mechanics and solid-state thermodynamics.

\phantomsection
\addcontentsline{toc}{subsection}{3.2 ML as an enabler for novel disorder-native capabilities}
\subsection*{3.2 ML as an enabler for novel disorder-native capabilities}

ML can move beyond accelerating conventional simulation workflows by making disorder itself the direct target of prediction, representation, generation, and optimization. This role is different from the acceleration strategies discussed earlier, where ML mainly lowers the cost of applying well-established simulation formalisms for faster microstate evaluation, broader configurational sampling, and more physically grounded thermodynamic closure. By contrast, in this category of data-driven approaches, ML offers fundamentally new ways to interact directly with chemical disorder prior to, during, and beyond conventional atomistic simulations. ML can first function as a disorder-aware decision layer, identifying when an ordered computational candidate should be analyzed cautiously, routed into explicit, high-fidelity disorder modeling, or reinterpreted as an ordered derivative of a known disordered family. It can further improve the representation of disorder by preserving the ordering and symmetry of coordination environments and describing key alchemical relationships that conventional ML representations of solid-state materials may obscure. Beyond prediction and representation, generative AI approaches can directly construct disordered crystals, representative configurational ensembles, and thermodynamic distributions, rather than only assessing atomistic configurations selected with an external sampling workflow. Lastly, kinetics-informed ML models can target barriers, rates, transition rules, and first-passage times that predict how disorder forms, relaxes, and becomes trapped under crucial synthesis and operating conditions. These disorder-native AI capabilities are best regarded as complementary extensions, instead of replacements for conventional ML-accelerated sampling workflows: they help clarify when disorder must be modeled, how it should be represented, what configurational distributions should be generated, and which kinetic steps may govern disorder experimentally.

\paragraph{Enabling disorder-aware workflow triage.}
Disorder-aware triage can offer a vital upstream AI capability that is different from simply accelerating an existing disorder modeling workflow. Instead of evaluating explicit microstates, sampling a known configurational ensemble, or using a thermodynamic model, the core prediction task is to decide whether an ordered computational candidate from high-throughput screening\cite{Merchant2023,Horton2025,Chen2024b} and generative inverse design\cite{Metni2026,Zeni2025,DeBreuck2025} should be trustworthy in the first place. This disorder-triage capability has become increasingly critical, as high-throughput screening and generative workflows still mainly operate on highly idealized, chemically ordered structures,\cite{Merchant2023,Chen2024b,Zeni2025} whereas experimentally reported materials often contain partial occupancies, occupational disorder, and symmetry-averaged crystal descriptions.\cite{Cheetham2024,Antypov2025,Jakob2026b} A disorder-aware triage tool can thus help estimate whether an ordered candidate downselected from high-throughput screening and inverse design workflows is trustworthy or, alternatively, it should be viewed as disorder-prone for further computational or experimental examination.

The most straightforward form of disorder-aware triage is to make crystallographic disorder itself the prediction target. For instance, ML classifiers trained on experimental inorganic crystal structures can learn chemical trends associated with crystallographic disorder and estimate how frequently such chemical disorder occurs in computational databases.\cite{Jakob2026b} This framing is crucial, as it can be utilized as a signal indicating when conventional convex-hull evaluation of ordered structures\cite{Riebesell2025,Merchant2023,Zeni2025} should be interpreted cautiously and followed by more detailed and rigorous disorder analysis.\cite{Cheetham2024,Jakob2026b} Relevant property-oriented ML workflows also show that experimentally reported chemical disorder can provide helpful structure--property information. As an example, an experimental structure--conductivity database of Li-containing compounds has been coupled with graph-based ML featurization to classify both ordered and disordered structures as possible superionic Li conductors, leading to screening and experimental validation of a newly identified superionic conductor (Li\textsubscript{9}B\textsubscript{19}S\textsubscript{33}).\cite{McHaffie2025} These studies show that disorder-aware learning and triage can promote both workflow routing and property screening by treating experimental disorder as valuable information in AI-driven materials discovery, rather than as representational noise.

Disorder-aware triage also requires crystallographic representations and comparison toolkits that can place ordered and disordered structures on a common basis. For example, a symmetry- and Wyckoff-sequence-based ML representation has recently been built to encode both ordered and disordered solids, including co-occupying atomic species and continuous site-stoichiometry information.\cite{Huang2026} Such a representation tackles a database-scale challenge that previous atomistic descriptors cannot solve cleanly: deciding whether two crystal data records describe genuinely distinct compounds, or alternatively, differently ordered or partially occupied representations of the same underlying disordered structure. This comparison issue has been particularly important for assessing novelty in AI-generated or high-throughput ordered structures.\cite{Cheetham2024,Falkowski2025,Martirossyan2025,Hagemann2026,Negishi2026} Moreover, order--(dis)order family trees based on group--subgroup relationships have been further designed to identify cases in which an apparently new ordered structure is more appropriately considered as an ordered child of an existing disordered parent.\cite{Yamazaki2026} Recent critiques of generative AI models reinforce the same point, showing that apparent predictions can overlap with compounds already represented in training data and that novelty evaluation should thus account for order--(dis)order structural relations, rather than being based only on data absence from a known database.\cite{Juelsholt2026} For disorder modeling, this warning is particularly consequential, as an AI-predicted ordered crystal structure should not automatically be regarded as a new material if it is either configurationally, symmetrically, or compositionally related to an experimentally reported disordered phase.

Overall, disorder-aware workflow triage offers a vital front-end capability that complements, rather than replaces, explicit disorder modeling. In other words, these methods do not determine the equilibrium distribution of site-occupancy microstates, nor do they bypass the need for SQS, cluster expansion, Monte Carlo, finite-ensemble averaging, and MLIP-enhanced sampling when detailed disorder thermodynamics is required. Their essential value is instead to make AI-driven materials screening or generation workflows more physically grounded, before costly follow-up calculations and experiments are performed. By identifying when disorder should be considered explicitly, when ordered predictions require reinterpretation, and when apparently novel crystal structures are better viewed as different representations of known chemical disorder, this triage layer can help promote experimentally grounded decisions in AI-driven materials discovery.

\paragraph{Enabling ordering-sensitive and alchemical representations.}
After a crystal candidate has been identified as potentially disorder-relevant, the next ML problem is to represent its ordering degrees of freedom without removing the distinctions that make different microstates physically meaningful. This capability differs from using an MLIP or GNN purely as an efficient evaluator of a fixed configuration. In conventional microstate evaluation, the site occupations are already specified, and the model primarily infers energies, forces, stresses, or properties of these explicit atomic arrangements.\cite{Friederich2021,Unke2021,Ko2023,Kalita2025,Chen2021d} In disorder-sensitive representation learning, by contrast, the key question is whether ML models can preserve the symmetry, site occupation, local coordination, short-range correlation, and chemical-identity information needed to distinguish configurations that may share the same composition and average lattice. This requirement is important because relative microstate stability, local motif preference, and short-range order can all depend heavily on subtle differences that are easily lost if the representation is too compositionally coarse.

\begin{figure}
\phantomsection
\begin{center}
\includegraphics[max size={\textwidth}{\textheight}]{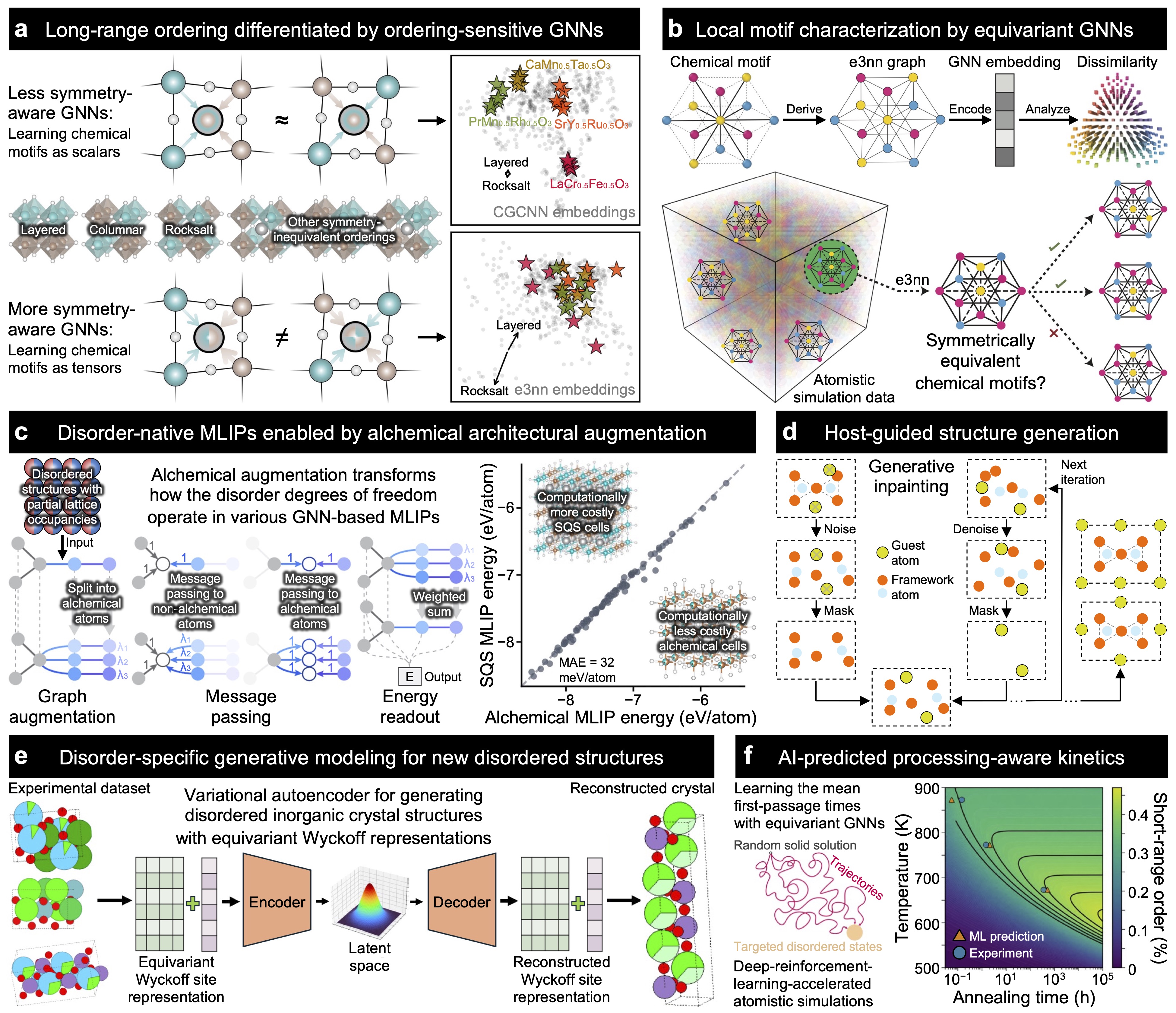}
\caption{
\textbf{ML as an enabler of disorder-native capabilities beyond conventional atomistic simulation workflows.}
(a) Symmetry-equivariant GNNs can better distinguish long-range orderings that conventional invariant models can view as nearly equivalent, enabling ordering-dependent energetics to be learned more faithfully.\cite{Peng2024b}
(b) Equivariant GNNs identify symmetrically equivalent chemical motifs and embed them into a dissimilarity space for quantifying short-range order beyond traditional Warren–Cowley parameters.\cite{Sheriff2023,Sheriff2024}
(c) Alchemical augmentation can introduce continuous site-occupancy degrees of freedom into MLIPs, enabling compact disorder representations and low-cost disorder energetic prediction.\cite{Nam2025}
(d) Host-guided inpainting generation uses a symmetrized framework to place guest atoms, driving the diffusion-model-driven generation of crystal structures with intercalation chemistry.\cite{Zhong2025b}
(e) Disorder-native generative models encode partial occupancies using equivariant Wyckoff representations, enabling reliable generation of symmetry-consistent disordered structures.\cite{Petersen2025}
(f) Mean-first-passage-time learning combines equivariant GNNs and reinforcement learning to predict short-range-order kinetics beyond traditionally accessible scales.\cite{Chun2025}
Panels (a), (b), (c), (e), and (f) adapted from ref.\cite{Peng2024b}, refs.\cite{Sheriff2023,Sheriff2024}, ref.\cite{Nam2025}, ref.\cite{Petersen2025}, and ref.\cite{Chun2025}, respectively, CC BY 4.0. Panel (d) adapted with permission from ref.\cite{Zhong2025b}, copyrighted by the Royal Society of Chemistry.
}
\label{fig:ml_as_an_enabler}
\end{center}
\end{figure}

Ordering-aware representations first address the discrete symmetry problem that arises when various lattice occupational configurations are physically distinct but compositionally identical. Conventional symmetry-invariant GNNs can struggle to differentiate diverse symmetry-distinct orderings of the same crystalline composition, while symmetry-aware graph representations can better retain the critical ordering information needed for energy and property prediction\cite{Peng2024b} (Fig. \ref{fig:ml_as_an_enabler}a). The implication for disorder modeling is straightforward: an ML model cannot accurately rank different microstates if its representation maps physically inequivalent ordering patterns to nearly identical features, as these ordering differences might be exactly what control the relative energies and Boltzmann weights of the configurations. Related representation needs arise when the relevant disorder variable is not a single ordered microstate, but a spatially distributed short-range-order pattern. Multiscale ML-based analyses of metallic alloys have shown how chemical short-range order can be assessed at realistic length scales using symmetry-sensitive equivariant GNNs and related to local lattice distortions and experimentally relevant spatial statistics\cite{Sheriff2023,Sheriff2024} (Fig. \ref{fig:ml_as_an_enabler}b). In this view, long- and short-range chemical orderings can become ML-resolved state variables for learning, rather than merely perturbations around ideal random solid solutions.

Beyond preserving symmetry and ordering information, chemical disorder also requires ML representations that can capture relationships among different chemical species. In conventional one-hot encoding, each of $M$ elemental species is represented using a length-$M$ vector with one species-specific entry equal to one and all other entries equal to zero. While this representation of materials uniquely identifies each existing element, it contains no information about chemical similarity among different species. This limitation has motivated ML representations that learn relationships among chemical species continuously. As an example, alchemical variants of ML representations based on the smooth overlap of atomic positions\cite{Bartok2013} have shown that correlations between chemical species can be optimized from data, giving rise to learned chemical similarity maps resembling and extending periodic-table chemical trends.\cite{Willatt2018} This idea can be valuable for chemically disordered materials, since multicomponent compositional spaces are more tractable when elements are embedded in a continuous, similarity-aware manifold, rather than treated as uncorrelated categorical identities. Alchemical compression has further converted this concept into a practical simulation approach for chemically complex disorder by reducing the effective compositional dimensionality of high-entropy alloys. For example, an alchemically compressed representation has been developed to construct an MLIP spanning 25 transition metals, retaining semi-quantitative accuracy for alloy energetics and short-range-order trends while also enabling the extraction of data-driven Hume--Rothery-type design rules from the simulated short-range-order statistics.\cite{Lopanitsyna2023} This compressed chemical representation has also been extended to modeling surface segregation in high-entropy alloys, where distinct elements display markedly different propensities to enrich or deplete at the surface.\cite{Mazitov2024} Hence, ML-learned chemical manifolds can make vast, chemically complex disorder spaces more navigable, while still retaining chemically meaningful ordering trends and segregation tendencies in data-driven disorder modeling.

A more direct alchemical modification of GNN-based MLIPs can promote site identity itself from a discrete data label to an interpolable, differentiable, and optimizable model parameter.\cite{Nam2025} The novelty is not simply building MLIPs that predict energies faster, but enabling a new family of MLIPs where chemically disordered lattice sites become directly computable input variables. In conventional GNN-driven MLIPs, each site is assigned one definite chemical identity before message passing begins. By contrast, alchemical graph augmentation allows a crystallographic site to carry multiple possible chemical identities with continuous weights, and thus, the partial occupancy on this lattice site can be an interpolable and optimizable MLIP parameter (Fig. \ref{fig:ml_as_an_enabler}c), rather than a post-processing approximation.\cite{Nam2025} Such a formulation can be written as a minimal extension of the conventional graph-based MLIP notation in Eqs. \eqref{eqn:conventional_gnn_1}--\eqref{eqn:conventional_gnn_4}. We can consider a crystallographic site $i$ that can be represented by $K_i$ alchemical states, which are further indexed by $a=1,\ldots,K_i$. Each alchemical state has a definite chemical identity $\chi_{ia}$ and a non-negative alchemical weight $\lambda_{ia}$. For a partially occupied site, the alchemical weights imply the fractional contribution of each possible chemical identity. Thus, they are non-negative and sum to one:
\begin{equation}
\label{eqn:alchemical_weight_constraint}
\sum_{a=1}^{K_i}\lambda_{ia}=1, \quad
\lambda_{ia}\ge 0
\end{equation}
A non-disordered, fully occupied lattice site is simply recovered by setting $K_i=1$ and $\lambda_{i1}=1$. Through this alchemical graph augmentation, the original site $i$ is replaced by alchemical nodes $(i,a)$ that share the same atomic position $\mathbf{q}_i$ but are initialized using the corresponding chemical identities $\chi_{ia}$. Consequently, the message aggregation in Eqs. \eqref{eqn:conventional_gnn_1} and \eqref{eqn:conventional_gnn_2} is modified as:
\begin{equation}
\label{eqn:alchemical_gnn_1}
\widetilde{\boldsymbol{\mu}}_{ia}^{(\ell)}
=
\sum_{(j,c)\in\widetilde{\mathcal{N}}(i,a)}
\gamma_{ia,jc}
\mathcal{M}_{\theta}^{(\ell)}
\!\left(
\widetilde{\mathbf{u}}_{ia}^{(\ell)},
\widetilde{\mathbf{u}}_{jc}^{(\ell)},
\mathbf{b}_{ij}
\right)
\end{equation}
\begin{equation}
\label{eqn:alchemical_gnn_2}
\widetilde{\mathbf{u}}_{ia}^{(\ell+1)}
=
\mathcal{U}_{\theta}^{(\ell)}
\!\left(
\widetilde{\mathbf{u}}_{ia}^{(\ell)},
\widetilde{\boldsymbol{\mu}}_{ia}^{(\ell)}
\right)
\end{equation}
where tildes denote quantities defined on the alchemically augmented graph, $\widetilde{\mathcal{N}}(i,a)$ represents the neighbor list of the alchemical node $(i,a)$, $c$ indexes the alchemical state on neighboring site $j$, and $\gamma_{ia,jc}$ indicates a non-learned message weight that rescales selected message contributions according to the alchemical composition. An asymmetric weighting scheme consistent with the original graph message-passing process in the underlying non-alchemical MLIP is:
\begin{equation}
\label{eqn:alchemical_gnn_3}
\gamma_{ia,jc}
=
\begin{cases}
\lambda_{jc}, & K_i=1,\ K_j>1\\
1, & \mathrm{otherwise}
\end{cases}
\end{equation}
This weighting scheme allows a partially occupied neighboring site to contribute to the message passing based on its fractional chemical identity, while keeping the learned message and update functions, $\mathcal{M}_{\theta}^{(\ell)}$ and $\mathcal{U}_{\theta}^{(\ell)}$, unchanged from the underlying non-alchemical MLIP. The readout in Eq. \eqref{eqn:conventional_gnn_3} can be further replaced by a weighted sum over alchemical node contributions:
\begin{equation}
\label{eqn:alchemical_gnn_4}
\widehat{E}^{\mathrm{alc}}_{\theta}
\!\left(\{\lambda_{ia}\},\{\mathbf{q}_i\}\right)
=
\sum_i
\sum_{a=1}^{K_i}
\lambda_{ia}\,
\rho_{\theta}
\!\left(
\widetilde{\mathbf{u}}_{ia}^{(L)}
\right)
\end{equation}
The forces remain available by differentiating the MLIP energy with respect to atomic positions:
\begin{equation}
\label{eqn:alchemical_gnn_5}
\widehat{\mathbf{f}}^{\mathrm{alc}}_{i,\theta}
=
-\frac{\partial \widehat{E}^{\mathrm{alc}}_{\theta}}{\partial \mathbf{q}_i}
\end{equation}
As the energy is also differentiable with respect to the alchemical weights, the same automatic-differentiation machinery can provide useful compositional space gradients that highlight how changing the partial occupancy of a mixed lattice site would raise or lower the predicted energy. For the crystallographic site $i$, this compositional space gradient can be written as:
\begin{equation}
\label{eqn:alchemical_gnn_6}
\mathbf{g}^{\lambda}_{i,\theta}
=
\nabla_{\boldsymbol{\lambda}_i}
\widehat{E}^{\mathrm{alc}}_{\theta}
=
\left(
\frac{\partial \widehat{E}^{\mathrm{alc}}_{\theta}}{\partial \lambda_{i1}},
\ldots,
\frac{\partial \widehat{E}^{\mathrm{alc}}_{\theta}}{\partial \lambda_{iK_i}}
\right)
\end{equation}
where $\boldsymbol{\lambda}_i=(\lambda_{i1},\ldots,\lambda_{iK_i})$ shows the alchemical weights on site $i$, and $\mathbf{g}^{\lambda}_{i,\theta}$ is the corresponding compositional space gradient. These gradients make the stoichiometry of any disordered lattice site directly optimizable. Gradient-driven optimization can update the fractional occupancies in the direction that lowers the predicted energy, while projecting the updated weights back so that these stoichiometric parameters in disordered crystals remain non-negative and sum to one:
\begin{equation}
\label{eqn:alchemical_gnn_7}
\boldsymbol{\lambda}_i^{(t+1)}
=
\Pi_{\Delta_i}
\left[
\boldsymbol{\lambda}_i^{(t)}
-
\eta_{\lambda}
\mathbf{g}^{\lambda}_{i,\theta}
\right]
\end{equation}
where $t$ is the optimization step, $\eta_{\lambda}$ is the step size for the alchemical weights, $\Delta_i=\{\boldsymbol{\lambda}_i:\lambda_{ia}\ge 0,\sum_{a=1}^{K_i}\lambda_{ia}=1\}$ is the set of physically valid fractional occupancies on site $i$, and $\Pi_{\Delta_i}$ denotes the projection step that maps the updated weights back into this allowed set. Hence, alchemical graph augmentation fundamentally changes the role of chemical identity in GNN-based MLIPs. Instead of treating lattice site occupation as a fixed discrete input, it makes the partial occupancy an internal, differentiable degree of freedom that can be interpolated, optimized, and compared across ordered and disordered representations of the same parent structure. Because the learned message, update, and readout functions from underlying non-alchemical MLIPs are preserved, this alchemical graph augmentation strategy can, in principle, be applied across different GNN-driven MLIP backbones\cite{Batatia2022,Deng2023} and can benefit from the continuous improvements in pretrained universal MLIPs\cite{Riebesell2025} without requiring a brand-new alchemical model trained from scratch. This enables disorder-native operations, e.g., gradient-based compositional optimization, alchemical free-energy calculations for substitutions and vacancies, and relatively efficient classification of experimental long-range order--disorder competition in multicomponent double perovskites.\cite{Nam2025} In this sense, alchemical MLIPs convert partial occupancy that requires enumeration and post-processing for effective sampling into a differentiable, optimizable variable in these models.

Taken together, ordering-sensitive and alchemical representations have made representation design itself a disorder-native ML capability. Symmetry-aware GNNs help preserve distinctions among physically inequivalent ordered microstates. Short-range-order-sensitive representations expose local chemical correlations. Alchemical compression further makes chemically complex compositional spaces more navigable, and alchemical MLIPs transform partial occupancy into a directly interpolable and differentiable model parameter. Crucially, these advances are not only architectural refinements that improve prediction accuracy: they also determine which disorder variables remain visible to the model in the first place. At the same time, their value for disorder modeling must ultimately be judged by configurational and thermodynamic fidelity. Physically robust representations must therefore preserve microstate rankings, local ordering trends, order parameters, and ensemble-level observables at an energy scale relevant to Boltzmann weighting, rather than only achieving low errors on conventional held-out energy and force benchmarks.

\paragraph{Enabling disordered structure and distribution generation.}
After vital disorder parameters become visible using ordering-sensitive and alchemical representations, a subsequent disorder-native AI capability is to generate disordered structures or configurational distributions directly. This approach is different from using ML only within an externally defined ensemble workflow. In ML-boosted Monte Carlo, molecular dynamics, active learning, or finite-ensemble averaging, the sampling procedure is generally specified outside the model, while ML offers fast energies, properties, or uncertainty estimates.\cite{Zhong2025a,Guan2025,Fang2025,Deshmukh2024} By contrast, in generative modeling, the learned object can be a plausible disordered crystal description, a representative category of microstates, a benchmarked statistical distribution over disordered configurations, or an approximation to a thermodynamic object (such as a partition function). Thus, ML starts to help build the disorder ensemble itself, rather than only assessing configurations selected by another algorithm.

Generalized generative models for inorganic materials\cite{Metni2026,DeBreuck2025} have already shown how crystal-structure distributions can be learned from large-scale datasets, but they still need stronger ways to represent and validate partial occupancy and related order--disorder relationships. MatterGen has set a leading example: it acts as a diffusion model that simultaneously generates atom types, fractional coordinates, and lattice degrees of freedom, which can be fine-tuned toward chemical, symmetry, or scalar property constraints.\cite{Zeni2025} It represents a crucial advance for inverse materials design, including reported improvements in stable, unique, and new structure generation and an experimental validation example involving a compositionally disordered material. Nonetheless, while the MatterGen workflow has already attempted to account for disorder via order--disorder structural matching,\cite{Zeni2025} a subsequent crystallographic reanalysis\cite{Juelsholt2026} has argued that the Ta--Cr--O validation example is more appropriately interpreted as a known disordered rutile solid-solution family rather than a genuinely novel compound, showing that disorder remains a difficult failure mode for current generative materials modeling. The central challenge is thus not only whether an AI model can generate thermodynamically low-energy ordered structures, but whether it can recognize when an ordered prediction should be better interpreted as merely an approximation to an already known disordered phase. Conditional generation may partly narrow this gap when the parent lattice or host framework is known. As a representative example, crystal host-guided generation combines inpainting generation with universal MLIP relaxation to generate missing or uncertain structural components within various host frameworks, which is naturally related to partial occupancy, intercalation chemistry, and other defective or fractionally occupied frameworks\cite{Zhong2025b} (Fig. \ref{fig:ml_as_an_enabler}d). Thus, while generative models can provide the foundation for learning crystal-structure distributions, disorder-native crystal generation requires tools that can separate new disordered materials from ordered representations of known disorder in solid materials.

Disorder-specific generative modeling thus requires representations that can directly capture partial occupancies, rather than only producing ordered approximations. One route toward this target is to generate Wyckoff-level objects, as shown by WyckoffDiff,\cite{Kelvinius2025} which uses symmetry-based prototypes to preserve crystallographic symmetries by construction. Dis-GEN also offers an essential example of this shift by designing a crystallographic equivariant representation that accommodates partial site occupancies and further ensures symmetry consistency\cite{Petersen2025} (Fig. \ref{fig:ml_as_an_enabler}e). Since this generative framework is trained on experimental structures from the Inorganic Crystal Structure Database\cite{Belsky2002,Hellenbrandt2004} and directly operates on disordered crystallographic descriptions,\cite{Antypov2025} it moves generative modeling closer to the way disordered solids are reported experimentally. The conceptual advance is therefore not simply the generation of another ordered configuration, but the generative modeling of symmetry-consistent partial-occupancy structures that can represent disordered materials more compactly than explicit enumeration of different ordered microstates. More recently, DMFlow has extended this direction by using a unified representation for ordered and disordered crystal structures, combined with simplex-constrained flow matching to generate physically valid disorder weights.\cite{Wu2026} These methods have established the representational basis for disorder-native generation, while also raising the next question of whether generated partial-occupancy descriptions capture the correct configurational statistics in disordered solids.

Generative models for chemical disorder further require benchmarks that assess whether the generated configurations reproduce the correct statistical distribution of site occupations, rather than only whether individual structures appear valid or novel. Dismai-Bench provides a helpful step in this direction through its fixed-composition Fe\textsubscript{60}Ni\textsubscript{20}Cr\textsubscript{20} alloy datasets, where generative models are evaluated for their ability to reproduce face-centered-cubic alloy configurations with various degrees of short-range order.\cite{Yong2024} As the composition and lattice type are preselected, AI-generated alloy structures can be compared directly against known configurational distributions using cluster fingerprints and short-range-order statistics, making this benchmark more relevant to chemical disorder than conventional validity or novelty metrics. This example highlights that disorder-native generative benchmarks should examine not only whether a model can generate plausible atomic configurations, but whether it can reproduce the occupational correlations that define the underlying configurational ensemble in realistic chemically disordered materials.

Beyond disordered crystal-structure generation, a more ambitious disorder-native generative target is the thermodynamic distribution of microstates. Autoregressive samplers adapted to the semi-grand canonical ensemble provide one lattice-based realization of this idea: SEGAL learns condition-dependent microstate probabilities at predefined temperature and chemical potential, driving free-energy and phase-stability estimation while remaining compatible with an arbitrary internal-energy model.\cite{Damewood2022} More recent any-order autoregressive and marginalization models can further improve scalability by enabling out-painting from smaller to larger lattices and memory-efficient direct training, extending thermodynamic generation toward larger crystal supercells while capturing phase transitions and critical behavior.\cite{Du2026} Furthermore, JANUS has extended thermodynamic generative modeling of chemically disordered crystal structures beyond purely occupational lattice sampling by jointly treating discrete chemical identities, vacancies, atomic displacements, and cell-volume fluctuations within semi-grand-canonical and grand-canonical ensembles.\cite{Blessing2026} In alloy benchmarks, JANUS has been shown to reproduce reference mixing free energies, phase boundaries, and short-range-order statistics. Another vital example is an inverse variational autoencoder that has been developed to iteratively generate atomistic configurations, evaluate them via atomistic calculations, and update the model toward estimation of the partition function.\cite{Karcz2026} A feature of this approach is that it does not require a pre-existing training dataset, but instead builds the relevant configurational sampling set through the generative-learning loop itself. For U--Pu mixed oxides, this framework has been used to compute point-defect formation energies and concentrations, thereby showing how local chemical environments influence defect behaviors. These examples highlight how thermodynamic generative modeling can be valuable for modeling chemically disordered materials: the central goal is to recover ensemble statistics, e.g., defect concentrations, free energies, short-range-order tendencies, and averaged properties, without exhaustively enumerating all the microstates that contribute to the partition function.

Overall, generative AI can move disorder simulations from evaluating predefined atomistic configurations toward constructing the vital disordered structures, configurational distributions, and thermodynamic objects that should be analyzed in the first place. Nevertheless, it is critical to stress that generating physically admissible or plausible-looking disordered configurations is not equivalent to sampling the correct thermodynamic ensemble. Symmetry, composition, and other structural constraints may ensure that generated configurations are physically permissible. Predictive thermodynamic modeling further requires the generated distribution to reproduce the appropriate relative statistical weights, along with physically meaningful long- and short-range-order trends. A promising route is conceptualized by Boltzmann generators, for which tractable generated probabilities allow samples to be statistically reweighted toward the target Boltzmann distribution.\cite{Noe2019} Recent scalable implementations have extended this strategy to large crystalline materials for accurate free-energy and phase-stability calculations.\cite{Schebek2026} Generally speaking, when the correct ensemble is not guaranteed by the generative model itself, generated configurations should instead be regarded as proposals for subsequent energy evaluation or incorporated into a statistically consistent sampling or reweighting scheme to recover the target distribution. Thus, generative AI is best viewed as a valuable tool for constructing and prioritizing configurational spaces, rather than as a full replacement for thermodynamically consistent sampling.

\paragraph{Enabling processing-aware disorder kinetic prediction.}
After disordered crystal structures and related thermodynamic distributions are generated, another disorder-native AI capability is to capture the kinetic quantities that control how chemical disorder forms, relaxes, and becomes trapped. This role is different from using ML models (e.g., MLIPs) only to run longer molecular dynamics trajectories: faster PES exploration can extend accessible length and time scales, but direct MLIP-driven molecular dynamics can still be too expensive for activated diffusion events, vacancy-mediated site exchanges, and processing-scale relaxation, as these processes may occur on time scales far longer than affordable atomistic trajectories.\cite{Henkelman2020} By contrast, when ML learns migration barriers, jump probabilities, transition rules, kinetic rates, or mean first-passage times, the learned model becomes a reduced kinetic description of disorder evolution. This distinction is important, as disorder is not always an equilibrium endpoint: it may alternatively be a history-dependent state affected by annealing, quenching, and other common operating conditions.

One route toward processing-aware disorder kinetics is to learn the local transition rules that determine diffusion-induced chemical ordering. A neural-network-based kinetic framework has been established to predict path-dependent migration barriers and simulate individual vacancy-mediated atomistic jumps using an efficient on-lattice structural and chemical representation.\cite{Xing2024} Applied to refractory NbMoTa alloys, this approach has revealed a temperature regime in which B2-type chemical ordering is maximized and connected such a behavior to diffusion multiplicity and heterogeneous atomic mobility. This example highlights how ML-driven kinetic models can be more than fast evaluators of static configurations by learning the barrier landscape and jump statistics that determine which ordering pathways can actually develop through diffusion.

A complementary approach is to learn the time required for a chemically disordered system to relax toward a desired ordering state. Deep reinforcement learning coupled with temporal-difference learning has been used to predict the mean first-passage times for diffusive relaxation and short-range-order generation in the medium-entropy alloy CrCoNi\cite{Chun2025} (Fig. \ref{fig:ml_as_an_enabler}f). In this ML-powered modeling framework, reaction models encode vacancy-diffusion events, reinforcement learning promotes the search for lower-energy ordering trajectories, and a learned time estimator restores the kinetic information needed to connect these trajectories to physical relaxation times. By connecting the dependence of short-range-order formation to temperature, time, and vacancy concentration, this approach helps separate thermodynamic driving forces from kinetic trapping. It helps tackle a key question that static energetics alone cannot answer: not only which ordering pattern is favorable, but how long it takes to form under a particular processing condition.

These kinetic models can be especially essential because experimentally observed chemical disorder often reflects competition between thermodynamic preference and kinetic accessibility. Annealing temperature, quench rate, defect concentration, diffusion barriers, and local chemical environment can jointly determine whether a material may reach equilibrium short-range order, stay metastable, or freeze into a non-equilibrium distribution of local motifs. Static calculations and equilibrium sampling can predict favorable configurations or ensemble averages, but kinetic models are critical for determining whether those configurations are reachable on realistic time scales and whether they would be retained after processing. A limitation of these kinetic models is that small errors in barriers, attempt frequencies, or transition probabilities might compound into much larger prediction errors in long-time evolution. Thus, robust validation requires more than short-time trajectory agreement: these ML-driven kinetic models should be able to capture experimentally known diffusion coefficients, ordering time scales, rare-event statistics, and key order--disorder trends across compositions, temperatures, and other processing conditions.

Taken together, these disorder-native capabilities have transformed AI from a cost-reduction tool for existing workflows into a fundamentally new approach to identify, represent, generate, and dynamically encode the physical objects and design principles that define chemical disorder. Disorder-aware triage helps unveil when ordered predictions require explicit disorder treatment. Ordering-sensitive and alchemical representations preserve the essential structural and chemical variables that largely control microstate ranking, local motif preference, and ensemble statistics. Generative models construct partial-occupancy structures and configurational distributions that can serve as valuable inputs for disorder prediction. Kinetic models eventually connect disorder states to various processing pathways through which they can form, relax, and become trapped. The core value of these disorder-native AI capabilities is not to completely replace SQS, cluster expansion, Monte Carlo, and MLIP-powered sampling, but to make such established tools more targeted, disorder-aware, and experimentally grounded. The central challenge ahead is therefore integration: disorder-native AI approaches must be coupled with statistically rigorous sampling, accurate energy models, uncertainty quantification, and kinetic validation to ensure that disorder can be predicted efficiently and reliably across thermodynamic and processing conditions.

\phantomsection
\addcontentsline{toc}{section}{4. Grand challenges and future perspectives}
\section*{4. Grand challenges and future perspectives}

In summary, traditional and AI-driven modeling approaches have jointly expanded the chemical complexity and atomistic length and time scales accessible for chemically disordered materials. Conventional methods have made chemical disorder computationally tractable by replacing the impractical enumeration of all site-occupancy microstates with complementary approximations (Figs. \ref{fig:conventional_method_comparison} and \ref{fig:conventional_method_demonstration}): effective average media, fitted lattice-model Hamiltonians, representative quasi-random supercells, finite-size ensembles, stochastic sampling, and macroscopic thermodynamic models. These conventional methods have created the physical language of disorder simulations by transforming partial site occupancies, local correlations, configurational entropies, and phase equilibria into directly computable quantities. Nonetheless, these traditional modeling methods remain constrained by a representation--sampling bottleneck: as chemical complexity, sublattice diversity, local structural relaxation, and finite-temperature effects increase, relevant microstates become harder to select, evaluate, weight, and validate without introducing bias. To address this critical bottleneck, AI can play two complementary roles: it can accelerate established modeling strategies (Fig. \ref{fig:ml_as_an_accelerator}), and it can further enable new disorder-native ML workflows, representations, and predictions that have been challenging to realize using conventional methods alone (Fig. \ref{fig:ml_as_an_enabler}). As an accelerator, ML largely reduces the cost of microstate evaluation, expands configurational sampling, and promotes atomistic-to-thermodynamic closure through energy prediction models, uncertainty-boosted searches, and data-driven phase-diagram assessments. As an enabler, ML also creates new disorder-native capabilities, such as disorder-aware workflow triage, ordering-sensitive and alchemical representations, generative construction of disordered lattice structures and thermodynamic distributions, and fast kinetic prediction of processing-dependent orderings. Overall, these methodological developments mark a key transition from merely making disorder computable to making the modeling broader, faster, more accurate, and more directly connected to critical thermodynamic and kinetic quantities that govern realistic chemical disorder.

Despite this methodological transition, faster and more flexible simulation workflows do not necessarily guarantee more faithful prediction of chemical disorder. While traditional atomistic and thermodynamic formalisms are relatively mature, many emerging AI-driven approaches for disorder modeling have so far been demonstrated on simplified lattices, limited chemical spaces, idealized benchmark systems, and confined configurational datasets (Fig. \ref{fig:computational_method_development_timeline}). Consequently, the broader value of these AI-assisted modeling workflows will largely depend on whether they can transfer to more realistic, complex disordered materials, where multiple sublattices, strong local relaxation, entangled internal degrees of freedom, finite-temperature entropies, kinetic trapping, and heterogeneous environments may all jointly control the relevant configurational ensembles. A useful disorder-native AI model should thus capture the free energies, local motif populations, transition temperatures, kinetic accessibility, and important ensemble-averaged observables that determine the experimental behaviors and properties of chemically disordered materials.

\begin{figure}
\phantomsection
\begin{center}
\includegraphics[max size={\textwidth}{\textheight}]{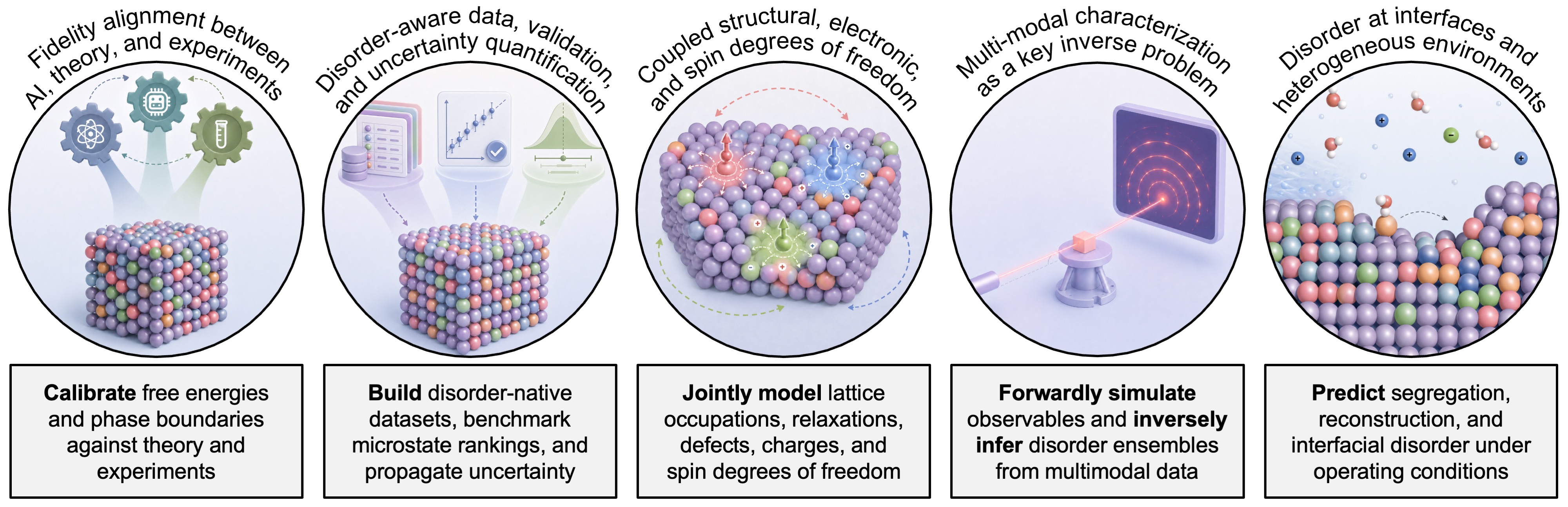}
\caption{
\textbf{Grand challenges and future directions for AI-powered modeling of chemically disordered materials.}
Schematic summary of five key challenges for making AI-driven disorder simulations more predictive, transferable, and experimentally grounded, and the lower boxes translate each challenge into representative actionable research directions. These essential future research directions include: (1) aligning AI with statistical-mechanics theory and experimental observables at the free-energy level; (2) developing disorder-aware datasets, validation protocols, and uncertainty estimates to capture realistic local motifs, microstate rankings, and ensemble properties; (3) extending disorder modeling beyond lattice site occupations to coupled structural, charge, and magnetic degrees of freedom; (4) leveraging multi-modal characterization data as an inverse problem for identifying physically plausible disorder ensembles; and (5) expanding disorder modeling to surfaces, interfaces, and heterogeneous environments involving reservoirs, adsorbates, solvents, and dynamically reactive conditions. Taken together, these directions collectively point toward physically robust AI-accelerated workflows that can reliably elucidate, evaluate, control, and optimize thermodynamically meaningful disorder ensembles for various desired material behaviors and properties.
}
\label{fig:grand_challenges_and_future_perspectives}
\end{center}
\end{figure}

This gap between faster modeling and more trustworthy prediction defines the future agenda for AI-assisted disorder modeling (Fig. \ref{fig:grand_challenges_and_future_perspectives}). The next step is not simply to substitute conventional formalisms with larger ML models, but to integrate these data-driven strategies with free-energy calibration, experimental constraints, uncertainty quantification, additional degrees of freedom, and physically robust ensembles that expand beyond bulk crystals. Such tasks can be viewed as increasingly realistic tests for AI-driven modeling, transitioning from model and data calibration to complex physical variables, experimental inversion, and heterogeneous environments.

\phantomsection
\addcontentsline{toc}{subsection}{4.1 Phase-level fidelity calibration and theory--experiment alignment}
\subsection*{4.1 Phase-level fidelity calibration and theory--experiment alignment}

The first goal in this forward-looking agenda is to make experimentally constrained free-energy behaviors the central targets of AI-assisted atomistic disorder modeling, rather than relying only on energy and force labels. Traditional modeling formalisms and modern ML accelerators have substantially expanded the capability to represent, assess, sample, and interpret configurational ensembles, but this advance does not by itself guarantee predictive alignment with experiments. A model that reproduces DFT energies and forces and promotes a statistically rigorous sampling workflow may still result in incorrect thermodynamics if small residual biases reorder thermally accessible microstates or distort their Boltzmann weights. This distinction is important because experiments typically resolve composition--temperature phase boundaries and short-range-order trends. These experimental observables depend on collective free-energy competition of various microstates across the entire ensemble, rather than on a single relaxed microstate.

Free-energy fidelity should thus be calibrated at the ensemble level, where small PES errors can be amplified into large phase-level errors. In disordered solids, phase stability is determined by how configurational statistics and finite-temperature corrections jointly transform microstate energetics into experimentally observed thermodynamics. The configurational entropy term sets the statistical weight of competing site-occupancy microstates, while vibrational free energies further shift the stability of various ordered and disordered phases. For instance, first-principles phase-stability analysis of the Li\textsubscript{2}S--P\textsubscript{2}S\textsubscript{5} system has shown that Li configurational disorder and further vibrational corrections are crucial for reproducing experimental trends.\cite{Kam2023} Therefore, the key requirement is not only to predict a local PES, but to learn the finite-temperature free-energy hierarchy that determines which disorder states are the most thermodynamically accessible.

A practical solution is observable-driven calibration, where atomistic ML models are refined against experimentally essential ensemble properties. For molecular simulations, differentiable and reweighting-based strategies have already demonstrated that this idea is technically feasible. Differentiable trajectory reweighting can learn MLIPs directly from thermodynamic, structural, and mechanical observables.\cite{Thaler2021} Another differentiable molecular force-field platform has made energies, forces, ensemble averages, and free energies accessible to automatic differentiation.\cite{Wang2023c} Relevant differentiable simulation approaches have shown that dynamical observables (such as transport and spectroscopic data) can serve as calibration targets.\cite{Han2025} Electrolyte-centered studies have shown a similar principle by demonstrating that bottom-up quantum-chemical training can be facilitated by experimental alignment against macroscopic observables.\cite{Gong2025,Yang2026,Wang2026b} Although such examples do not directly study the thermodynamics of chemical disorder, they offer a key lesson for this field: atomistic ML models can remain microscopically grounded while being calibrated against key experimental observables beyond their original energy and force training labels.

For chemically disordered solids, phase diagrams and CALPHAD-derived thermodynamics offer a natural interface for making theory--experiment alignment operational. Rather than using MLIPs only as upstream generators of thermodynamic inputs or leveraging CALPHAD only as a downstream phase-diagram tool, a future workflow should close the loop between them, where MLIPs and other atomistic ML models can provide local configurational energetics, relaxations, and short-range-order information, while experimental phase equilibria can constrain whether these atomistic inputs can predict the correct phase-level free-energy behaviors.\cite{Zhu2025b,Kam2026} A recently developed thermodynamic interatomic potential framework provides a direct realization of this direction by extending a universal MLIP from offering configuration-level potential energies to constructing thermodynamically consistent, differentiable Gibbs free-energy models that can be calibrated against higher-fidelity calculations or experiments.\cite{Nam2026} For disordered alloys, such an approach couples the learned free-energy model with semi-grand-canonical sampling to predict composition-dependent phase behavior. Through differentiable reweighting, the model can then be calibrated against experimental phase equilibria, as demonstrated by improving the predicted Au--Pt miscibility gap toward the experimentally measured phase boundary. Recent experiment-calibrated uncertainty strategies have sharpened this argument further, as the model uncertainty should be evaluated not only by how closely an MLIP reproduces a selected DFT reference, but also by how reliably its results reflect experimentally measurable reality.\cite{Kellner2026} Related simulation-to-experiment distribution-alignment strategies for generative AI models point toward a similar direction by treating the refinement target as an experimentally consistent ensemble, rather than an unconstrained simulation prior.\cite{Nelson2026} In this view, experimental properties of disordered solids should no longer be regarded only as the final validation targets, but as active calibration metrics that can continuously refine atomistic ML models to enable trustworthy disorder prediction.

\phantomsection
\addcontentsline{toc}{subsection}{4.2 Disorder-native data, benchmarks, and uncertainty quantification}
\subsection*{4.2 Disorder-native data, benchmarks, and uncertainty quantification}

Once experimental behaviors become the calibration targets, the next goal is to develop datasets that faithfully represent the disorder ensembles being examined. Critically, dataset construction is not merely a technical task of adding more structures: it is a physical assumption about which microstates, local orderings, distortions, and thermodynamic regimes the AI models are allowed to learn. Prior large-scale databases for materials informatics\cite{Horton2025,Kirklin2015,Zhuang2026} and MLIP training,\cite{Deng2023,Merchant2023,Schmidt2024,Kaplan2025,Mazitov2025a,Levine2026,Malosso2026,BarrosoLuque2026,Peng2026} together with relevant ML prediction benchmarks,\cite{Riebesell2025,Dunn2020,Chiang2025} have been transformative for general atomistic modeling. Nevertheless, these fundamental resources are not designed specifically for disorder thermodynamics, short-range-order statistics, and finite-temperature behaviors. Recent disorder-focused efforts\cite{Wang2026a,Wang2025b,Li2024b,Yong2024} have begun to shift toward disorder-aware data construction and benchmark development, but they remain specialized by material families, disorder classes, target properties, and model types, rather than forming a generalized ML training and validation framework for disorder prediction. Future datasets should thus include not only relaxed ordered prototypes, but also symmetrically different orderings, short-range-order structures, off-lattice relaxations, and other crucial configurations that matter in disordered solids. As such expanded coverage can quickly become combinatorial, disorder-native dataset construction will also need selection strategies that avoid redundant configurations while preserving physically meaningful local environments. Kernel-based or structural-similarity metrics,\cite{Cheng2020,Qi2024} for instance, can prune redundant orderings and identify representative supercells for efficient data generation, making disorder-native datasets broader without making their construction too exhaustive.

Disorder-native datasets also require thermodynamically faithful labels, as disorder stability is often controlled by meV/atom-scale free-energy differences. Hence, systematic label bias can be as damaging as missing configurations. An error that appears minor for individual structures may reorder competing microstates, distort Boltzmann weights, and shift predicted short-range-order tendencies or phase boundaries. The bias can arise from the selection of DFT functionals, as well as from magnetic initialization, charge-state assignments, relaxation protocols, and other factors.\cite{Peng2024a,Kaplan2025,Warford2026} Recent thermodynamic workflows\cite{Horton2025,Kingsbury2022} and MLIP studies\cite{Ko2025,Kim2025a,Kim2026a} have indicated that multi-fidelity learning can greatly improve thermodynamic consistency across chemistries. For disorder modeling, these strategies can be especially helpful, as broad low-fidelity coverage can be paired with targeted higher-fidelity labels for capturing critical configurations that control ordering and phase stability. Overall, new disorder-native datasets should be judged not only by structural diversity and chemical coverage, but also by whether their labels have enough fidelity for the microstate rankings and free-energy trends that AI models are expected to learn.

Benchmarking efforts for AI-assisted disorder modeling must measure disorder fidelity, not only average prediction errors. Mean absolute errors in energies and forces remain valuable, but they do not necessarily imply whether a model preserves the configurational physics of disorder. Such a limitation has already appeared in short-range-order-centered MLIP studies.\cite{Cao2025,Sheriff2026} As an example, in CrCoNi, two training strategies have led to similar energy root-mean-square errors of $5.3$ and $5.6$ meV/atom at $1800$ K but have differed in their ability to reproduce chemical and structural short-range order in the liquid phase, showing that traditional scalar errors can mask differences in how faithfully various MLIP models reproduce local atomic ordering.\cite{Cao2025} Broader universal MLIP benchmarks similarly show that models can perform well near common training distributions, while failing for defects, migration barriers, and non-equilibrium structures.\cite{Chiang2025,Deng2025b} A practical validation hierarchy for disorder-aware AI can move from its model outputs, through configurational and ensemble physics, to experiments. At the model level, conventional metrics (e.g., energy or force errors and structural validity) remain useful baseline tests. Representation-level diagnostics\cite{Edamadaka2025,Chorna2026,Li2026} can further reveal whether models with similar scalar errors encode local chemical or structural motifs and their order--disorder relationships differently. Configurational validation should then assess relative microstate energies and rankings, mixing enthalpies, and local motif preferences, while ensemble-level validation should examine Warren--Cowley short-range-order parameters, pair and multi-site correlations, free energies, order--disorder transition temperatures, and phase boundaries. Finally, experimentally grounded validation should assess and compare forward-predicted diffraction, spectroscopic, and other measurable signatures with observations. Good performance at one level does not necessarily ensure reliability at the next. Hence, the required validation should ultimately match the scientific claim being made. Future disorder-native benchmarks should further test transfer across ordered and disordered structures, random and short-range-ordered states, temperature regimes, and chemical complexities.

Uncertainty quantification should accompany every step of the disorder modeling workflow. A useful question is whether an AI model is interpolating in the training-set disorder ensembles or extrapolating to unseen chemistries. Recent studies\cite{Dai2025,Grasselli2025,Frombgen2026} have widely shown that uncertainty is vital for reliable atomistic simulations. However, uncertainty estimates are not automatically calibrated: prior MLIP work has shown systematic misalignment between predicted uncertainty intervals and true errors before calibration, which has motivated explicit calibration on held-out configurations before uncertainties are propagated.\cite{Ho2026a} For disordered materials, this uncertainty should indicate how confidence levels change when predictions are aggregated from individual microstates into order parameters, free energies, phase boundaries, and macroscopic properties. A useful uncertainty estimate should identify which regions of the configurational ensembles are underrepresented, which label sources may be biased, and which experiments, adversarial stress tests, and higher-fidelity calculations would reduce ambiguity. Hence, disorder-native datasets, benchmarks, and uncertainty estimates should not be treated as supporting infrastructure, but as qualification criteria that assess when ML predictions are trustworthy for disordered solids.

\phantomsection
\addcontentsline{toc}{subsection}{4.3 Disorder beyond site occupations: defect, charge, and magnetism}
\subsection*{4.3 Disorder beyond site occupations: defect, charge, and magnetism}

In solid materials, site occupations and atomic positions alone may not be sufficient to faithfully describe chemical disorder, since structural relaxation, together with defects, charge states, and magnetic degrees of freedom, can be strongly coupled to the occupational configuration. When these coupled variables alter the relative energetics or statistical weights of different microstates, a purely occupational description can therefore become inadequate. Defects illustrate this point, as vacancies, interstitials, and non-stoichiometry can influence not only which species occupy a parent lattice, but also which lattice sites are available, which local atomic motifs are stable, and which migration pathways are accessible under synthesis and operating conditions.\cite{Klein2025} This distinction is especially critical for chemically complex solids, where defect formation energies become distributions over local orderings and chemical environments.\cite{Zhang2023,Tang2025,Ibrahim2026,DeSilva2025,Potter2026} Recent data-driven approaches for defect energetics\cite{Witman2023,Baldassarri2023,Way2025} and defect-driven structural reconstructions\cite{MosqueraLois2024,Yang2025,Wang2026c} suggest that these processes can increasingly be modeled at larger scales. Future disorder-aware AI models should treat defects as part of the same ensembles as chemical disorder, rather than as isolated perturbations after the substitutional disorder has already been resolved.

Charge orderings and long-range interactions provide a second frontier where the relevant disorder variables are not fully specified by atomic positions and species labels. For instance, in mixed-valence solids, multiple charge-ordering patterns may be compatible with similar nuclear arrangements, and electronic configurational entropy can strongly influence phase behaviors.\cite{Zhou2006} A recent study has shown that MLIP-driven structural optimization can assign Fe\textsuperscript{2+}/Fe\textsuperscript{3+} states in Na\textsubscript{\textit{x}}FePO\textsubscript{4} incorrectly and lead to erroneous ordering energetics, whereas embedding charge-state information into the MLIP representations improves the results.\cite{Petersen2026} Charge-informed MLIP simulations\cite{Deng2023} point toward a related direction and have already been useful for modeling phase transformations in mixed-valence battery cathode materials.\cite{Zhong2025a} More broadly, recent long-range MLIP developments have introduced charge equilibration,\cite{Ko2021,Gao2025} latent Ewald summation,\cite{Cheng2025,King2025,Kim2025b,Zhong2025c,Kim2026b} and reciprocal-space\cite{Ramasubramanian2025,Ji2025,Guo2026a,Zhang2026a} or attention-based nonlocal representations\cite{Frank2026,Qu2026} to model atomistic systems where local disorder controls charge transfer, long-range fields, and dielectric response. These architectures are not, by themselves, disorder sampling frameworks, but they provide key machinery for disorder modeling when charge distribution and polarization are critical.

Magnetism offers a third frontier where the disorder ensembles might include spin variables, moment amplitudes, and spin--lattice coupling in addition to lattice occupations. Notably, spin Hamiltonians (e.g., magnetic cluster expansions) offer a lattice-based route that extends cluster-expansion formalisms from lattice-site occupations to localized spin variables and encodes the key exchange, anisotropy, and higher-order magnetic couplings permitted by symmetries.\cite{Drautz2004,Thomas2018,Kitchaev2020} In such a framework, first-principles energies are mapped onto effective spin Hamiltonians on a parent lattice,\cite{Li2021} which can be further sampled through Monte Carlo to predict magnetic phase diagrams\cite{Puchala2023} and composition-dependent phase boundaries in magnetic alloy systems.\cite{Kitchaev2021} The strength of this approach is its interpretability and statistical-mechanics rigor when the magnetic problems map cleanly onto well-defined lattices, but its core limitation emerges when chemical disorder, local relaxation, moment stability, and off-lattice distortions evolve together. To tackle this limitation, recent spin-aware MLIPs\cite{Kostiuchenko2024,Yu2024a,Yu2024b,Yang2024,Xu2025b,Zheng2026,Ho2026b} and Hamiltonian-learning models\cite{Zhong2023b,Zhong2026} push this logic toward continuous atomistic modeling by treating lattice coordinates, magnetic moments, spin--orbit coupling, and electronic-structure Hamiltonians within their learned representations. For disordered magnets, the critical opportunity is thus to move beyond a fixed-lattice magnetic Hamiltonian and toward AI models in which site occupations, relaxations, moment amplitudes, and spin orientations can co-evolve within the same finite-temperature disorder ensembles.

Across these three coupled degrees of freedom, the appropriate treatment depends strongly on the underlying disorder representation. Effective-medium approaches such as VCA and CPA average over local environments, and thus they do not explicitly resolve the full distribution of configuration-specific structural, charge, and magnetic responses. Cluster expansions primarily capture site occupations on a fixed parent lattice, although extended state spaces can incorporate vacancies, charge states, or spin variables yet may simultaneously suffer from limited scalability when multiple species are included. Moreover, explicit supercell methods can recover structural relaxation, charge redistribution, and magnetic responses within each calculated configuration, whereas off-lattice Monte Carlo and molecular-dynamics-based approaches can directly couple occupational changes to continuous relaxation. In all cases, defects, charge states, or spins must be represented explicitly when they are active degrees of freedom in the underlying disorder.

Taken together, defects, charge states, and magnetic variables can expand disorder modeling from a problem of assigning chemical species to lattice sites into a broader problem of sampling coupled structural, electronic, and spin degrees of freedom. The long-term target for disorder simulations is therefore a unified finite-temperature framework in which chemical swaps, defect reactions, atomic relaxations, charge redistribution, long-range electrostatic responses, and spin updates are performed consistently on a shared PES. Such a framework would allow AI-assisted simulations to predict favorable occupational patterns together with defect populations, charge states, and magnetic textures that are physically accessible under realistic conditions.

\phantomsection
\addcontentsline{toc}{subsection}{4.4 Multi-modal characterization as an inverse problem for disorder}
\subsection*{4.4 Multi-modal characterization as an inverse problem for disorder}

The experiment--simulation gap discussed above also makes disorder characterization an inverse problem: rather than predicting experimental results from known atomistic ensembles, one must infer which latent ensembles of atomistic configurations may have led to the measured signals. Different experimental techniques, including diffraction,\cite{Schlegel2025,Coury2023} scattering,\cite{Szymanski2023,Deng2025a} spectroscopies,\cite{Joress2023,Morris2026} imaging,\cite{Lun2021,Moniri2023,Vogl2025} and chemical mapping,\cite{He2024} each probe distinct aspects of chemical disorder, such as average symmetries, pair correlations, charge states, and element-specific local coordination environments. Consequently, none of these modalities uniquely determines the full distribution of microscopic microstates on its own. Notably, this non-uniqueness has long been recognized in the neighboring field of structural disorder,\cite{Drabold2009,Biswas2009,Berthier2011,Jones2020,Berthier2023,Liu2025a,Wolf2025,Madanchi2025} where non-crystalline structures are typically derived from complex experimental data through underdetermined inverse problems, rather than reconstructed atom by atom.\cite{Cliffe2010} Thus, the central challenge is not only to detect whether disorder exists, but to infer which configurational ensembles are consistent with measurements.

This inverse problem can be tackled through two complementary strategies: forward-model-constrained inference and learned inverse generation. First, for the forward direction, candidate ensembles generated by first-principles atomistic simulations or MLIP-powered configurational sampling can be translated into predicted experimental observables. This process is important, since experimental signatures are ensemble projections. For instance, cation-disordered rocksalt materials require both short-range order and bond-length relaxation to reproduce experimental pair distribution functions.\cite{Szymanski2023} This forward-modeling logic extends to other ensemble-sensitive experimental observables: NMR spectra can be constructed by averaging simulated signals over weighted sets of ordered microstates,\cite{Moran2019} while Li--vacancy lattice configurational models can be related to measured voltage profiles, accessible capacities, and transport behaviors in disordered rocksalts.\cite{Zhong2025a,Ji2019,Zhong2023a} Moreover, while extended X-ray absorption fine structure is locally sensitive, complex concentrated alloys can yield non-unique fits.\cite{Joress2023} Likewise, diffuse electron-diffraction features may not uniquely reveal short-range ordering,\cite{Coury2023,Walsh2023,Walsh2024} but the combination of microscopy, diffraction, and atomistic modeling has recently been shown to better constrain the hidden local motifs.\cite{Moniri2023,Vogl2025,Attiaoui2026} Second, for the inverse direction, recently proposed differentiable and generative frameworks have shown how experimental signals can be utilized to refine or generate atomistic structures directly.\cite{Kwon2024,Guo2024,Riesel2024,Lai2025,Guo2025,Li2025,Anker2025,Guo2026b} These inverse-generation demonstrations have centered on ordered and structurally disordered systems, rather than chemically disordered ensembles, but they can offer useful templates for chemical-disorder-specific methods. Together, these tools can generate and eliminate competing ensemble hypotheses for refining against complex multi-modal data.

Beyond inferring and generating candidate structures, AI-driven modeling can help integrate multi-modal measurements into probabilistic constraints on chemical disorder. ML models have been developed to predict or fit complex spectroscopic signatures from atomistic structures and infer structural descriptors from measured spectra.\cite{Kharel2025,Kulaev2026,Grizzi2026} Related methods designed for electron microscopic data can turn large experimental datasets into more quantitative constraints on local structures.\cite{Spurgeon2021,Crozier2025} More broadly, multi-modal and cross-modality frameworks point toward future workflows in which the combined data streams from diffraction, spectroscopy, and microscopy can be analyzed jointly, rather than regarded as independent evidence.\cite{NaNarong2025,Corrao2025,Zhu2026b,Johnson2026} In this setting, AI adds value by propagating constraints across modalities to find the key disorder ensembles.

In summary, AI-driven interpretation of multi-modal experimental data should be regarded as disorder ensemble identification, instead of deterministic reconstruction of a ``true'' atomistic configuration. Even when a structural model reproduces a diffraction pattern with high apparent agreement with experimental data, recent ML analyses have shown that high spectral similarity does not necessarily guarantee structural correctness.\cite{Segal2026} Therefore, a reliable workflow needs to combine AI inference with uncertainty quantification, thermodynamic constraints, and physical priors, together with robust forward and inverse models that elucidate which structural features are truly identifiable. In this regard, characterization becomes an active component of disorder modeling: simulations supply physically plausible microstates, while AI updates the ensembles toward those most consistent with physical constraints and characterization signatures.

\phantomsection
\addcontentsline{toc}{subsection}{4.5 Beyond bulk disorder: interfaces and heterogeneous environments}
\subsection*{4.5 Beyond bulk disorder: interfaces and heterogeneous environments}

Looking ahead, the next stage of disorder modeling must move beyond equilibrium bulk parent lattices toward the heterogeneous environments where interfacial disorder can control material behaviors and properties. Many functional responses are governed not by the bulk phases alone, but by surfaces, grain boundaries, and hetero-interfaces. These environments break translational symmetries, alter local coordination environments, reshape electrostatics and strains, and create local chemical potentials that can possibly lead to segregation, reconstruction, vacancy ordering, amorphization, and interfacial phases that are absent from the bulk.\cite{Kristoffersen2022,Mazitov2024} This issue is especially clear in electrochemistry and heterogeneous catalysis,\cite{Kuznetsov2020,Yuan2022,Zheng2024,Xin2026b,Zhang2026b} where surface reconstruction might occur through leaching, corrosion, redeposition, ion insertion, adsorbate-induced restructuring, and dissolution--precipitation equilibria.\cite{Peng2022b,Shen2022,Peng2023} In these cases, a nominally well-defined material surface or interface is often better understood as a condition-dependent ensemble of metastable states rather than as a single static termination, as discussed in more detail in a recent perspective article on data-driven modeling of electrocatalyst stability and surface reconstruction.\cite{Peng2025} Hence, the relevant disorder configurational ensembles should expand from bulk site occupations to the heterogeneous, reactive, and metastable environments in which materials actually function.

At the modeling level, this expansion calls for grand-canonical, reactive, and kinetics-aware treatments of disorder as a coupled ensemble problem. Surface or interfacial disorder depends not only on site occupations, but also on adsorbate coverage, solvent or electrolyte environment, electrochemical potential, and ion exchange with external environments. While first-principles thermodynamic tools, including surface phase\cite{Reuter2001} and Pourbaix diagrams,\cite{Rong2015} can provide helpful equilibrium baselines, their predictions are limited by the surface terminations, reconstructions, coverages, and metastable states included in the candidate sets.\cite{Du2025,Riccius2025} More fundamentally, these environmental variables can influence the accessible configurational space itself, as atoms may migrate between the bulk and the surface, vacancies may be generated or annihilated, adsorbates or electrolyte species may stabilize otherwise energetically unfavorable motifs, and metastable reconstructions may persist because of kinetic barriers.\cite{Peng2025,Zhang2026} Therefore, extending bulk disorder modeling methods to heterogeneous environments requires more than larger supercells or more complex occupations on a fixed parent lattice: it requires advanced hybrid on-lattice/off-lattice sampling strategies,\cite{Du2025,Du2023} MLIPs validated for defective and low-coordination environments,\cite{SchwalbeKoda2025} alchemical or ordering-sensitive ML representations for local composition or segregation,\cite{Peng2024b,Mazitov2024} and generative AI models that can propose plausible surface microstates beyond human-selected templates.\cite{Ronne2024} For electrochemical interfaces, these workflows must further account for solvation effects, electrolyte ions, electric fields, and charge-transfer processes,\cite{Govindarajan2025,Wang2025c} as these parameters might reshape both the thermodynamic stability and kinetic accessibility of interfacial disorder. Thus, the key goal is not simply to sample more configurations, but to identify which disordered motifs are persistent and kinetically reachable under realistic heterogeneous conditions.

Taken together, these grand challenges point toward a broader methodological transition in atomistic disorder modeling. The central goal is no longer only to transform partial occupancies into tractable configurations, or to accelerate the assessment of many candidate microstates, but to unveil the disorder ensembles that are thermodynamically meaningful, kinetically accessible, experimentally identifiable, and relevant to desired material behaviors and properties. Reaching this target will require AI-accelerated workflows that remain grounded in statistical mechanics, calibrated by experimental observables, qualified by uncertainty estimates, and flexible enough to account for chemical, structural, electronic, magnetic, interfacial, and processing-dependent degrees of freedom. If developed in such a direction, AI will not merely make existing disorder simulations faster: it will help transform chemical disorder from a representational obstacle into a controllable design variable for materials discovery across chemical and functional spaces.

\phantomsection
\section*{Conflicts of interest}

\noindent
There are no conflicts of interest to declare.

\phantomsection
\section*{Data availability}

\noindent
No new data were generated or analyzed as part of this Review.

\phantomsection
\section*{Acknowledgments}

\noindent
J.P. acknowledges the start-up funding from the University at Buffalo, as well as support from the ChatGPT for Academic Researchers program and the ACS Petroleum Research Fund under Doctoral New Investigator Grant 70302-DNI5. P.Z. acknowledges the NUS Presidential Young Professorship and NUS Artificial Intelligence Institute Seed Fund (H2FY2025). 

\clearpage
\phantomsection
\addcontentsline{toc}{section}{References}
\bibliographystyle{rsc}
\bibliography{refs}

@article{Simonov2020,
author = {Simonov, Arkadiy and Goodwin, Andrew L},
doi = {10.1038/s41570-020-00228-3},
issn = {2397-3358},
journal = {Nat. Rev. Chem.},
number = {12},
pages = {657--673},
title = {{Designing disorder into crystalline materials}},
url = {https://doi.org/10.1038/s41570-020-00228-3},
volume = {4},
year = {2020}
}

@article{Shockley1938,
author = {Nix, Foster C and Shockley, William},
doi = {10.1103/RevModPhys.10.1},
journal = {Rev. Mod. Phys.},
month = {jan},
number = {1},
pages = {1--71},
publisher = {American Physical Society},
title = {{Order-disorder transformations in alloys}},
url = {https://link.aps.org/doi/10.1103/RevModPhys.10.1},
volume = {10},
year = {1938}
}

@article{Peng2024a,
author = {Peng, Jiayu and Damewood, James and G{\'{o}}mez-Bombarelli, Rafael},
issn = {2666-3864},
journal = {Cell Rep. Phys. Sci.},
number = {5},
pages = {101942},
title = {{Data-driven physics-informed descriptors of cation ordering in multicomponent perovskite oxides}},
volume = {5},
year = {2024}
}

@article{Young2023,
author = {Young, Samuel D and Chen, Jiadong and Sun, Wenhao and Goldsmith, Bryan R and Pilania, Ghanshyam},
doi = {10.1021/acs.chemmater.3c00943},
issn = {0897-4756},
journal = {Chem. Mater.},
month = {aug},
number = {15},
pages = {5975--5987},
publisher = {American Chemical Society},
title = {{Thermodynamic stability and anion ordering of perovskite oxynitrides}},
url = {https://doi.org/10.1021/acs.chemmater.3c00943},
volume = {35},
year = {2023}
}

@article{Zhang2023,
author = {Zhang, Xie and Kang, Jun and Wei, Su-Huai},
doi = {10.1038/s43588-023-00403-8},
issn = {2662-8457},
journal = {Nat. Comput. Sci.},
number = {3},
pages = {210--220},
title = {{Defect modeling and control in structurally and compositionally complex materials}},
url = {https://doi.org/10.1038/s43588-023-00403-8},
volume = {3},
year = {2023}
}

@article{Han2024a,
author = {Han, Liuliu and Zhu, Shuya and Rao, Ziyuan and Scheu, Christina and Ponge, Dirk and Ludwig, Alfred and Zhang, Hongbin and Gutfleisch, Oliver and Hahn, Horst and Li, Zhiming and Raabe, Dierk},
doi = {10.1038/s41578-024-00720-y},
issn = {2058-8437},
journal = {Nat. Rev. Mater.},
number = {12},
pages = {846--865},
title = {{Multifunctional high-entropy materials}},
url = {https://doi.org/10.1038/s41578-024-00720-y},
volume = {9},
year = {2024}
}

@article{Hsu2024,
author = {Hsu, Wei-Lin and Tsai, Che-Wei and Yeh, An-Chou and Yeh, Jien-Wei},
doi = {10.1038/s41570-024-00602-5},
issn = {2397-3358},
journal = {Nat. Rev. Chem.},
number = {6},
pages = {471--485},
title = {{Clarifying the four core effects of high-entropy materials}},
url = {https://doi.org/10.1038/s41570-024-00602-5},
volume = {8},
year = {2024}
}

@article{Ding2026,
author = {Ding, Jun},
doi = {10.1038/s41578-025-00887-y},
issn = {2058-8437},
journal = {Nat. Rev. Mater.},
number = {2},
pages = {82--83},
title = {{Order or disorder, that is the question in high-entropy alloys}},
url = {https://doi.org/10.1038/s41578-025-00887-y},
volume = {11},
year = {2026}
}

@article{Kang2024,
author = {Kang, Seongkoo and Lee, Suwon and Lee, Hakwoo and Kang, Yong-Mook},
doi = {10.1038/s41570-024-00622-1},
issn = {2397-3358},
journal = {Nat. Rev. Chem.},
number = {8},
pages = {587--604},
title = {{Manipulating disorder within cathodes of alkali-ion batteries}},
url = {https://doi.org/10.1038/s41570-024-00622-1},
volume = {8},
year = {2024}
}

@article{Hirai2026,
author = {Hirai, Daigorou},
doi = {10.1021/acs.chemmater.5c02598},
issn = {0897-4756},
journal = {Chem. Mater.},
month = {jan},
number = {2},
pages = {559--571},
publisher = {American Chemical Society},
title = {{Strongly correlated high-entropy materials: Electronic and phononic states under disorder}},
url = {https://doi.org/10.1021/acs.chemmater.5c02598},
volume = {38},
year = {2026}
}

@article{Cheng2026,
archivePrefix = {arXiv},
author = {Cheng, Eric Jianfeng and Hong, Min and Zeng, Zhiquan and Liu, Chuanyu and Wang, Qian and Meadowcroft, Melissa Jane and Badilita, Vlad and Costa, Carlos Miguel and Lanceros-M{\'{e}}ndez, Senentxu and Li, Ying and Ma, Chenyao and Gong, Weibo and Martha, Surendra and Soon, Aloysius and Ma, Piao and Zhang, Di and Ariyanto, Teguh and Sudaryanto and Kakinuma, Hiroshi and Zhao, Jie and Peng, Jiayu and Ou, Pengfei and Lu, Jun and Orimo, Shin-ichi and Li, Hao},
journal = {Preprint at arXiv},
pages = {10.48550/arXiv.2606.24480},
eprint = {2606.24480},
primaryClass = {cond-mat.mtrl-sci},
title = {{Breaking bottlenecks in solid electrolyte discovery with large artificial intelligence models}},
url = {https://arxiv.org/abs/2606.24480},
year = {2026}
}

@article{Jang2026,
author = {Jang, Seong-Hoon and Yao, Yiwen and Liu, Chuanyu and Zhang, Linda and Zhang, Di and Jia, Xue and Tran, Hung Ba and Cheng, Eric Jianfeng and Sato, Ryuhei and Ohashi, Yusuke and Sato, Toyoto and Hashimoto, Yusuke and Allendorf, Mark D and Artrith, Nongnuch and Baricco, Marcello and Borgschulte, Andreas and Broom, Darren P and Cao, Ang and Chen, Benjamin Wei Jie and Chen, Lixin and Chen, Ping and Cho, Eun Seon and Deledda, Stefano and Ding, Zhao and Dornheim, Martin and Felderhoff, Michael and Filinchuk, Yaroslav and Froudakis, George E and Gao, Mingxia and Gennett, Thomas and Guo, Zaiping and Hamada, Ikutaro and Hattrick-Simpers, Jason and Hauback, Bj{\o}rn C and Hirscher, Michael and Jensen, Torben R and Jia, Baohua and Kim, Hyoung Seop and Kondo, Takahiro and Kutsukake, Kentaro and Li, Xiao-Yan and Liu, Tongliang and Ma, Piao and Mao, Jianfeng and Mohtadi, Rana and Oh, Hyunchul and Paskevicius, Mark and Pickard, Chris J and Pundt, Astrid and Ramirez-Cuesta, Anibal J and Saitoh, Hiroyuki and Shi, Kaihang and Soon, Aloysius and Sun, Chenghua and Wolverton, Chris and Yabu, Hiroshi and Yang, Weijie and Yao, Zhenpeng and Yu, Xuebin and Zou, Jianxin and Hu, Shouyi and Zhou, Panpan and Lin, Xi and Hu, Zhigang and Zhou, Zhenhao and Ou, Pengfei and Peng, Jiayu and Orimo, Shin-ichi and Li, Hao},
doi = {10.1021/acsenergylett.6c01856},
issn = {2380-8195},
journal = {ACS Energy Lett.},
pages = {10.1021/acsenergylett.6c01856},
month = {aug},
title = {{Building a physics-aware AI ecosystem for solid-state hydrogen storage materials}},
url = {https://doi.org/10.1021/acsenergylett.6c01856},
year = {2026}
}

@article{Chen2021a,
author = {Chen, Shuai and Aitken, Zachary H and Pattamatta, Subrahmanyam and Wu, Zhaoxuan and Yu, Zhi Gen and Srolovitz, David J and Liaw, Peter K and Zhang, Yong-Wei},
doi = {10.1038/s41467-021-25264-5},
issn = {2041-1723},
journal = {Nat. Commun.},
number = {1},
pages = {4953},
title = {{Simultaneously enhancing the ultimate strength and ductility of high-entropy alloys via short-range ordering}},
url = {https://doi.org/10.1038/s41467-021-25264-5},
volume = {12},
year = {2021}
}

@article{Zhang2022,
author = {Zhang, Yanwen and Osetsky, Yuri N and Weber, William J},
doi = {10.1021/acs.chemrev.1c00387},
issn = {0009-2665},
journal = {Chem. Rev.},
month = {jan},
number = {1},
pages = {789--829},
publisher = {American Chemical Society},
title = {{Tunable chemical disorder in concentrated alloys: Defect physics and radiation performance}},
url = {https://doi.org/10.1021/acs.chemrev.1c00387},
volume = {122},
year = {2022}
}

@article{Lun2021,
author = {Lun, Zhengyan and Ouyang, Bin and Kwon, Deok-Hwang and Ha, Yang and Foley, Emily E and Huang, Tzu-Yang and Cai, Zijian and Kim, Hyunchul and Balasubramanian, Mahalingam and Sun, Yingzhi and Huang, Jianping and Tian, Yaosen and Kim, Haegyeom and McCloskey, Bryan D and Yang, Wanli and Cl{\'{e}}ment, Rapha{\"{e}}le J and Ji, Huiwen and Ceder, Gerbrand},
doi = {10.1038/s41563-020-00816-0},
issn = {1476-4660},
journal = {Nat. Mater.},
number = {2},
pages = {214--221},
title = {{Cation-disordered rocksalt-type high-entropy cathodes for Li-ion batteries}},
url = {https://doi.org/10.1038/s41563-020-00816-0},
volume = {20},
year = {2021}
}

@article{Zeng2022,
author = {Zeng, Yan and Ouyang, Bin and Liu, Jue and Byeon, Young-Woon and Cai, Zijian and Miara, Lincoln J and Wang, Yan and Ceder, Gerbrand},
doi = {10.1126/science.abq1346},
journal = {Science},
month = {dec},
number = {6626},
pages = {1320--1324},
publisher = {American Association for the Advancement of Science},
title = {{High-entropy mechanism to boost ionic conductivity}},
url = {https://doi.org/10.1126/science.abq1346},
volume = {378},
year = {2022}
}

@article{Peng2021,
author = {Peng, Jiayu and Damewood, James K and Karaguesian, Jessica and G{\'{o}}mez-Bombarelli, Rafael and Shao-Horn, Yang},
doi = {10.1016/j.joule.2021.11.011},
issn = {2542-4351},
journal = {Joule},
number = {12},
pages = {3069--3071},
title = {{Navigating multimetallic catalyst space with Bayesian optimization}},
url = {https://www.sciencedirect.com/science/article/pii/S2542435121005377},
volume = {5},
year = {2021}
}

@article{Peng2025,
author = {Peng, Jiayu},
doi = {10.1063/5.0271797},
issn = {0021-9606},
journal = {J. Chem. Phys.},
month = {jul},
number = {4},
pages = {040902},
title = {{Toward data-driven predictive modeling of electrocatalyst stability and surface reconstruction}},
url = {https://doi.org/10.1063/5.0271797},
volume = {163},
year = {2025}
}

@article{Hass1984,
author = {Hass, K C and Lempert, R J and Ehrenreich, H},
doi = {10.1103/PhysRevLett.52.77},
journal = {Phys. Rev. Lett.},
month = {jan},
number = {1},
pages = {77--80},
publisher = {American Physical Society},
title = {{Effects of chemical and structural disorder in semiconducting pseudobinary alloys}},
url = {https://link.aps.org/doi/10.1103/PhysRevLett.52.77},
volume = {52},
year = {1984}
}

@article{Sokolovskiy2012,
author = {Sokolovskiy, V V and Buchelnikov, V D and Zagrebin, M A and Entel, P and Sahoo, S and Ogura, M},
doi = {10.1103/PhysRevB.86.134418},
journal = {Phys. Rev. B},
month = {oct},
number = {13},
pages = {134418},
publisher = {American Physical Society},
title = {{First-principles investigation of chemical and structural disorder in magnetic Ni2Mn1+xSn1-x Heusler alloys}},
url = {https://link.aps.org/doi/10.1103/PhysRevB.86.134418},
volume = {86},
year = {2012}
}

@article{Moniri2023,
author = {Moniri, Saman and Yang, Yao and Ding, Jun and Yuan, Yakun and Zhou, Jihan and Yang, Long and Zhu, Fan and Liao, Yuxuan and Yao, Yonggang and Hu, Liangbing and Ercius, Peter and Miao, Jianwei},
issn = {1476-4687},
journal = {Nature},
number = {7992},
pages = {564--569},
title = {{Three-dimensional atomic structure and local chemical order of medium- and high-entropy nanoalloys}},
volume = {624},
year = {2023}
}

@book{HumeRothery1969,
address = {London},
author = {Hume-Rothery, W and Smallman, R and Haworth, C},
publisher = {Metals and Metallurgy Trust},
title = {{The Structure of Metals and Alloys}},
year = {1969}
}

@article{Zacharias2025,
author = {Zacharias, Marios and Even, Jacky},
doi = {10.1021/acsenergylett.5c02706},
journal = {ACS Energy Lett.},
month = {nov},
number = {11},
pages = {5635--5643},
publisher = {American Chemical Society},
title = {{Local structural disorder in crystalline materials}},
url = {https://doi.org/10.1021/acsenergylett.5c02706},
volume = {10},
year = {2025}
}

@article{Pena2001,
author = {Pe{\~{n}}a, M A and Fierro, J L G},
doi = {10.1021/cr980129f},
issn = {0009-2665},
journal = {Chem. Rev.},
month = {jul},
number = {7},
pages = {1981--2018},
publisher = {American Chemical Society},
title = {{Chemical structures and performance of perovskite oxides}},
url = {https://doi.org/10.1021/cr980129f},
volume = {101},
year = {2001}
}

@article{Cheng1987,
author = {Cheng, Yang-Tse and Johnson, William L},
doi = {10.1126/science.235.4792.997},
journal = {Science},
number = {4792},
pages = {997--1002},
title = {{Disordered materials: A survey of amorphous solids}},
url = {https://www.science.org/doi/abs/10.1126/science.235.4792.997},
volume = {235},
year = {1987}
}

@article{Drabold2009,
author = {Drabold, D A},
doi = {10.1140/epjb/e2009-00080-0},
issn = {1434-6036},
journal = {Eur. Phys. J. B},
number = {1},
pages = {1--21},
title = {{Topics in the theory of amorphous materials}},
url = {https://doi.org/10.1140/epjb/e2009-00080-0},
volume = {68},
year = {2009}
}

@article{Biswas2009,
author = {Biswas, Parthapratim and Tafen, D N and Inam, F and Cai, Bin and Drabold, D A},
doi = {10.1088/0953-8984/21/8/084207},
journal = {J. Phys. Condens. Matter},
month = {jan},
number = {8},
pages = {084207},
title = {{Materials modeling by design: applications to amorphous solids}},
url = {https://doi.org/10.1088/0953-8984/21/8/084207},
volume = {21},
year = {2009}
}

@article{Berthier2011,
author = {Berthier, Ludovic and Biroli, Giulio},
doi = {10.1103/RevModPhys.83.587},
journal = {Rev. Mod. Phys.},
month = {jun},
number = {2},
pages = {587--645},
publisher = {American Physical Society},
title = {{Theoretical perspective on the glass transition and amorphous materials}},
url = {https://link.aps.org/doi/10.1103/RevModPhys.83.587},
volume = {83},
year = {2011}
}

@article{Jones2020,
author = {Jones, Eric B and Stevanovi{\'{c}}, Vladan},
doi = {10.1038/s41524-020-0329-2},
issn = {2057-3960},
journal = {npj Comput. Mater.},
number = {1},
pages = {56},
title = {{The glassy solid as a statistical ensemble of crystalline microstates}},
url = {https://doi.org/10.1038/s41524-020-0329-2},
volume = {6},
year = {2020}
}

@article{Berthier2023,
author = {Berthier, Ludovic and Reichman, David R},
doi = {10.1038/s42254-022-00548-x},
issn = {2522-5820},
journal = {Nat. Rev. Phys.},
number = {2},
pages = {102--116},
title = {{Modern computational studies of the glass transition}},
url = {https://doi.org/10.1038/s42254-022-00548-x},
volume = {5},
year = {2023}
}

@article{Liu2025a,
author = {Liu, Yuanbin and Madanchi, Ata and Anker, Andy S and Simine, Lena and Deringer, Volker L},
doi = {10.1038/s41578-024-00754-2},
issn = {2058-8437},
journal = {Nat. Rev. Mater.},
number = {3},
pages = {228--241},
title = {{The amorphous state as a frontier in computational materials design}},
url = {https://doi.org/10.1038/s41578-024-00754-2},
volume = {10},
year = {2025}
}

@article{Wolf2025,
author = {Wolf, Laszlo and Novick, Andrew and Stevanovi{\'{c}}, Vladan},
doi = {10.1063/5.0243888},
issn = {0021-8979},
journal = {J. Appl. Phys.},
number = {9},
pages = {095101},
title = {{Modeling glasses from first principles using random structure sampling}},
url = {https://doi.org/10.1063/5.0243888},
volume = {137},
year = {2025}
}

@article{Madanchi2025,
author = {Madanchi, Ata and Azek, Emna and Zongo, Karim and B{\'{e}}land, Laurent K and Mousseau, Normand and Simine, Lena},
doi = {10.1021/acsphyschemau.4c00063},
journal = {ACS Phys. Chem. Au},
month = {jan},
number = {1},
pages = {3--16},
publisher = {American Chemical Society},
title = {{Is the future of materials amorphous? Challenges and opportunities in simulations of amorphous materials}},
url = {https://doi.org/10.1021/acsphyschemau.4c00063},
volume = {5},
year = {2025}
}

@article{Ma2026,
author = {Ma, Chenyao and Zhang, Linda and Chen, Yuheng and Du, Wei and Fang, Shangwen and Jiang, Zihao and Liu, Chuanyu and Ma, Xinyu and Su, Rui and Wang, Gang and Yu, Muyao and Zhong, Dong and Zhu, Jie and Gong, Weibo and Gu, Huan and Li, Limin and Shen, Chen and Wu, Rui and Wu, Zhenghao and Xu, Kan and Zhou, Min and He, Donglin and Huang, Xiayun and Jiang, Shan and Ou, Pengfei and Peng, Jiayu and Zhang, Yuwei and Zhao, Jie and Zhang, Di and Ma, Piao and Li, Zheng-Hao and Li, Hao},
doi = {10.1021/jacsau.6c01014},
issn = {2691-3704},
journal = {JACS Au},
month = {aug},
pages = {10.1021/jacsau.6c01014},
title = {{Empowering polymeric materials discovery by artificial intelligence}},
url = {https://doi.org/10.1021/jacsau.6c01014},
year = {2026}
}

@article{Song2017,
author = {Song, Hongquan and Tian, Fuyang and Hu, Qing-Miao and Vitos, Levente and Wang, Yandong and Shen, Jiang and Chen, Nanxian},
doi = {10.1103/PhysRevMaterials.1.023404},
journal = {Phys. Rev. Mater.},
month = {jul},
number = {2},
pages = {023404},
publisher = {American Physical Society},
title = {{Local lattice distortion in high-entropy alloys}},
url = {https://link.aps.org/doi/10.1103/PhysRevMaterials.1.023404},
volume = {1},
year = {2017}
}

@article{Barber2025,
author = {Barber, John P and Deary, William J and Titus, Andrew N and Bejger, Gerald R and Almishal, Saeed S I and Rost, Christina M},
doi = {10.1039/D5MH01033K},
issn = {2051-6347},
journal = {Mater. Horiz.},
number = {24},
pages = {10453--10477},
publisher = {The Royal Society of Chemistry},
title = {{Disorder by design: Unveiling local structure and functional insights in high entropy oxides}},
url = {http://dx.doi.org/10.1039/D5MH01033K},
volume = {12},
year = {2025}
}

@article{Schlegel2025,
author = {Schlegel, Nicolas and Punke, Stefanie and Clausen, Christian M and Friis-Jensen, Ulrik and Sapnik, Adam F and Stoian, Dragos and Aalling-Frederiksen, Olivia and Gautam, Divyansh and Rossmeisl, Jan and Pittkowski, Rebecca K and Arenz, Matthias and Jensen, Kirsten M {\O}},
doi = {10.1021/acs.chemmater.4c02470},
issn = {0897-4756},
journal = {Chem. Mater.},
month = {feb},
number = {3},
pages = {939--953},
publisher = {American Chemical Society},
title = {{Tracking the formation of high entropy solid solutions and high entropy intermetallics by in situ X-ray diffraction and spectroscopy}},
url = {https://doi.org/10.1021/acs.chemmater.4c02470},
volume = {37},
year = {2025}
}

@article{Wyatt2025,
author = {Wyatt, Brian C and Yang, Yinan and Micha{\l}owski, Pawe{\l} P and Parker, Tetiana and Morency, Yamil{\'{e}}e and Urban, Francesca and Kadagishvili, Givi and Tanwar, Manushree and Muhoza, Sixbert P and Nemani, Srinivasa Kartik and Bedford, Annabelle and Fang, Hui and Hood, Zachary D and Jang, Junwoo and Kamath, Krutarth and Wright, Bethany G and Disko, Rebecca and Thakur, Anupma and Han, Sanguk and Ghosh, Neil and Xu, Xianfan and Fakhraai, Zahra and Gogotsi, Yury and Vojvodic, Aleksandra and Jiang, De-en and Anasori, Babak},
doi = {10.1126/science.adv4415},
journal = {Science},
number = {6764},
pages = {1054--1058},
title = {{Order-to-disorder transition due to entropy in layered and 2D carbides}},
url = {https://www.science.org/doi/abs/10.1126/science.adv4415},
volume = {389},
year = {2025}
}

@article{Huang2024,
author = {Huang, Zhennan and Li, Tangyuan and Li, Boyang and Dong, Qi and Smith, Jacob and Li, Shuke and Xu, Lin and Wang, Guofeng and Chi, Miaofang and Hu, Liangbing},
doi = {10.1021/jacs.3c12048},
issn = {0002-7863},
journal = {J. Am. Chem. Soc.},
month = {jan},
number = {3},
pages = {2167--2173},
publisher = {American Chemical Society},
title = {{Tailoring local chemical ordering via elemental tuning in high-entropy alloys}},
url = {https://doi.org/10.1021/jacs.3c12048},
volume = {146},
year = {2024}
}

@article{Vogl2025,
author = {Vogl, Lilian M and Chen, Shunda and Schweizer, Peter and Jin, Xiaochen and Yu, Shui-Qing and Liu, Jifeng and Li, Tianshu and Minor, Andrew M},
doi = {10.1126/science.adu0719},
journal = {Science},
number = {6767},
pages = {1342--1346},
title = {{Identification of short-range ordering motifs in semiconductors}},
url = {https://www.science.org/doi/abs/10.1126/science.adu0719},
volume = {389},
year = {2025}
}

@article{Chen2021b,
author = {Chen, Xuefei and Wang, Qi and Cheng, Zhiying and Zhu, Mingliu and Zhou, Hao and Jiang, Ping and Zhou, Lingling and Xue, Qiqi and Yuan, Fuping and Zhu, Jing and Wu, Xiaolei and Ma, En},
doi = {10.1038/s41586-021-03428-z},
issn = {1476-4687},
journal = {Nature},
number = {7856},
pages = {712--716},
title = {{Direct observation of chemical short-range order in a medium-entropy alloy}},
url = {https://doi.org/10.1038/s41586-021-03428-z},
volume = {592},
year = {2021}
}

@article{Chen2023a,
author = {Chen, Wei and Li, Lin and Zhu, Qiang and Zhuang, Houlong},
doi = {10.1557/s43577-023-00575-8},
issn = {1938-1425},
journal = {MRS Bull.},
number = {7},
pages = {762--768},
title = {{Chemical short-range order in complex concentrated alloys}},
url = {https://doi.org/10.1557/s43577-023-00575-8},
volume = {48},
year = {2023}
}

@article{Wang2009,
author = {Wang, Yan Ping and Li, Bang Sheng and Fu, Heng Zhi},
doi = {10.1002/adem.200900057},
journal = {Adv. Eng. Mater.},
number = {8},
pages = {641--644},
title = {{Solid solution or intermetallics in a high-entropy alloy}},
url = {https://advanced.onlinelibrary.wiley.com/doi/abs/10.1002/adem.200900057},
volume = {11},
year = {2009}
}

@article{Cui2022,
author = {Cui, Mingjin and Yang, Chunpeng and Hwang, Sooyeon and Yang, Menghao and Overa, Sean and Dong, Qi and Yao, Yonggang and Brozena, Alexandra H and Cullen, David A and Chi, Miaofang and Blum, Thomas F and Morris, David and Finfrock, Zou and Wang, Xizheng and Zhang, Peng and Goncharov, Vitaliy G and Guo, Xiaofeng and Luo, Jian and Mo, Yifei and Jiao, Feng and Hu, Liangbing},
doi = {10.1126/sciadv.abm4322},
journal = {Sci. Adv.},
month = {sep},
number = {4},
pages = {eabm4322},
publisher = {American Association for the Advancement of Science},
title = {{Multi-principal elemental intermetallic nanoparticles synthesized via a disorder-to-order transition}},
url = {https://doi.org/10.1126/sciadv.abm4322},
volume = {8},
year = {2022}
}

@article{Wang2022,
author = {Wang, Hang and He, Quan-Feng and Yang, Yong},
doi = {10.1007/s12598-021-01926-7},
issn = {1867-7185},
journal = {Rare Met.},
number = {6},
pages = {1989--2001},
title = {{High-entropy intermetallics: From alloy design to structural and functional properties}},
url = {https://doi.org/10.1007/s12598-021-01926-7},
volume = {41},
year = {2022}
}

@article{Liu2024a,
author = {Liu, Zheng and Zhang, Laiqi},
doi = {10.1016/j.jallcom.2024.173716},
issn = {0925-8388},
journal = {J. Alloys Compd.},
pages = {173716},
title = {{Phase formation criteria for structurally ordered high-entropy intermetallics}},
url = {https://www.sciencedirect.com/science/article/pii/S0925838824003025},
volume = {981},
year = {2024}
}

@article{Pei2026,
author = {Pei, Zongrui and Gong, Yilun and Singh, Prashant and Li, Yue and K{\"{o}}rmann, Fritz and Xie, Qingge and Wang, Kun and Wu, Xiaoxiang and Mu, Sai and Gao, Michael C and Liaw, Peter K and Tong, Yang and Zhang, Fan and Wang, Yang and Li, Rui},
doi = {10.1016/j.cossms.2026.101254},
issn = {1359-0286},
journal = {Curr. Opin. Solid State Mater. Sci.},
pages = {101254},
title = {{Can chemical short-range order be transformed into a practical alloy-engineering tool?}},
url = {https://www.sciencedirect.com/science/article/pii/S135902862600001X},
volume = {41},
year = {2026}
}

@incollection{Voter2007,
address = {Dordrecht},
author = {Voter, Arthur F},
booktitle = {Radiation Effects in Solids},
editor = {Sickafus, Kurt E and Kotomin, Eugene A and Uberuaga, Blas P},
isbn = {978-1-4020-5295-8},
pages = {1--23},
publisher = {Springer Netherlands},
title = {{Introduction to the kinetic Monte Carlo method}},
year = {2007}
}

@article{Shen2021,
author = {Shen, Zeqi and Du, Jun-Ping and Shinzato, Shuhei and Sato, Yuji and Yu, Peijun and Ogata, Shigenobu},
doi = {https://doi.org/10.1016/j.commatsci.2021.110670},
issn = {0927-0256},
journal = {Comput. Mater. Sci.},
pages = {110670},
title = {{Kinetic Monte Carlo simulation framework for chemical short-range order formation kinetics in a multi-principal-element alloy}},
url = {https://www.sciencedirect.com/science/article/pii/S0927025621003979},
volume = {198},
year = {2021}
}

@article{Xu2023,
author = {Xu, Biao and Ma, Shihua and Huang, Shasha and Zhang, Jun and Xiong, Yaoxu and Fu, Haijun and Xiang, Xuepeng and Zhao, Shijun},
doi = {10.1103/PhysRevMaterials.7.033605},
journal = {Phys. Rev. Mater.},
month = {mar},
number = {3},
pages = {033605},
publisher = {American Physical Society},
title = {{Influence of short-range order on diffusion in multiprincipal element alloys from long-time atomistic simulations}},
url = {https://link.aps.org/doi/10.1103/PhysRevMaterials.7.033605},
volume = {7},
year = {2023}
}

@article{Li2024,
author = {Li, Yangen and Du, Jun-Ping and Shinzato, Shuhei and Ogata, Shigenobu},
doi = {10.1038/s41524-024-01322-6},
issn = {2057-3960},
journal = {npj Comput. Mater.},
number = {1},
pages = {134},
title = {{Tunable interstitial and vacancy diffusivity by chemical ordering control in CrCoNi medium-entropy alloy}},
url = {https://doi.org/10.1038/s41524-024-01322-6},
volume = {10},
year = {2024}
}

@article{Mees2012,
author = {Mees, Maarten J and Pourtois, Geoffrey and Neyts, Erik C and Thijsse, Barend J and Stesmans, Andr{\'{e}}},
doi = {10.1103/PhysRevB.85.134301},
journal = {Phys. Rev. B},
month = {apr},
number = {13},
pages = {134301},
publisher = {American Physical Society},
title = {{Uniform-acceptance force-bias Monte Carlo method with time scale to study solid-state diffusion}},
url = {https://link.aps.org/doi/10.1103/PhysRevB.85.134301},
volume = {85},
year = {2012}
}

@article{Bal2014,
author = {Bal, Kristof M and Neyts, Erik C},
doi = {10.1063/1.4902136},
issn = {0021-9606},
journal = {J. Chem. Phys.},
month = {nov},
number = {20},
pages = {204104},
title = {{On the time scale associated with Monte Carlo simulations}},
url = {https://doi.org/10.1063/1.4902136},
volume = {141},
year = {2014}
}

@article{Ferrari2023,
author = {Ferrari, Alberto and K{\"{o}}rmann, Fritz and Asta, Mark and Neugebauer, J{\"{o}}rg},
issn = {2662-8457},
journal = {Nat. Comput. Sci.},
number = {3},
pages = {221--229},
title = {{Simulating short-range order in compositionally complex materials}},
volume = {3},
year = {2023}
}

@article{Santodonato2018,
author = {Santodonato, L J and Liaw, P K and Unocic, R R and Bei, H and Morris, J R},
doi = {10.1038/s41467-018-06757-2},
issn = {2041-1723},
journal = {Nat. Commun.},
number = {1},
pages = {4520},
title = {{Predictive multiphase evolution in Al-containing high-entropy alloys}},
url = {https://doi.org/10.1038/s41467-018-06757-2},
volume = {9},
year = {2018}
}

@article{EkborgTanner2024,
author = {Ekborg-Tanner, Pernilla and Rosander, Petter and Fransson, Erik and Erhart, Paul},
doi = {10.1103/PRXEnergy.3.042001},
journal = {PRX Energy},
month = {oct},
number = {4},
pages = {042001},
publisher = {American Physical Society},
title = {{Construction and sampling of alloy cluster expansions---A tutorial}},
url = {https://link.aps.org/doi/10.1103/PRXEnergy.3.042001},
volume = {3},
year = {2024}
}

@article{Brahlek2022,
author = {Brahlek, Matthew and Gazda, Maria and Keppens, Veerle and Mazza, Alessandro R and McCormack, Scott J and Mielewczyk-Gry{\'{n}}, Aleksandra and Musico, Brianna and Page, Katharine and Rost, Christina M and Sinnott, Susan B and Toher, Cormac and Ward, Thomas Z and Yamamoto, Ayako},
doi = {10.1063/5.0122727},
issn = {2166-532X},
journal = {APL Mater.},
number = {11},
pages = {110902},
title = {{What is in a name: Defining “high entropy” oxides}},
url = {https://doi.org/10.1063/5.0122727},
volume = {10},
year = {2022}
}

@article{Aamlid2023,
author = {Aamlid, Solveig S and Oudah, Mohamed and Rottler, J{\"{o}}rg and Hallas, Alannah M},
doi = {10.1021/jacs.2c11608},
issn = {0002-7863},
journal = {J. Am. Chem. Soc.},
month = {mar},
number = {11},
pages = {5991--6006},
publisher = {American Chemical Society},
title = {{Understanding the role of entropy in high entropy oxides}},
url = {https://doi.org/10.1021/jacs.2c11608},
volume = {145},
year = {2023}
}

@article{Bragg1934,
author = {Bragg, William Lawrence and Williams, Evan James},
doi = {10.1098/rspa.1934.0132},
issn = {0950-1207},
journal = {Proc. R. Soc. London. A. Math. Phys. Sci.},
number = {855},
pages = {699--730},
title = {{The effect of thermal agitation on atomic arrangement in alloys}},
url = {https://doi.org/10.1098/rspa.1934.0132},
volume = {145},
year = {1934}
}

@article{Bragg1935,
author = {Bragg, William Lawrence and Williams, Evan James},
doi = {10.1098/rspa.1935.0165},
issn = {0080-4630},
journal = {Proc. R. Soc. London. A. Math. Phys. Sci.},
number = {874},
pages = {540--566},
title = {{The effect of thermal agitation on atomic arrangement in alloys—II}},
url = {https://doi.org/10.1098/rspa.1935.0165},
volume = {151},
year = {1935}
}

@article{Williams1935,
author = {Williams, Evan James},
doi = {10.1098/rspa.1935.0188},
issn = {0080-4630},
journal = {Proc. R. Soc. London. A. Math. Phys. Sci.},
number = {875},
pages = {231--252},
title = {{The effect of thermal agitation on atomic arrangement in alloys-III}},
url = {https://doi.org/10.1098/rspa.1935.0188},
volume = {152},
year = {1935}
}

@article{Toher2019,
author = {Toher, Cormac and Oses, Corey and Hicks, David and Curtarolo, Stefano},
doi = {10.1038/s41524-019-0206-z},
issn = {2057-3960},
journal = {npj Comput. Mater.},
number = {1},
pages = {69},
title = {{Unavoidable disorder and entropy in multi-component systems}},
url = {https://doi.org/10.1038/s41524-019-0206-z},
volume = {5},
year = {2019}
}

@article{Dey2024,
author = {Dey, Dibyendu and Liang, Liangbo and Yu, Liping},
doi = {10.1021/jacs.4c00209},
issn = {0002-7863},
journal = {J. Am. Chem. Soc.},
month = {feb},
number = {8},
pages = {5142--5151},
publisher = {American Chemical Society},
title = {{Mixed Enthalpy–Entropy Descriptor for the Rational Design of Synthesizable High-Entropy Materials Over Vast Chemical Spaces}},
url = {https://doi.org/10.1021/jacs.4c00209},
volume = {146},
year = {2024}
}

@article{Almishal2025,
author = {Almishal, Saeed S I and Furst, Matthew and Tan, Yueze and Sivak, Jacob T and Bejger, Gerald and Petruska, Joseph and Ayyagari, Sai Venkata Gayathri and Srikanth, Dhiya and Alem, Nasim and Rost, Christina M and Sinnott, Susan B and Chen, Long-Qing and Maria, Jon-Paul},
doi = {10.1038/s41467-025-63567-z},
issn = {2041-1723},
journal = {Nat. Commun.},
number = {1},
pages = {8211},
title = {{Thermodynamics-inspired high-entropy oxide synthesis}},
url = {https://doi.org/10.1038/s41467-025-63567-z},
volume = {16},
year = {2025}
}

@article{Sivak2025,
author = {Sivak, Jacob T and Almishal, Saeed S I and Caucci, Mary Kathleen and Tan, Yueze and Srikanth, Dhiya and Petruska, Joseph and Furst, Matthew and Chen, Long-Qing and Rost, Christina M and Maria, Jon-Paul and Sinnott, Susan B},
doi = {10.1103/PhysRevLett.134.216101},
journal = {Phys. Rev. Lett.},
month = {may},
number = {21},
pages = {216101},
publisher = {American Physical Society},
title = {{Discovering high-entropy oxides with a machine-learning interatomic potential}},
url = {https://link.aps.org/doi/10.1103/PhysRevLett.134.216101},
volume = {134},
year = {2025}
}

@article{Dicks2026,
author = {Dicks, Oliver A and Aamlid, Solveig S and Hallas, Alannah M and Rottler, Joerg},
doi = {10.1016/j.commatsci.2026.114581},
issn = {0927-0256},
journal = {Comput. Mater. Sci.},
pages = {114581},
title = {{Expanding the search space of high entropy oxides and predicting synthesizability using machine learning interatomic potentials}},
url = {https://www.sciencedirect.com/science/article/pii/S092702562600100X},
volume = {267},
year = {2026}
}

@article{Xing2024,
author = {Xing, Bin and Rupert, Timothy J and Pan, Xiaoqing and Cao, Penghui},
doi = {10.1038/s41467-024-47927-9},
issn = {2041-1723},
journal = {Nat. Commun.},
number = {1},
pages = {3879},
title = {{Neural network kinetics for exploring diffusion multiplicity and chemical ordering in compositionally complex materials}},
url = {https://doi.org/10.1038/s41467-024-47927-9},
volume = {15},
year = {2024}
}

@article{Han2024b,
author = {Han, Ying and Chen, Hangman and Sun, Yongwen and Liu, Jian and Wei, Shaolou and Xie, Bijun and Zhang, Zhiyu and Zhu, Yingxin and Li, Meng and Yang, Judith and Chen, Wen and Cao, Penghui and Yang, Yang},
doi = {10.1038/s41467-024-49606-1},
issn = {2041-1723},
journal = {Nat. Commun.},
number = {1},
pages = {6486},
title = {{Ubiquitous short-range order in multi-principal element alloys}},
url = {https://doi.org/10.1038/s41467-024-49606-1},
volume = {15},
year = {2024}
}

@article{Bacurau2024,
author = {Bacurau, Vin{\'{i}}cius P and Moreira, Pedro A F P and Bertoli, Gustavo and Andreoli, Angelo F and Mazzer, Eric and de Assis, Fl{\'{a}}vio F and Gargarella, Piter and Koga, Guilherme and Stumpf, Guilherme C and Figueroa, Santiago J A and Widom, Michael and Kaufman, Michael and Fantin, Andrea and Cao, Yifan and Freitas, Rodrigo and Miracle, Daniel and Coury, Francisco G},
doi = {10.1038/s41467-024-52018-w},
issn = {2041-1723},
journal = {Nat. Commun.},
number = {1},
pages = {7815},
title = {{Comprehensive analysis of ordering in CoCrNi and CrNi2 alloys}},
url = {https://doi.org/10.1038/s41467-024-52018-w},
volume = {15},
year = {2024}
}

@article{Islam2025,
author = {Islam, Mahmudul and Sheriff, Killian and Cao, Yifan and Freitas, Rodrigo},
doi = {10.1038/s41467-025-64733-z},
issn = {2041-1723},
journal = {Nat. Commun.},
number = {1},
pages = {8926},
title = {{Nonequilibrium chemical short-range order in metallic alloys}},
url = {https://doi.org/10.1038/s41467-025-64733-z},
volume = {16},
year = {2025}
}

@article{Zhong2025a,
author = {Zhong, Peichen and Deng, Bowen and Anand, Shashwat and Mishra, Tara and Ceder, Gerbrand},
doi = {10.1103/mk2d-tjyj},
journal = {Phys. Rev. Mater.},
month = {oct},
number = {10},
pages = {105404},
publisher = {American Physical Society},
title = {{Modeling phase transformations in Mn-rich disordered rocksalt cathodes with machine-learning interatomic potentials}},
url = {https://link.aps.org/doi/10.1103/mk2d-tjyj},
volume = {9},
year = {2025}
}

@article{Chun2025,
archivePrefix = {arXiv},
author = {Hoje Chun and Hao Tang and Bin Xing and Rafael Gomez-Bombarelli and Ju Li},
journal = {Preprint at arXiv},
pages = {10.48550/arXiv.2411.17839},
eprint = {2411.17839},
primaryClass = {cond-mat.mtrl-sci},
title = {{Deep learning of mean first passage time scape: Chemical short-range order and kinetics of diffusive relaxation}},
url = {https://arxiv.org/abs/2411.17839},
year = {2025}
}

@article{Szymanski2023,
author = {Szymanski, Nathan J and Lun, Zhengyan and Liu, Jue and Self, Ethan C and Bartel, Christopher J and Nanda, Jagjit and Ouyang, Bin and Ceder, Gerbrand},
doi = {10.1021/acs.chemmater.2c03827},
issn = {0897-4756},
journal = {Chem. Mater.},
month = {jul},
number = {13},
pages = {4922--4934},
publisher = {American Chemical Society},
title = {{Modeling short-range order in disordered rocksalt cathodes by pair distribution function analysis}},
url = {https://doi.org/10.1021/acs.chemmater.2c03827},
volume = {35},
year = {2023}
}

@article{Deng2025a,
archivePrefix = {arXiv},
author = {Deng, Hao and Qi, Jue-Yi and Xia, Qin-Han and Li, Jinshan and Zhang, Xie},
journal = {Preprint at arXiv},
pages = {10.48550/arXiv.2506.05684},
eprint = {2506.05684},
primaryClass = {cond-mat.mtrl-sci},
title = {{Nature of nonanalytic chemical short-range order in metallic alloys}},
url = {https://arxiv.org/abs/2506.05684},
year = {2025}
}

@article{Joress2023,
author = {Joress, Howie and Ravel, Bruce and Anber, Elaf and Hollenbach, Jonathan and Sur, Debashish and Hattrick-Simpers, Jason and Taheri, Mitra L and DeCost, Brian},
doi = {10.1016/j.matt.2023.09.010},
issn = {2590-2385},
journal = {Matter},
number = {11},
pages = {3763--3781},
title = {{Why is EXAFS for complex concentrated alloys so hard? Challenges and opportunities for measuring ordering with X-ray absorption spectroscopy}},
url = {https://www.sciencedirect.com/science/article/pii/S2590238523004666},
volume = {6},
year = {2023}
}

@article{Morris2026,
author = {Morris, David and Yao, Yonggang and Zhang, Peng},
doi = {10.1021/acsnano.5c21782},
issn = {1936-0851},
journal = {ACS Nano},
month = {mar},
number = {11},
pages = {8959--8970},
publisher = {American Chemical Society},
title = {{When atoms choose their neighbors: Element-specific views of local chemical order in high-entropy alloys}},
url = {https://doi.org/10.1021/acsnano.5c21782},
volume = {20},
year = {2026}
}

@article{He2024,
author = {He, Mengwei and Davids, William J and Breen, Andrew J and Ringer, Simon P},
doi = {10.1038/s41563-024-01912-1},
issn = {1476-4660},
journal = {Nat. Mater.},
number = {9},
pages = {1200--1207},
title = {{Quantifying short-range order using atom probe tomography}},
url = {https://doi.org/10.1038/s41563-024-01912-1},
volume = {23},
year = {2024}
}

@article{Cowley1950,
author = {Cowley, J M},
doi = {10.1103/PhysRev.77.669},
journal = {Phys. Rev.},
month = {mar},
number = {5},
pages = {669--675},
publisher = {American Physical Society},
title = {{An approximate theory of order in alloys}},
url = {https://link.aps.org/doi/10.1103/PhysRev.77.669},
volume = {77},
year = {1950}
}

@article{Norman1951,
author = {Norman, N and Warren, B E},
doi = {10.1063/1.1699988},
issn = {0021-8979},
journal = {J. Appl. Phys.},
number = {4},
pages = {483--486},
title = {{X‐ray measurement of short range order in Ag‐Au}},
url = {https://doi.org/10.1063/1.1699988},
volume = {22},
year = {1951}
}

@article{Raabe2023,
author = {Raabe, Dierk and Mianroodi, Jaber Rezaei and Neugebauer, J{\"{o}}rg},
doi = {10.1038/s43588-023-00412-7},
issn = {2662-8457},
journal = {Nat. Comput. Sci.},
number = {3},
pages = {198--209},
title = {{Accelerating the design of compositionally complex materials via physics-informed artificial intelligence}},
url = {https://doi.org/10.1038/s43588-023-00412-7},
volume = {3},
year = {2023}
}

@article{Jiao2010,
author = {Jiao, Yang and Stillinger, Frank H. and Torquato, Salvatore},
doi = {10.1103/PhysRevE.81.011105},
journal = {Phys. Rev. E},
month = {jan},
number = {1},
pages = {011105},
publisher = {American Physical Society},
title = {{Geometrical ambiguity of pair statistics: Point configurations}},
url = {https://link.aps.org/doi/10.1103/PhysRevE.81.011105},
volume = {81},
year = {2010}
}

@article{Coury2023,
author = {Coury, Francisco Gil and Miller, Cody and Field, Robert and Kaufman, Michael},
doi = {10.1038/s41586-023-06530-6},
issn = {1476-4687},
journal = {Nature},
number = {7984},
pages = {742--747},
title = {{On the origin of diffuse intensities in fcc electron diffraction patterns}},
url = {https://doi.org/10.1038/s41586-023-06530-6},
volume = {622},
year = {2023}
}

@article{Walsh2023,
author = {Walsh, Flynn and Zhang, Mingwei and Ritchie, Robert O and Minor, Andrew M and Asta, Mark},
doi = {10.1038/s41563-023-01570-9},
issn = {1476-4660},
journal = {Nat. Mater.},
number = {8},
pages = {926--929},
title = {{Extra electron reflections in concentrated alloys do not necessitate short-range order}},
url = {https://doi.org/10.1038/s41563-023-01570-9},
volume = {22},
year = {2023}
}

@article{Walsh2024,
author = {Walsh, Flynn and Zhang, Mingwei and Ritchie, Robert O and Asta, Mark and Minor, Andrew M},
doi = {10.1126/sciadv.adn9673},
journal = {Sci. Adv.},
number = {31},
pages = {eadn9673},
title = {{Multiple origins of extra electron diffractions in fcc metals}},
url = {https://www.science.org/doi/abs/10.1126/sciadv.adn9673},
volume = {10},
year = {2024}
}

@article{Yuan2026,
author = {Yuan, Eric C.-Y. and Liu, Yunsheng and Chen, Junmin and Zhong, Peichen and Raja, Sanjeev and Kreiman, Tobias and Vargas, Santiago and Xu, Wenbin and Head-Gordon, Martin and Yang, Chao and Blau, Samuel M and Cheng, Bingqing and Krishnapriyan, Aditi and Head-Gordon, Teresa},
doi = {10.1038/s41570-025-00793-5},
issn = {2397-3358},
journal = {Nat. Rev. Chem.},
pages = {212--230},
title = {{Foundation models for atomistic simulation of chemistry and materials}},
url = {https://doi.org/10.1038/s41570-025-00793-5},
volume = {10},
year = {2026}
}

@article{Metni2026,
author = {Metni, Houssam and Ruple, Laura and Walters, Lauren N and Torresi, Luca and Teufel, Jonas and Schopmans, Henrik and {\"{O}}streicher, Jona and Zhang, Yumeng and Neubert, Marlen and Koide, Yuri and Steiner, Kevin and Link, Paul and B{\"{a}}r, Lukas and Petrova, Mariana and Ceder, Gerbrand and Friederich, Pascal},
doi = {10.1002/adma.202523620},
issn = {0935-9648},
journal = {Adv. Mater.},
month = {feb},
number = {18},
pages = {e23620},
publisher = {John Wiley & Sons, Ltd},
title = {{Generative models for crystalline materials}},
url = {https://doi.org/10.1002/adma.202523620},
volume = {38},
year = {2026}
}

@article{Peng2022a,
author = {Peng, Jiayu and Schwalbe-Koda, Daniel and Akkiraju, Karthik and Xie, Tian and Giordano, Livia and Yu, Yang and Eom, C John and Lunger, Jaclyn R and Zheng, Daniel J and Rao, Reshma R and Muy, Sokseiha and Grossman, Jeffrey C and Reuter, Karsten and G{\'{o}}mez-Bombarelli, Rafael and Shao-Horn, Yang},
issn = {2058-8437},
journal = {Nat. Rev. Mater.},
pages = {991--1009},
title = {{Human- and machine-centred designs of molecules and materials for sustainability and decarbonization}},
volume = {7},
year = {2022}
}

@article{Cheetham2024,
author = {Cheetham, Anthony K and Seshadri, Ram},
doi = {10.1021/acs.chemmater.4c00643},
issn = {0897-4756},
journal = {Chem. Mater.},
month = {apr},
number = {8},
pages = {3490--3495},
publisher = {American Chemical Society},
title = {{Artificial intelligence driving materials discovery? Perspective on the article: Scaling deep learning for materials discovery}},
url = {https://doi.org/10.1021/acs.chemmater.4c00643},
volume = {36},
year = {2024}
}

@article{vanderVen2018,
author = {van der Ven, A and Thomas, J C and Puchala, B and Natarajan, A R},
doi = {https://doi.org/10.1146/annurev-matsci-070317-124443},
issn = {1545-4118},
journal = {Annu. Rev. Mater. Res.},
pages = {27--55},
publisher = {Annual Reviews},
title = {{First-principles statistical mechanics of multicomponent crystals}},
type = {Journal Article},
url = {https://www.annualreviews.org/content/journals/10.1146/annurev-matsci-070317-124443},
volume = {48},
year = {2018}
}

@article{Xie2022,
author = {Xie, Jun-Zhong and Zhou, Xu-Yuan and Jiang, Hong},
doi = {10.1063/5.0106788},
issn = {0021-9606},
journal = {J. Chem. Phys.},
month = {nov},
number = {20},
pages = {200901},
title = {{Perspective on optimal strategies of building cluster expansion models for configurationally disordered materials}},
url = {https://doi.org/10.1063/5.0106788},
volume = {157},
year = {2022}
}

@article{Ghazisaeidi2026,
author = {Ghazisaeidi, Maryam and Arr{\'{o}}yave, Raymundo and Varvenne, C{\'{e}}line},
doi = {https://doi.org/10.1146/annurev-matsci-072924-115031},
issn = {1545-4118},
journal = {Annu. Rev. Mater. Res.},
pages = {311--338},
publisher = {Annual Reviews},
title = {{Modeling high-entropy alloys: Challenges, progress, and future directions}},
type = {Journal Article},
url = {https://www.annualreviews.org/content/journals/10.1146/annurev-matsci-072924-115031},
volume = {56},
year = {2026}
}

@article{Wang2023a,
author = {Wang, Lin and He, Zhengda and Ouyang, Bin},
doi = {https://doi.org/10.1016/j.commatsci.2023.112513},
issn = {0927-0256},
journal = {Comput. Mater. Sci.},
pages = {112513},
title = {{Data driven design of compositionally complex energy materials}},
url = {https://www.sciencedirect.com/science/article/pii/S0927025623005074},
volume = {230},
year = {2023}
}

@article{Hart2021,
author = {Hart, Gus L W and Mueller, Tim and Toher, Cormac and Curtarolo, Stefano},
doi = {10.1038/s41578-021-00340-w},
issn = {2058-8437},
journal = {Nat. Rev. Mater.},
number = {8},
pages = {730--755},
title = {{Machine learning for alloys}},
url = {https://doi.org/10.1038/s41578-021-00340-w},
volume = {6},
year = {2021}
}

@article{Liu2023,
author = {Liu, Xianglin and Zhang, Jiaxin and Pei, Zongrui},
doi = {https://doi.org/10.1016/j.pmatsci.2022.101018},
issn = {0079-6425},
journal = {Prog. Mater. Sci.},
pages = {101018},
title = {{Machine learning for high-entropy alloys: Progress, challenges and opportunities}},
url = {https://www.sciencedirect.com/science/article/pii/S0079642522000998},
volume = {131},
year = {2023}
}

@article{Nordheim1931,
author = {Nordheim, Lothar},
doi = {10.1002/andp.19314010507},
issn = {0003-3804},
journal = {Ann. Phys.},
month = {jan},
number = {5},
pages = {607--640},
publisher = {John Wiley & Sons, Ltd},
title = {{Zur Elektronentheorie der Metalle. I}},
url = {https://doi.org/10.1002/andp.19314010507},
volume = {401},
year = {1931}
}

@article{Soven1967,
author = {Soven, Paul},
doi = {10.1103/PhysRev.156.809},
journal = {Phys. Rev.},
month = {apr},
number = {3},
pages = {809--813},
publisher = {American Physical Society},
title = {{Coherent-potential model of substitutional disordered alloys}},
url = {https://link.aps.org/doi/10.1103/PhysRev.156.809},
volume = {156},
year = {1967}
}

@book{Kaufman1970,
address = {New York},
author = {Larry Kaufman and Harold Bernstein},
publisher = {Academic Press},
title = {{Computer Calculation of Phase Diagrams with Special Reference to Refractory Metals}},
year = {1970}
}

@article{Sanchez1984,
author = {Sanchez, J M and Ducastelle, F and Gratias, D},
doi = {10.1016/0378-4371(84)90096-7},
issn = {0378-4371},
journal = {Phys. A: Stat. Mech. Appl.},
number = {1},
pages = {334--350},
title = {{Generalized cluster description of multicomponent systems}},
url = {https://www.sciencedirect.com/science/article/pii/0378437184900967},
volume = {128},
year = {1984}
}

@article{vandeWalle2002a,
author = {van de Walle, A and Asta, M and Ceder, G},
doi = {10.1016/S0364-5916(02)80006-2},
issn = {0364-5916},
journal = {Calphad},
number = {4},
pages = {539--553},
title = {{The alloy theoretic automated toolkit: A user guide}},
url = {https://www.sciencedirect.com/science/article/pii/S0364591602800062},
volume = {26},
year = {2002}
}

@article{Angqvist2019,
author = {{\AA}ngqvist, Mattias and Mu{\~{n}}oz, William A and Rahm, J Magnus and Fransson, Erik and Durniak, C{\'{e}}line and Rozyczko, Piotr and Rod, Thomas H and Erhart, Paul},
doi = {10.1002/adts.201900015},
issn = {2513-0390},
journal = {Adv. Theory Simul.},
month = {jul},
number = {7},
pages = {1900015},
publisher = {John Wiley & Sons, Ltd},
title = {{ICET – A Python library for constructing and sampling alloy cluster expansions}},
url = {https://doi.org/10.1002/adts.201900015},
volume = {2},
year = {2019}
}

@article{BarrosoLuque2022,
author = {Barroso-Luque, Luis and Zhong, Peichen and Yang, Julia H and Xie, Fengyu and Chen, Tina and Ouyang, Bin and Ceder, Gerbrand},
doi = {10.1103/PhysRevB.106.144202},
journal = {Phys. Rev. B},
month = {oct},
number = {14},
pages = {144202},
publisher = {American Physical Society},
title = {{Cluster expansions of multicomponent ionic materials: Formalism and methodology}},
url = {https://link.aps.org/doi/10.1103/PhysRevB.106.144202},
volume = {106},
year = {2022}
}

@article{Zunger1990,
author = {Zunger, Alex and Wei, S.-H. and Ferreira, L G and Bernard, James E},
doi = {10.1103/PhysRevLett.65.353},
journal = {Phys. Rev. Lett.},
month = {jul},
number = {3},
pages = {353--356},
publisher = {American Physical Society},
title = {{Special quasirandom structures}},
url = {https://link.aps.org/doi/10.1103/PhysRevLett.65.353},
volume = {65},
year = {1990}
}

@article{Wei1990,
author = {Wei, S.-H. and Ferreira, L G and Bernard, James E and Zunger, Alex},
doi = {10.1103/PhysRevB.42.9622},
journal = {Phys. Rev. B},
month = {nov},
number = {15},
pages = {9622--9649},
publisher = {American Physical Society},
title = {{Electronic properties of random alloys: Special quasirandom structures}},
url = {https://link.aps.org/doi/10.1103/PhysRevB.42.9622},
volume = {42},
year = {1990}
}

@article{Yang2016,
author = {Yang, Kesong and Oses, Corey and Curtarolo, Stefano},
doi = {10.1021/acs.chemmater.6b01449},
issn = {0897-4756},
journal = {Chem. Mater.},
month = {sep},
number = {18},
pages = {6484--6492},
publisher = {American Chemical Society},
title = {{Modeling off-stoichiometry materials with a high-throughput ab-initio approach}},
url = {https://doi.org/10.1021/acs.chemmater.6b01449},
volume = {28},
year = {2016}
}

@article{Singh2021,
author = {Singh, Rahul and Sharma, Aayush and Singh, Prashant and Balasubramanian, Ganesh and Johnson, Duane D},
doi = {10.1038/s43588-020-00006-7},
issn = {2662-8457},
journal = {Nat. Comput. Sci.},
number = {1},
pages = {54--61},
title = {{Accelerating computational modeling and design of high-entropy alloys}},
url = {https://doi.org/10.1038/s43588-020-00006-7},
volume = {1},
year = {2021}
}

@article{Lopanitsyna2023,
author = {Lopanitsyna, Nataliya and Fraux, Guillaume and Springer, Maximilian A and De, Sandip and Ceriotti, Michele},
doi = {10.1103/PhysRevMaterials.7.045802},
journal = {Phys. Rev. Mater.},
month = {apr},
number = {4},
pages = {045802},
publisher = {American Physical Society},
title = {{Modeling high-entropy transition metal alloys with alchemical compression}},
url = {https://link.aps.org/doi/10.1103/PhysRevMaterials.7.045802},
volume = {7},
year = {2023}
}

@article{Peng2024b,
archivePrefix = {arXiv},
author = {Peng, Jiayu and Damewood, James and Karaguesian, Jessica and Lunger, Jaclyn R and G{\'{o}}mez-Bombarelli, Rafael},
journal = {Preprint at arXiv},
pages = {10.48550/arXiv.2409.13851},
eprint = {2409.13851},
primaryClass = {cond-mat.mtrl-sci},
title = {{Learning ordering in crystalline materials with symmetry-aware graph neural networks}},
url = {https://arxiv.org/abs/2409.13851},
year = {2024}
}

@article{Petersen2025,
archivePrefix = {arXiv},
author = {Petersen, Martin Hoffmann and Zhu, Ruiming and Dai, Haiwen and Aggarwal, Savyasanchi and Nong, Wei and Chen, Andy Paul and Bhowmik, Arghya and Garcia-Lastra, Juan Maria and Hippalgaonkar, Kedar},
journal = {Preprint at arXiv},
pages = {10.48550/arXiv.2507.18275},
eprint = {2507.18275},
primaryClass = {cond-mat.mtrl-sci},
title = {{Dis-GEN: Disordered crystal structure generation}},
url = {https://arxiv.org/abs/2507.18275},
year = {2025}
}

@article{Nam2025,
author = {Nam, Juno and Peng, Jiayu and G{\'{o}}mez-Bombarelli, Rafael},
issn = {2041-1723},
journal = {Nat. Commun.},
number = {1},
pages = {4350},
title = {{Interpolation and differentiation of alchemical degrees of freedom in machine learning interatomic potentials}},
volume = {16},
year = {2025}
}

@article{Divilov2025,
author = {Divilov, Simon and Eckert, Hagen and Thiel, Scott D and Griesemer, Sean D and Friedrich, Rico and Anderson, Nicholas H and Mehl, Michael J and Hicks, David and Esters, Marco and Hotz, Nico and Campilongo, Xiomara and Calzolari, Arrigo and Curtarolo, Stefano},
doi = {10.1007/s44210-025-00058-2},
issn = {2731-5827},
journal = {High Entropy Alloys Mater.},
number = {1},
pages = {178--187},
title = {{AFLOW4: Heading toward disorder}},
url = {https://doi.org/10.1007/s44210-025-00058-2},
volume = {3},
year = {2025}
}

@article{Hohenberg1964,
author = {Hohenberg, Pierre and Kohn, Walter},
journal = {Phys. Rev.},
number = {3B},
pages = {B864},
publisher = {APS},
title = {{Inhomogeneous electron gas}},
volume = {136},
year = {1964}
}

@article{Kohn1965,
author = {Kohn, Walter and Sham, Lu Jeu},
journal = {Phys. Rev.},
number = {4A},
pages = {A1133},
publisher = {APS},
title = {{Self-consistent equations including exchange and correlation effects}},
volume = {140},
year = {1965}
}

@article{Ramer2000,
author = {Ramer, Nicholas J and Rappe, Andrew M},
doi = {10.1103/PhysRevB.62.R743},
journal = {Phys. Rev. B},
month = {jul},
number = {2},
pages = {R743--R746},
publisher = {American Physical Society},
title = {{Virtual-crystal approximation that works: Locating a compositional phase boundary in PbZr1-xTixO3}},
url = {https://link.aps.org/doi/10.1103/PhysRevB.62.R743},
volume = {62},
year = {2000}
}

@article{Eckhardt2014,
author = {Eckhardt, C and Hummer, K and Kresse, G},
doi = {10.1103/PhysRevB.89.165201},
journal = {Phys. Rev. B},
month = {apr},
number = {16},
pages = {165201},
publisher = {American Physical Society},
title = {{Indirect-to-direct gap transition in strained and unstrained SnxGe1-x alloys}},
url = {https://link.aps.org/doi/10.1103/PhysRevB.89.165201},
volume = {89},
year = {2014}
}

@article{Bellaiche2000,
author = {Bellaiche, L and Vanderbilt, David},
doi = {10.1103/PhysRevB.61.7877},
journal = {Phys. Rev. B},
month = {mar},
number = {12},
pages = {7877--7882},
publisher = {American Physical Society},
title = {{Virtual crystal approximation revisited: Application to dielectric and piezoelectric properties of perovskites}},
url = {https://link.aps.org/doi/10.1103/PhysRevB.61.7877},
volume = {61},
year = {2000}
}

@article{Winkler2002,
author = {Winkler, Bj{\"{o}}rn and Pickard, Chris and Milman, Victor},
doi = {10.1016/S0009-2614(02)01029-1},
issn = {0009-2614},
journal = {Chem. Phys. Lett.},
number = {3},
pages = {266--270},
title = {{Applicability of a quantum mechanical 'virtual crystal approximation' to study Al/Si-disorder}},
url = {https://www.sciencedirect.com/science/article/pii/S0009261402010291},
volume = {362},
year = {2002}
}

@article{Iniguez2003,
author = {{\'{I}}{\~{n}}iguez, Jorge and Vanderbilt, David and Bellaiche, L},
doi = {10.1103/PhysRevB.67.224107},
journal = {Phys. Rev. B},
month = {jun},
number = {22},
pages = {224107},
publisher = {American Physical Society},
title = {{First-principles study of (BiScO3)1-x-(PbTiO3)x piezoelectric alloys}},
url = {https://link.aps.org/doi/10.1103/PhysRevB.67.224107},
volume = {67},
year = {2003}
}

@article{Payne2025,
author = {Payne, Bartholomew T and Juelsholt, Mikkel and P{\'{e}}rez-Osorio, Miguel A and Melvin, Dominic L R and Cuello, Gabriel J and Suard, Emmanuelle and Irving, Daniel J M and Rees, Nicholas H and Feaviour, Mark and Petrucco, Enrico and Day, Stephen P and Rees, Gregory J and Bruce, Peter G},
doi = {10.1039/D5EE01612F},
issn = {1754-5692},
journal = {Energy Environ. Sci.},
number = {19},
pages = {8876--8888},
publisher = {The Royal Society of Chemistry},
title = {{How multi-length scale disorder shapes ion transport in lithium argyrodites}},
url = {http://dx.doi.org/10.1039/D5EE01612F},
volume = {18},
year = {2025}
}

@article{Yonezawa1973,
author = {Yonezawa, Fumiko and Morigaki, Kazuo},
doi = {10.1143/PTPS.53.1},
issn = {0375-9687},
journal = {Prog. Theor. Phys. Suppl.},
month = {jan},
pages = {1--76},
title = {{Coherent potential approximation. Basic concepts and applications}},
url = {https://doi.org/10.1143/PTPS.53.1},
volume = {53},
year = {1973}
}

@article{Soven1970,
author = {Soven, Paul},
doi = {10.1103/PhysRevB.2.4715},
journal = {Phys. Rev. B},
month = {dec},
number = {12},
pages = {4715--4722},
publisher = {American Physical Society},
title = {{Application of the coherent potential approximation to a system of muffin-tin potentials}},
url = {https://link.aps.org/doi/10.1103/PhysRevB.2.4715},
volume = {2},
year = {1970}
}

@article{Shiba1971,
author = {Shiba, Hiroyuki},
doi = {10.1143/PTP.46.77},
issn = {0033-068X},
journal = {Prog. Theor. Phys.},
month = {jul},
number = {1},
pages = {77--94},
title = {{A reformulation of the coherent potential approximation and its applications}},
url = {https://doi.org/10.1143/PTP.46.77},
volume = {46},
year = {1971}
}

@article{Johnson1986,
author = {Johnson, D D and Nicholson, D M and Pinski, F J and Gyorffy, B L and Stocks, G M},
doi = {10.1103/PhysRevLett.56.2088},
journal = {Phys. Rev. Lett.},
month = {may},
number = {19},
pages = {2088--2091},
publisher = {American Physical Society},
title = {{Density-functional theory for random alloys: Total energy within the coherent-potential approximation}},
url = {https://link.aps.org/doi/10.1103/PhysRevLett.56.2088},
volume = {56},
year = {1986}
}

@article{Johnson1990,
author = {Johnson, D D and Nicholson, D M and Pinski, F J and Gy{\"{o}}rffy, B L and Stocks, G M},
doi = {10.1103/PhysRevB.41.9701},
journal = {Phys. Rev. B},
month = {may},
number = {14},
pages = {9701--9716},
publisher = {American Physical Society},
title = {{Total-energy and pressure calculations for random substitutional alloys}},
url = {https://link.aps.org/doi/10.1103/PhysRevB.41.9701},
volume = {41},
year = {1990}
}

@article{Singh2015,
author = {Singh, Prashant and Smirnov, A V and Johnson, D D},
doi = {10.1103/PhysRevB.91.224204},
journal = {Phys. Rev. B},
month = {jun},
number = {22},
pages = {224204},
publisher = {American Physical Society},
title = {{Atomic short-range order and incipient long-range order in high-entropy alloys}},
url = {https://link.aps.org/doi/10.1103/PhysRevB.91.224204},
volume = {91},
year = {2015}
}

@article{Gyorffy1972,
author = {Gyorffy, B L},
doi = {10.1103/PhysRevB.5.2382},
journal = {Phys. Rev. B},
month = {mar},
number = {6},
pages = {2382--2384},
publisher = {American Physical Society},
title = {{Coherent-potential approximation for a nonoverlapping-muffin-tin-potential model of random substitutional alloys}},
url = {https://link.aps.org/doi/10.1103/PhysRevB.5.2382},
volume = {5},
year = {1972}
}

@article{AlcazarRuano2025,
author = {{Alc{\'{a}}zar Ruano}, Pedro L and Mart{\'{i}}nez, Dunkan and Arroyo-Gasc{\'{o}}n, Olga and Baba, Yuriko and Quereda, Jorge and Dom{\'{i}}nguez-Adame, Francisco},
doi = {10.1103/58pr-k9vt},
journal = {Phys. Rev. B},
month = {aug},
number = {6},
pages = {064204},
publisher = {American Physical Society},
title = {{Disordered transition metal dichalcogenides: A coherent potential approximation study}},
url = {https://link.aps.org/doi/10.1103/58pr-k9vt},
volume = {112},
year = {2025}
}

@article{Rowlands2008,
author = {Rowlands, D A and Zhang, X.-G. and Gonis, A},
doi = {10.1103/PhysRevB.78.115119},
journal = {Phys. Rev. B},
month = {sep},
number = {11},
pages = {115119},
publisher = {American Physical Society},
title = {{Reformulation of the nonlocal coherent-potential approximation as a unique reciprocal-space theory of disorder}},
url = {https://link.aps.org/doi/10.1103/PhysRevB.78.115119},
volume = {78},
year = {2008}
}

@book{Chandler1987,
address = {New York},
author = {David Chandler},
publisher = {Oxford University Press},
title = {{Introduction to Modern Statistical Mechanics}},
year = {1987}
}

@article{vandeWalle2002b,
author = {van de Walle, A and Ceder, G},
doi = {10.1103/RevModPhys.74.11},
journal = {Rev. Mod. Phys.},
month = {jan},
number = {1},
pages = {11--45},
publisher = {American Physical Society},
title = {{The effect of lattice vibrations on substitutional alloy thermodynamics}},
url = {https://link.aps.org/doi/10.1103/RevModPhys.74.11},
volume = {74},
year = {2002}
}

@article{Ceder1993,
author = {Ceder, G},
doi = {10.1016/0927-0256(93)90005-8},
issn = {0927-0256},
journal = {Comput. Mater. Sci.},
number = {2},
pages = {144--150},
title = {{A derivation of the Ising model for the computation of phase diagrams}},
url = {https://www.sciencedirect.com/science/article/pii/0927025693900058},
volume = {1},
year = {1993}
}

@article{Garbulsky1994,
author = {Garbulsky, G D and Ceder, G},
doi = {10.1103/PhysRevB.49.6327},
journal = {Phys. Rev. B},
month = {mar},
number = {9},
pages = {6327--6330},
publisher = {American Physical Society},
title = {{Effect of lattice vibrations on the ordering tendencies in substitutional binary alloys}},
url = {https://link.aps.org/doi/10.1103/PhysRevB.49.6327},
volume = {49},
year = {1994}
}

@phdthesis{Garbulsky1996,
title = {{Ground-state structure and vibrational free energy in first-principles models of substitutional-alloy thermodynamics}},
author = {Garbulsky, Gerardo Dami{\'a}n},
year = {1996},
school = {Massachusetts Institute of Technology}
}

@phdthesis{vanderVen2000,
author = {Van der Ven, Anton},
title = {{First principles investigation of the thermodynamic and kinetic properties of lithium transition metal oxides}},
year = {2000},
school = {Massachusetts Institute of Technology},
}

@article{Laks1992,
author = {Laks, David B and Ferreira, L G and Froyen, Sverre and Zunger, Alex},
doi = {10.1103/PhysRevB.46.12587},
journal = {Phys. Rev. B},
month = {nov},
number = {19},
pages = {12587--12605},
publisher = {American Physical Society},
title = {{Efficient cluster expansion for substitutional systems}},
url = {https://link.aps.org/doi/10.1103/PhysRevB.46.12587},
volume = {46},
year = {1992}
}

@article{Zarkevich2004,
author = {Zarkevich, Nikolai A and Johnson, D D},
doi = {10.1103/PhysRevLett.92.255702},
journal = {Phys. Rev. Lett.},
month = {jun},
number = {25},
pages = {255702},
publisher = {American Physical Society},
title = {{Reliable first-principles alloy thermodynamics via truncated cluster expansions}},
url = {https://link.aps.org/doi/10.1103/PhysRevLett.92.255702},
volume = {92},
year = {2004}
}

@article{Sanchez2010,
author = {Sanchez, J M},
doi = {10.1103/PhysRevB.81.224202},
journal = {Phys. Rev. B},
month = {jun},
number = {22},
pages = {224202},
publisher = {American Physical Society},
title = {{Cluster expansion and the configurational theory of alloys}},
url = {https://link.aps.org/doi/10.1103/PhysRevB.81.224202},
volume = {81},
year = {2010}
}

@article{Puchala2023,
author = {Puchala, Brian and Thomas, John C and Natarajan, Anirudh Raju and Goiri, Jon Gabriel and Behara, Sesha Sai and Kaufman, Jonas L and {Van der Ven}, Anton},
doi = {10.1016/j.commatsci.2022.111897},
issn = {0927-0256},
journal = {Comput. Mater. Sci.},
pages = {111897},
title = {{CASM — A software package for first-principles based study of multicomponent crystalline solids}},
url = {https://www.sciencedirect.com/science/article/pii/S0927025622006085},
volume = {217},
year = {2023}
}

@article{BarrosoLuque2024,
author = {Barroso-Luque, Luis and Ceder, Gerbrand},
doi = {10.1038/s41524-024-01338-y},
issn = {2057-3960},
journal = {npj Comput. Mater.},
number = {1},
pages = {158},
title = {{The cluster decomposition of the configurational energy of multicomponent alloys}},
url = {https://doi.org/10.1038/s41524-024-01338-y},
volume = {10},
year = {2024}
}

@article{Sanchez2019,
author = {Sanchez, J M},
doi = {10.1103/PhysRevB.99.134206},
journal = {Phys. Rev. B},
month = {apr},
number = {13},
pages = {134206},
publisher = {American Physical Society},
title = {{Renormalized interactions in truncated cluster expansions}},
url = {https://link.aps.org/doi/10.1103/PhysRevB.99.134206},
volume = {99},
year = {2019}
}

@article{Tepesch1995,
author = {Tepesch, P D and Garbulsky, G D and Ceder, G},
doi = {10.1103/PhysRevLett.74.2272},
journal = {Phys. Rev. Lett.},
month = {mar},
number = {12},
pages = {2272--2275},
publisher = {American Physical Society},
title = {{Model for configurational thermodynamics in ionic systems}},
url = {https://link.aps.org/doi/10.1103/PhysRevLett.74.2272},
volume = {74},
year = {1995}
}

@article{Seko2009,
author = {Seko, Atsuto and Koyama, Yukinori and Tanaka, Isao},
doi = {10.1103/PhysRevB.80.165122},
journal = {Phys. Rev. B},
month = {oct},
number = {16},
pages = {165122},
publisher = {American Physical Society},
title = {{Cluster expansion method for multicomponent systems based on optimal selection of structures for density-functional theory calculations}},
url = {https://link.aps.org/doi/10.1103/PhysRevB.80.165122},
volume = {80},
year = {2009}
}

@article{vandeWalle2009,
author = {van de Walle, Axel},
doi = {10.1016/j.calphad.2008.12.005},
issn = {0364-5916},
journal = {Calphad},
number = {2},
pages = {266--278},
title = {{Multicomponent multisublattice alloys, nonconfigurational entropy and other additions to the Alloy Theoretic Automated Toolkit}},
url = {https://www.sciencedirect.com/science/article/pii/S0364591608001314},
volume = {33},
year = {2009}
}

@article{Zhang2016,
author = {Zhang, Xi and Sluiter, Marcel H F},
doi = {10.1007/s11669-015-0427-x},
issn = {1863-7345},
journal = {J. Phase Equilib. Diffus.},
number = {1},
pages = {44--52},
title = {{Cluster expansions for thermodynamics and kinetics of multicomponent alloys}},
url = {https://doi.org/10.1007/s11669-015-0427-x},
volume = {37},
year = {2016}
}

@article{BarrosoLuque2021,
author = {Barroso-Luque, Luis and Yang, Julia H and Ceder, Gerbrand},
doi = {10.1103/PhysRevB.104.224203},
journal = {Phys. Rev. B},
month = {dec},
number = {22},
pages = {224203},
publisher = {American Physical Society},
title = {{Sparse expansions of multicomponent oxide configuration energy using coherency and redundancy}},
url = {https://link.aps.org/doi/10.1103/PhysRevB.104.224203},
volume = {104},
year = {2021}
}

@article{Rigamonti2024,
author = {Rigamonti, Santiago and Troppenz, Maria and Kuban, Martin and H{\"{u}}bner, Axel and Draxl, Claudia},
doi = {10.1038/s41524-024-01363-x},
issn = {2057-3960},
journal = {npj Comput. Mater.},
number = {1},
pages = {195},
title = {{CELL: A Python package for cluster expansion with a focus on complex alloys}},
url = {https://doi.org/10.1038/s41524-024-01363-x},
volume = {10},
year = {2024}
}

@article{Mueller2009,
author = {Mueller, Tim and Ceder, Gerbrand},
doi = {10.1103/PhysRevB.80.024103},
journal = {Phys. Rev. B},
month = {jul},
number = {2},
pages = {024103},
publisher = {American Physical Society},
title = {{Bayesian approach to cluster expansions}},
url = {https://link.aps.org/doi/10.1103/PhysRevB.80.024103},
volume = {80},
year = {2009}
}

@article{Nelson2013,
author = {Nelson, Lance J and Ozoliņ{\v{s}}, Vidvuds and Reese, C Shane and Zhou, Fei and Hart, Gus L W},
doi = {10.1103/PhysRevB.88.155105},
journal = {Phys. Rev. B},
month = {oct},
number = {15},
pages = {155105},
publisher = {American Physical Society},
title = {{Cluster expansion made easy with Bayesian compressive sensing}},
url = {https://link.aps.org/doi/10.1103/PhysRevB.88.155105},
volume = {88},
year = {2013}
}

@article{Zhong2022,
author = {Zhong, Peichen and Chen, Tina and Barroso-Luque, Luis and Xie, Fengyu and Ceder, Gerbrand},
doi = {10.1103/PhysRevB.106.024203},
journal = {Phys. Rev. B},
month = {jul},
number = {2},
pages = {024203},
publisher = {American Physical Society},
title = {{An l0l2-norm regularized regression model for construction of robust cluster expansions in multicomponent systems}},
url = {https://link.aps.org/doi/10.1103/PhysRevB.106.024203},
volume = {106},
year = {2022}
}

@article{Huang2017,
author = {Huang, Wenxuan and Urban, Alexander and Rong, Ziqin and Ding, Zhiwei and Luo, Chuan and Ceder, Gerbrand},
doi = {10.1038/s41524-017-0032-0},
issn = {2057-3960},
journal = {npj Comput. Mater.},
number = {1},
pages = {30},
title = {{Construction of ground-state preserving sparse lattice models for predictive materials simulations}},
url = {https://doi.org/10.1038/s41524-017-0032-0},
volume = {3},
year = {2017}
}

@article{Ouyang2019,
author = {Ouyang, Bin and Chakraborty, Tanmoy and Kim, Namhoon and Perry, Nicola H and Mueller, Tim and Aluru, Narayana R and Ertekin, Elif},
doi = {10.1021/acs.chemmater.8b04285},
issn = {0897-4756},
journal = {Chem. Mater.},
month = {may},
number = {9},
pages = {3144--3153},
publisher = {American Chemical Society},
title = {{Cluster expansion framework for the Sr(Ti1–xFex)O3–x/2 (0 < x < 1) mixed ionic electronic conductor: Properties based on realistic configurations}},
url = {https://doi.org/10.1021/acs.chemmater.8b04285},
volume = {31},
year = {2019}
}

@article{Ji2019,
author = {Ji, Huiwen and Urban, Alexander and Kitchaev, Daniil A and Kwon, Deok-Hwang and Artrith, Nongnuch and Ophus, Colin and Huang, Wenxuan and Cai, Zijian and Shi, Tan and Kim, Jae Chul and Kim, Haegyeom and Ceder, Gerbrand},
doi = {10.1038/s41467-019-08490-w},
issn = {2041-1723},
journal = {Nat. Commun.},
number = {1},
pages = {592},
title = {{Hidden structural and chemical order controls lithium transport in cation-disordered oxides for rechargeable batteries}},
url = {https://doi.org/10.1038/s41467-019-08490-w},
volume = {10},
year = {2019}
}

@article{vandeWalle2002c,
author = {van de Walle, A and Ceder, G},
doi = {10.1361/105497102770331596},
issn = {1054-9714},
journal = {J. Phase Equilib.},
number = {4},
pages = {348},
title = {{Automating first-principles phase diagram calculations}},
url = {https://doi.org/10.1361/105497102770331596},
volume = {23},
year = {2002}
}

@article{Wang2020,
author = {Wang, Kang and Cheng, Du and Fu, Chu-Liang and Zhou, Bi-Cheng},
doi = {10.1103/PhysRevMaterials.4.013606},
journal = {Phys. Rev. Mater.},
month = {jan},
number = {1},
pages = {013606},
publisher = {American Physical Society},
title = {{First-principles investigation of the phase stability and early stages of precipitation in Mg-Sn alloys}},
url = {https://link.aps.org/doi/10.1103/PhysRevMaterials.4.013606},
volume = {4},
year = {2020}
}

@article{Cheng2023,
author = {Cheng, Du and Wang, Kang and Zhou, Bi-Cheng},
doi = {10.1016/j.actamat.2022.118443},
issn = {1359-6454},
journal = {Acta Mater.},
pages = {118443},
title = {{Crystal structure and stability of phases in Mg-Zn alloys: A comprehensive first-principles study}},
url = {https://www.sciencedirect.com/science/article/pii/S1359645422008205},
volume = {242},
year = {2023}
}

@article{Xie2023,
author = {Xie, Fengyu and Zhong, Peichen and Barroso-Luque, Luis and Ouyang, Bin and Ceder, Gerbrand},
doi = {10.1016/j.commatsci.2022.112000},
issn = {0927-0256},
journal = {Comput. Mater. Sci.},
pages = {112000},
title = {{Semigrand-canonical Monte-Carlo simulation methods for charge-decorated cluster expansions}},
url = {https://www.sciencedirect.com/science/article/pii/S092702562200711X},
volume = {218},
year = {2023}
}

@article{Wolverton1998,
author = {Wolverton, C and Zunger, Alex},
doi = {10.1103/PhysRevLett.81.606},
journal = {Phys. Rev. Lett.},
month = {jul},
number = {3},
pages = {606--609},
publisher = {American Physical Society},
title = {{First-principles prediction of vacancy order-disorder and intercalation battery voltages in LixCoO2}},
url = {https://link.aps.org/doi/10.1103/PhysRevLett.81.606},
volume = {81},
year = {1998}
}

@article{Chen2023b,
author = {Chen, Tina and Yang, Julia and Barroso-Luque, Luis and Ceder, Gerbrand},
doi = {10.1021/acsenergylett.2c02141},
journal = {ACS Energy Lett.},
month = {jan},
number = {1},
pages = {314--319},
publisher = {American Chemical Society},
title = {{Removing the two-phase transition in spinel LiMn2O4 through cation disorder}},
url = {https://doi.org/10.1021/acsenergylett.2c02141},
volume = {8},
year = {2023}
}

@article{Guo2023,
author = {Guo, Xingyu and Chen, Chi and Ong, Shyue Ping},
doi = {10.1021/acs.chemmater.2c02839},
issn = {0897-4756},
journal = {Chem. Mater.},
month = {feb},
number = {4},
pages = {1537--1546},
publisher = {American Chemical Society},
title = {{Intercalation chemistry of the disordered rocksalt Li3V2O5 anode from cluster expansions and machine learning interatomic potentials}},
url = {https://doi.org/10.1021/acs.chemmater.2c02839},
volume = {35},
year = {2023}
}

@article{Fu2024,
author = {Fu, Chu-Liang and Gorrey, Rajendra Prasad and Zhou, Bi-Cheng},
doi = {10.1016/j.actamat.2024.120138},
issn = {1359-6454},
journal = {Acta Mater.},
pages = {120138},
title = {{A cluster-based computational thermodynamics framework with intrinsic chemical short-range order: Part I. Configurational contribution}},
url = {https://www.sciencedirect.com/science/article/pii/S1359645424004890},
volume = {277},
year = {2024}
}

@article{Roy2026,
author = {Roy, Abhinav and Sieradzki, Karl and Waters, Michael J and Rondinelli, James M and McCue, Ian},
doi = {10.1016/j.scriptamat.2025.117137},
issn = {1359-6462},
journal = {Scr. Mater.},
pages = {117137},
title = {{Percolation diagrams derived from first-principles investigation of chemical short-range order in binary alloys}},
url = {https://www.sciencedirect.com/science/article/pii/S1359646225005974},
volume = {274},
year = {2026}
}

@article{Liu2026a,
author = {Liu, Tzu-chen and Torrisi, Steven B and Wolverton, Chris},
doi = {10.1002/smll.202514811},
issn = {1613-6810},
journal = {Small},
month = {mar},
number = {22},
pages = {e14811},
publisher = {John Wiley & Sons, Ltd},
title = {{Short-range order and LixTM4-x probability maps for disordered rocksalt cathodes}},
url = {https://doi.org/10.1002/smll.202514811},
volume = {22},
year = {2026}
}

@article{Aamlid2024,
author = {Aamlid, Solveig S and Mugiraneza, Sam and Gonz{\'{a}}lez-Rivas, Mario U and King, Graham and Hallas, Alannah M and Rottler, J{\"{o}}rg},
doi = {10.1021/acs.chemmater.4c01702},
issn = {0897-4756},
journal = {Chem. Mater.},
month = {oct},
number = {19},
pages = {9636--9645},
publisher = {American Chemical Society},
title = {{Short-range order and local distortions in entropy stabilized oxides}},
url = {https://doi.org/10.1021/acs.chemmater.4c01702},
volume = {36},
year = {2024}
}

@article{Zhong2024,
author = {Zhong, Peichen and Gupta, Sunny and Deng, Bowen and Jun, KyuJung and Ceder, Gerbrand},
doi = {10.1021/acsenergylett.4c00799},
journal = {ACS Energy Lett.},
month = {jun},
number = {6},
pages = {2775--2781},
publisher = {American Chemical Society},
title = {{Effect of cation disorder on lithium transport in halide superionic conductors}},
url = {https://doi.org/10.1021/acsenergylett.4c00799},
volume = {9},
year = {2024}
}

@article{Huang2023,
author = {Huang, Liliang and Zhong, Peichen and Ha, Yang and Cai, Zijian and Byeon, Young-Woon and Huang, Tzu-Yang and Sun, Yingzhi and Xie, Fengyu and Hau, Han-Ming and Kim, Haegyeom and Balasubramanian, Mahalingam and McCloskey, Bryan D and Yang, Wanli and Ceder, Gerbrand},
doi = {10.1002/aenm.202202345},
issn = {1614-6832},
journal = {Adv. Energy Mater.},
month = {jan},
number = {4},
pages = {2202345},
publisher = {John Wiley & Sons, Ltd},
title = {{Optimizing Li-excess cation-disordered rocksalt cathode design through partial Li deficiency}},
url = {https://doi.org/10.1002/aenm.202202345},
volume = {13},
year = {2023}
}

@article{Zhong2020,
author = {Zhong, Peichen and Cai, Zijian and Zhang, Yaqian and Giovine, Raynald and Ouyang, Bin and Zeng, Guobo and Chen, Yu and Cl{\'{e}}ment, Rapha{\"{e}}le and Lun, Zhengyan and Ceder, Gerbrand},
doi = {10.1021/acs.chemmater.0c04109},
issn = {0897-4756},
journal = {Chem. Mater.},
month = {dec},
number = {24},
pages = {10728--10736},
publisher = {American Chemical Society},
title = {{Increasing capacity in disordered rocksalt cathodes by mg doping}},
url = {https://doi.org/10.1021/acs.chemmater.0c04109},
volume = {32},
year = {2020}
}

@article{Yang2022,
author = {Yang, Julia H and Chen, Tina and Barroso-Luque, Luis and Jadidi, Zinab and Ceder, Gerbrand},
doi = {10.1038/s41524-022-00818-3},
issn = {2057-3960},
journal = {npj Comput. Mater.},
number = {1},
pages = {133},
title = {{Approaches for handling high-dimensional cluster expansions of ionic systems}},
url = {https://doi.org/10.1038/s41524-022-00818-3},
volume = {8},
year = {2022}
}

@article{Zhong2023a,
author = {Zhong, Peichen and Xie, Fengyu and Barroso-Luque, Luis and Huang, Liliang and Ceder, Gerbrand},
doi = {10.1103/PRXEnergy.2.043005},
journal = {PRX Energy},
month = {oct},
number = {4},
pages = {043005},
publisher = {American Physical Society},
title = {{Modeling intercalation chemistry with multiredox reactions by sparse lattice models in disordered rocksalt cathodes}},
url = {https://link.aps.org/doi/10.1103/PRXEnergy.2.043005},
volume = {2},
year = {2023}
}

@article{Muller2025,
author = {M{\"{u}}ller, Yann L and Natarajan, Anirudh Raju},
issn = {2057-3960},
journal = {npj Comput. Mater.},
number = {1},
pages = {60},
title = {{Constructing multicomponent cluster expansions with machine-learning and chemical embedding}},
volume = {11},
year = {2025}
}

@article{Wang2026a,
author = {Wang, Lin and Shen, Bo and He, Zheng-Da and Ye, Zihao and Zeng, Yan and Mirkin, Chad A and Ouyang, Bin},
doi = {10.1038/s41467-026-69585-9},
issn = {2041-1723},
journal = {Nat. Commun.},
pages = {3093},
title = {{Universal framework for efficient estimation of stability in multi-principal element alloys}},
url = {https://doi.org/10.1038/s41467-026-69585-9},
volume = {17},
year = {2026}
}

@article{Drautz2019,
author = {Drautz, Ralf},
doi = {10.1103/PhysRevB.99.014104},
journal = {Phys. Rev. B},
month = {jan},
number = {1},
pages = {014104},
publisher = {American Physical Society},
title = {{Atomic cluster expansion for accurate and transferable interatomic potentials}},
url = {https://link.aps.org/doi/10.1103/PhysRevB.99.014104},
volume = {99},
year = {2019}
}

@article{Darby2023,
author = {Darby, James P and Kov{\'{a}}cs, D{\'{a}}vid P and Batatia, Ilyes and Caro, Miguel A and Hart, Gus L. W. and Ortner, Christoph and Cs{\'{a}}nyi, G{\'{a}}bor},
doi = {10.1103/PhysRevLett.131.028001},
journal = {Phys. Rev. Lett.},
month = {jul},
number = {2},
pages = {028001},
publisher = {American Physical Society},
title = {{Tensor-reduced atomic density representations}},
url = {https://link.aps.org/doi/10.1103/PhysRevLett.131.028001},
volume = {131},
year = {2023}
}

@article{Vorotnikov2025,
author = {Vorotnikov, Igor and Romashov, Fedor and Rybin, Nikita and Rakhuba, Maxim and Novikov, Ivan S},
doi = {10.1063/5.0300163},
issn = {0021-9606},
journal = {J. Chem. Phys.},
month = {dec},
number = {24},
pages = {244112},
title = {{Low-rank matrix and tensor approximations for compression of machine-learning interatomic potentials}},
url = {https://doi.org/10.1063/5.0300163},
volume = {163},
year = {2025}
}

@article{Blum2004,
author = {Blum, Volker and Zunger, Alex},
doi = {10.1103/PhysRevB.70.155108},
journal = {Phys. Rev. B},
month = {oct},
number = {15},
pages = {155108},
publisher = {American Physical Society},
title = {{Mixed-basis cluster expansion for thermodynamics of bcc alloys}},
url = {https://link.aps.org/doi/10.1103/PhysRevB.70.155108},
volume = {70},
year = {2004}
}

@article{Wang2023b,
author = {Wang, Kang and Cheng, Du and Zhou, Bi-Cheng},
doi = {10.1038/s41524-023-01029-0},
issn = {2057-3960},
journal = {npj Comput. Mater.},
number = {1},
pages = {75},
title = {{Generalization of the mixed-space cluster expansion method for arbitrary lattices}},
url = {https://doi.org/10.1038/s41524-023-01029-0},
volume = {9},
year = {2023}
}

@article{Zhou2006,
author = {Zhou, Fei and Maxisch, Thomas and Ceder, Gerbrand},
doi = {10.1103/PhysRevLett.97.155704},
journal = {Phys. Rev. Lett.},
month = {oct},
number = {15},
pages = {155704},
publisher = {American Physical Society},
title = {{Configurational electronic entropy and the phase diagram of mixed-valence oxides: The case of LixFePO4}},
url = {https://link.aps.org/doi/10.1103/PhysRevLett.97.155704},
volume = {97},
year = {2006}
}

@article{vandeWalle2013,
author = {van de Walle, A and Tiwary, P and de Jong, M and Olmsted, D L and Asta, M and Dick, A and Shin, D and Wang, Y and Chen, L.-Q. and Liu, Z.-K.},
doi = {10.1016/j.calphad.2013.06.006},
issn = {0364-5916},
journal = {Calphad},
pages = {13--18},
title = {{Efficient stochastic generation of special quasirandom structures}},
url = {https://www.sciencedirect.com/science/article/pii/S0364591613000540},
volume = {42},
year = {2013}
}

@article{Kadzielawa2026,
archivePrefix = {arXiv},
author = {Kadzielawa, Andrzej P.},
journal = {Preprint at arXiv},
pages = {10.48550/arXiv.2602.10872},
eprint = {2602.10872},
primaryClass = {cond-mat.dis-nn},
title = {{On generating special quasirandom structures: Optimization for the DFT computational efficiency}},
url = {https://arxiv.org/abs/2602.10872},
year = {2026}
}

@article{Shin2006,
author = {Shin, Dongwon and Arr{\'{o}}yave, Raymundo and Liu, Zi-Kui and van de Walle, Axel},
doi = {10.1103/PhysRevB.74.024204},
journal = {Phys. Rev. B},
month = {jul},
number = {2},
pages = {024204},
publisher = {American Physical Society},
title = {{Thermodynamic properties of binary hcp solution phases from special quasirandom structures}},
url = {https://link.aps.org/doi/10.1103/PhysRevB.74.024204},
volume = {74},
year = {2006}
}

@article{Jiang2004,
author = {Jiang, Chao and Wolverton, C and Sofo, Jorge and Chen, Long-Qing and Liu, Zi-Kui},
doi = {10.1103/PhysRevB.69.214202},
journal = {Phys. Rev. B},
month = {jun},
number = {21},
pages = {214202},
publisher = {American Physical Society},
title = {{First-principles study of binary bcc alloys using special quasirandom structures}},
url = {https://link.aps.org/doi/10.1103/PhysRevB.69.214202},
volume = {69},
year = {2004}
}

@article{Jiang2016a,
author = {Jiang, Zhijun and Nahas, Yousra and Xu, Bin and Prosandeev, Sergey and Wang, Dawei and Bellaiche, Laurent},
doi = {10.1088/0953-8984/28/47/475901},
issn = {0953-8984},
journal = {J. Phys. Condens. Matter},
number = {47},
pages = {475901},
publisher = {IOP Publishing},
title = {{Special quasirandom structures for perovskite solid solutions}},
url = {https://dx.doi.org/10.1088/0953-8984/28/47/475901},
volume = {28},
year = {2016}
}

@incollection{Gao2016,
address = {Cham},
author = {Gao, Michael C and Niu, Changning and Jiang, Chao and Irving, Douglas L},
booktitle = {High-Entropy Alloys: Fundamentals and Applications},
doi = {10.1007/978-3-319-27013-5_10},
editor = {Gao, Michael C and Yeh, Jien-Wei and Liaw, Peter K and Zhang, Yong},
issn = {978-3-319-27013-5},
pages = {333--368},
publisher = {Springer},
title = {{Applications of Special Quasi-random Structures to High-Entropy Alloys}},
url = {https://doi.org/10.1007/978-3-319-27013-5_10},
year = {2016}
}

@article{Gehringer2023,
author = {Gehringer, Dominik and Fri{\'{a}}k, Martin and Holec, David},
doi = {10.1016/j.cpc.2023.108664},
issn = {0010-4655},
journal = {Comput. Phys. Commun.},
pages = {108664},
title = {{Models of configurationally-complex alloys made simple}},
url = {https://www.sciencedirect.com/science/article/pii/S0010465523000097},
volume = {286},
year = {2023}
}

@article{Lian2025,
author = {Lian, Ji-Chun and Li, Lei and Huang, Gui-Fang and Hu, Wangyu and Huang, Wei-Qing},
doi = {10.1103/vphm-rcgr},
journal = {Phys. Rev. B},
month = {jun},
number = {22},
pages = {224207},
publisher = {American Physical Society},
title = {{Highly efficient search strategy for special quasirandom structures}},
url = {https://link.aps.org/doi/10.1103/vphm-rcgr},
volume = {111},
year = {2025}
}

@article{Lebeda2026,
author = {Lebeda, Miroslav and Drahokoupil, Jan and Vl{\v{c}}{\'{a}}k, Petr and Svoboda, {\v{S}}imon and van de Walle, Axel},
doi = {10.1016/j.jocs.2026.102846},
issn = {1877-7503},
journal = {J. Comput. Sci.},
pages = {102846},
title = {{SimplySQS: An automated and reproducible workflow for special quasirandom structure generation with ATAT}},
url = {https://www.sciencedirect.com/science/article/pii/S1877750326000645},
volume = {96},
year = {2026}
}

@article{Liu2025b,
author = {Liu, Tzu-chen and Salgado-Casanova, Adolfo and Yubuchi, So and Baldassarri, Bianca and Aykol, Muratahan and Yoshida, Jun and Yamasaki, Hisatsugu and Zhu, Yizhou and Torrisi, Steven B and Wolverton, Chris},
doi = {10.1002/aenm.202503660},
issn = {1614-6832},
journal = {Adv. Energy Mater.},
month = {dec},
number = {47},
pages = {e03660},
publisher = {John Wiley & Sons, Ltd},
title = {{Tailored ordering enables high-capacity cathode materials}},
url = {https://doi.org/10.1002/aenm.202503660},
volume = {15},
year = {2025}
}

@article{Wang2025a,
author = {Wang, Lin and Wang, You and Martin, Jayden and Scivally, Elena and He, Zhengda and Kim, Do-hoon and Yeon, Dong-hee and Zeng, Yan and Chen, Dongchang and Ouyang, Bin},
doi = {10.1039/D5EB00104H},
journal = {EES Batter.},
number = {6},
pages = {1731--1739},
publisher = {RSC},
title = {{Origin of enhanced disorder in high entropy rocksalt type Li-ion battery cathodes}},
url = {http://dx.doi.org/10.1039/D5EB00104H},
volume = {1},
year = {2025}
}

@article{Schuler2024,
author = {Schuler, Thomas and Nastar, Maylise and Li, Kangming and Fu, Chu-Chun},
doi = {10.1016/j.actamat.2024.120074},
issn = {1359-6454},
journal = {Acta Mater.},
pages = {120074},
title = {{Towards accurate thermodynamics from random energy sampling}},
url = {https://www.sciencedirect.com/science/article/pii/S1359645424004257},
volume = {276},
year = {2024}
}

@article{Li2024a,
author = {Li, Kangming and Schuler, Thomas and Fu, Chu-Chun and Nastar, Maylise},
doi = {10.1016/j.actamat.2024.120355},
issn = {1359-6454},
journal = {Acta Mater.},
pages = {120355},
title = {{Vacancy formation free energy in concentrated alloys: Equilibrium vs. random sampling}},
url = {https://www.sciencedirect.com/science/article/pii/S1359645424007055},
volume = {281},
year = {2024}
}

@article{Seko2015,
author = {Seko, Atsuto and Tanaka, Isao},
doi = {10.1103/PhysRevB.91.024106},
journal = {Phys. Rev. B},
month = {jan},
number = {2},
pages = {024106},
publisher = {American Physical Society},
title = {{Special quasirandom structure in heterovalent ionic systems}},
url = {https://link.aps.org/doi/10.1103/PhysRevB.91.024106},
volume = {91},
year = {2015}
}

@article{Liu2016,
author = {Liu, Jian and Fern{\'{a}}ndez-Serra, Maria V and Allen, Philip B},
doi = {10.1103/PhysRevB.93.054207},
journal = {Phys. Rev. B},
month = {feb},
number = {5},
pages = {054207},
publisher = {American Physical Society},
title = {{Special quasiordered structures: Role of short-range order in the semiconductor alloy (GaN)1-x(ZnO)x}},
url = {https://link.aps.org/doi/10.1103/PhysRevB.93.054207},
volume = {93},
year = {2016}
}

@article{GrauCrespo2007,
author = {Grau-Crespo, R and Hamad, S and Catlow, C R A and de Leeuw, N H},
doi = {10.1088/0953-8984/19/25/256201},
issn = {0953-8984},
journal = {J. Phys. Condens. Matter},
number = {25},
pages = {256201},
title = {{Symmetry-adapted configurational modelling of fractional site occupancy in solids}},
url = {https://doi.org/10.1088/0953-8984/19/25/256201},
volume = {19},
year = {2007}
}

@article{Hart2008,
author = {Hart, Gus L W and Forcade, Rodney W},
doi = {10.1103/PhysRevB.77.224115},
journal = {Phys. Rev. B},
month = {jun},
number = {22},
pages = {224115},
publisher = {American Physical Society},
title = {{Algorithm for generating derivative structures}},
url = {https://link.aps.org/doi/10.1103/PhysRevB.77.224115},
volume = {77},
year = {2008}
}

@article{Sarker2018,
author = {Sarker, Pranab and Harrington, Tyler and Toher, Cormac and Oses, Corey and Samiee, Mojtaba and Maria, Jon-Paul and Brenner, Donald W and Vecchio, Kenneth S and Curtarolo, Stefano},
doi = {10.1038/s41467-018-07160-7},
issn = {2041-1723},
journal = {Nat. Commun.},
number = {1},
pages = {4980},
title = {{High-entropy high-hardness metal carbides discovered by entropy descriptors}},
url = {https://doi.org/10.1038/s41467-018-07160-7},
volume = {9},
year = {2018}
}

@article{Divilov2024,
author = {Divilov, Simon and Eckert, Hagen and Hicks, David and Oses, Corey and Toher, Cormac and Friedrich, Rico and Esters, Marco and Mehl, Michael J and Zettel, Adam C and Lederer, Yoav and Zurek, Eva and Maria, Jon-Paul and Brenner, Donald W and Campilongo, Xiomara and Filipovi{\'{c}}, Suzana and Fahrenholtz, William G and Ryan, Caillin J and DeSalle, Christopher M and Crealese, Ryan J and Wolfe, Douglas E and Calzolari, Arrigo and Curtarolo, Stefano},
doi = {10.1038/s41586-023-06786-y},
issn = {1476-4687},
journal = {Nature},
number = {7993},
pages = {66--73},
title = {{Disordered enthalpy–entropy descriptor for high-entropy ceramics discovery}},
url = {https://doi.org/10.1038/s41586-023-06786-y},
volume = {625},
year = {2024}
}

@article{Oses2023,
author = {Oses, Corey and Esters, Marco and Hicks, David and Divilov, Simon and Eckert, Hagen and Friedrich, Rico and Mehl, Michael J and Smolyanyuk, Andriy and Campilongo, Xiomara and van de Walle, Axel and Schroers, Jan and Kusne, A Gilad and Takeuchi, Ichiro and Zurek, Eva and Nardelli, Marco Buongiorno and Fornari, Marco and Lederer, Yoav and Levy, Ohad and Toher, Cormac and Curtarolo, Stefano},
doi = {https://doi.org/10.1016/j.commatsci.2022.111889},
issn = {0927-0256},
journal = {Comput. Mater. Sci.},
pages = {111889},
title = {{aflow++: A C++ framework for autonomous materials design}},
url = {https://www.sciencedirect.com/science/article/pii/S0927025622006000},
volume = {217},
year = {2023}
}

@article{Okhotnikov2016,
author = {Okhotnikov, Kirill and Charpentier, Thibault and Cadars, Sylvian},
doi = {10.1186/s13321-016-0129-3},
issn = {1758-2946},
journal = {J. Cheminform.},
number = {1},
pages = {17},
title = {{Supercell program: A combinatorial structure-generation approach for the local-level modeling of atomic substitutions and partial occupancies in crystals}},
url = {https://doi.org/10.1186/s13321-016-0129-3},
volume = {8},
year = {2016}
}

@article{Esters2021,
author = {Esters, Marco and Oses, Corey and Hicks, David and Mehl, Michael J and Jahn{\'{a}}tek, Michal and Hossain, Mohammad Delower and Maria, Jon-Paul and Brenner, Donald W and Toher, Cormac and Curtarolo, Stefano},
doi = {10.1038/s41467-021-25979-5},
issn = {2041-1723},
journal = {Nat. Commun.},
number = {1},
pages = {5747},
title = {{Settling the matter of the role of vibrations in the stability of high-entropy carbides}},
url = {https://doi.org/10.1038/s41467-021-25979-5},
volume = {12},
year = {2021}
}

@article{Calzolari2022,
author = {Calzolari, Arrigo and Oses, Corey and Toher, Cormac and Esters, Marco and Campilongo, Xiomara and Stepanoff, Sergei P and Wolfe, Douglas E and Curtarolo, Stefano},
doi = {10.1038/s41467-022-33497-1},
issn = {2041-1723},
journal = {Nat. Commun.},
number = {1},
pages = {5993},
title = {{Plasmonic high-entropy carbides}},
url = {https://doi.org/10.1038/s41467-022-33497-1},
volume = {13},
year = {2022}
}

@article{Esters2023,
author = {Esters, Marco and Smolyanyuk, Andriy and Oses, Corey and Hicks, David and Divilov, Simon and Eckert, Hagen and Campilongo, Xiomara and Toher, Cormac and Curtarolo, Stefano},
doi = {10.1016/j.actamat.2022.118594},
issn = {1359-6454},
journal = {Acta Mater.},
pages = {118594},
title = {{QH-POCC: Taming tiling entropy in thermal expansion calculations of disordered materials}},
url = {https://www.sciencedirect.com/science/article/pii/S1359645422009697},
volume = {245},
year = {2023}
}

@article{Toher2022,
author = {Toher, Cormac and Oses, Corey and Esters, Marco and Hicks, David and Kotsonis, George N and Rost, Christina M and Brenner, Donald W and Maria, Jon-Paul and Curtarolo, Stefano},
doi = {10.1557/s43577-022-00281-x},
issn = {1938-1425},
journal = {MRS Bull.},
number = {2},
pages = {194--202},
title = {{High-entropy ceramics: Propelling applications through disorder}},
url = {https://doi.org/10.1557/s43577-022-00281-x},
volume = {47},
year = {2022}
}

@article{Jiang2016b,
author = {Jiang, Chao and Uberuaga, Blas P},
doi = {10.1103/PhysRevLett.116.105501},
journal = {Phys. Rev. Lett.},
month = {mar},
number = {10},
pages = {105501},
publisher = {American Physical Society},
title = {{Efficient ab initio modeling of random multicomponent alloys}},
url = {https://link.aps.org/doi/10.1103/PhysRevLett.116.105501},
volume = {116},
year = {2016}
}

@article{Sorkin2021,
author = {Sorkin, V and Tan, Teck L and Yu, Z G and Zhang, Y W},
doi = {10.1016/j.commatsci.2020.110213},
issn = {0927-0256},
journal = {Comput. Mater. Sci.},
pages = {110213},
title = {{High-throughput calculations based on the small set of ordered structures method for non-equimolar high entropy alloys}},
url = {https://www.sciencedirect.com/science/article/pii/S0927025620307047},
volume = {188},
year = {2021}
}

@article{Kristoffersen2022,
doi = {10.1021/acs.jpcc.2c00478},
author = {Kristoffersen, Henrik H and Rossmeisl, Jan},
issn = {1932-7447},
journal = {J. Phys. Chem. C},
month = {apr},
number = {15},
pages = {6782--6790},
publisher = {American Chemical Society},
title = {{Local order in AgAuCuPdPt high-entropy alloy surfaces}},
volume = {126},
year = {2022}
}

@article{Novick2023,
author = {Novick, Andrew and Nguyen, Quan and Garnett, Roman and Toberer, Eric and Stevanovi{\'{c}}, Vladan},
doi = {10.1103/PhysRevMaterials.7.063801},
journal = {Phys. Rev. Mater.},
month = {jun},
number = {6},
pages = {063801},
publisher = {American Physical Society},
title = {{Simulating high-entropy alloys at finite temperatures: an uncertainty-based approach}},
url = {https://link.aps.org/doi/10.1103/PhysRevMaterials.7.063801},
volume = {7},
year = {2023}
}

@article{Kuner2024,
author = {Kuner, Matthew C and Rothchild, Eric and Asta, Mark D and Chrzan, Daryl C},
doi = {10.1016/j.commatsci.2024.112924},
issn = {0927-0256},
journal = {Comput. Mater. Sci.},
pages = {112924},
title = {{Ab initio property predictions of quinary solid solutions using small binary cells}},
url = {https://www.sciencedirect.com/science/article/pii/S0927025624001459},
volume = {238},
year = {2024}
}

@article{Moran2019,
author = {Moran, Robert F and McKay, David and Tornstrom, Paulynne C and Aziz, Alex and Fernandes, Arantxa and Grau-Crespo, Ricardo and Ashbrook, Sharon E},
doi = {10.1021/jacs.9b09036},
issn = {0002-7863},
journal = {J. Am. Chem. Soc.},
month = {nov},
number = {44},
pages = {17838--17846},
publisher = {American Chemical Society},
title = {{Ensemble-based modeling of the NMR spectra of solid solutions: Cation disorder in Y2(Sn,Ti)2O7}},
url = {https://doi.org/10.1021/jacs.9b09036},
volume = {141},
year = {2019}
}

@article{Shi2024,
author = {Shi, Panhua and Yang, Yiying and Yao, Baodian and Si, Jiaxuan and Wang, Yuexia},
doi = {10.1039/D4NA00202D},
journal = {Nanoscale Adv.},
number = {15},
pages = {3793--3800},
publisher = {RSC},
title = {{Unveiling the mechanism of tuning elemental distribution in high entropy alloys and its effect on thermal stability}},
url = {http://dx.doi.org/10.1039/D4NA00202D},
volume = {6},
year = {2024}
}

@article{Roy2024,
author = {Roy, Abhinav and Sieradzki, Karl and Rondinelli, James M and McCue, Ian D},
doi = {10.1103/PhysRevB.110.085420},
journal = {Phys. Rev. B},
month = {aug},
number = {8},
pages = {085420},
publisher = {American Physical Society},
title = {{Effect of chemical short-range order and percolation on passivation in binary alloys}},
url = {https://link.aps.org/doi/10.1103/PhysRevB.110.085420},
volume = {110},
year = {2024}
}

@article{Tang2025,
archivePrefix = {arXiv},
author = {Hao Tang and Hoje Chun and Rafael Gomez-Bombarelli and Yuri Mishin and Ju Li},
journal = {Preprint at arXiv},
pages = {10.48550/arXiv.2512.22370},
eprint = {2512.22370},
primaryClass = {cond-mat.mtrl-sci},
title = {{Thermal equilibrium vacancy concentration in an alloy with chemical short-range order}},
url = {https://arxiv.org/abs/2512.22370},
year = {2025}
}

@article{Jin2021,
author = {Jin, Xiaochen and Chen, Shunda and Li, Tianshu},
doi = {10.1103/PhysRevMaterials.5.104606},
journal = {Phys. Rev. Mater.},
month = {oct},
number = {10},
pages = {104606},
publisher = {American Physical Society},
title = {{Short-range order in SiSn alloy enriched by second-nearest-neighbor repulsion}},
url = {https://link.aps.org/doi/10.1103/PhysRevMaterials.5.104606},
volume = {5},
year = {2021}
}

@article{Oyeniran2026,
author = {Oyeniran, Noah and Hu, Chongze},
doi = {10.1039/d6mh00312e},
issn = {2051-6347},
journal = {Mater. Horiz.},
number = {17},
pages = {8560--8569},
month = {jun},
title = {{Ab initio Monte Carlo prediction of chemical order and disorder in multicomponent MXenes}},
url = {https://doi.org/10.1039/d6mh00312e},
volume = {13},
year = {2026}
}

@article{Kirkpatrick1983,
author = {Kirkpatrick, S and Gelatt, C D and Vecchi, M P},
doi = {10.1126/science.220.4598.671},
journal = {Science},
month = {may},
number = {4598},
pages = {671--680},
publisher = {American Association for the Advancement of Science},
title = {{Optimization by simulated annealing}},
url = {https://doi.org/10.1126/science.220.4598.671},
volume = {220},
year = {1983}
}

@article{Hukushima1996,
author = {Hukushima, Koji and Nemoto, Koji},
doi = {10.1143/JPSJ.65.1604},
issn = {0031-9015},
journal = {J. Phys. Soc. Jpn.},
month = {jun},
number = {6},
pages = {1604--1608},
publisher = {The Physical Society of Japan},
title = {{Exchange Monte Carlo method and application to spin glass simulations}},
url = {https://doi.org/10.1143/JPSJ.65.1604},
volume = {65},
year = {1996}
}

@article{Earl2005,
author = {Earl, David J and Deem, Michael W},
doi = {10.1039/B509983H},
issn = {1463-9076},
journal = {Phys. Chem. Chem. Phys.},
number = {23},
pages = {3910--3916},
publisher = {The Royal Society of Chemistry},
title = {{Parallel tempering: Theory, applications, and new perspectives}},
url = {http://dx.doi.org/10.1039/B509983H},
volume = {7},
year = {2005}
}

@article{Kasamatsu2019,
author = {Kasamatsu, Shusuke and Sugino, Osamu},
doi = {10.1088/1361-648X/aaf75c},
issn = {0953-8984},
journal = {J. Phys. Condens. Matter},
number = {8},
pages = {085901},
publisher = {IOP Publishing},
title = {{Direct coupling of first-principles calculations with replica exchange Monte Carlo sampling of ion disorder in solids}},
url = {https://doi.org/10.1088/1361-648X/aaf75c},
volume = {31},
year = {2019}
}

@article{Zhu2026a,
author = {Zhu, Kai and Trizio, Enrico and Zhang, Jintu and Hu, Renling and Jiang, Linlong and Hou, Tingjun and Bonati, Luigi},
doi = {10.1021/acs.chemrev.5c00700},
issn = {0009-2665},
journal = {Chem. Rev.},
month = {jan},
number = {1},
pages = {671--713},
publisher = {American Chemical Society},
title = {{Enhanced sampling in the age of machine learning: Algorithms and applications}},
url = {https://doi.org/10.1021/acs.chemrev.5c00700},
volume = {126},
year = {2026}
}

@article{Neyts2012,
author = {Neyts, Erik C and Bogaerts, Annemie},
doi = {10.1007/s00214-012-1320-x},
issn = {1432-2234},
journal = {Theor. Chem. Acc.},
number = {2},
pages = {1320},
title = {{Combining molecular dynamics with Monte Carlo simulations: Implementations and applications}},
url = {https://doi.org/10.1007/s00214-012-1320-x},
volume = {132},
year = {2012}
}

@article{Dolezal2025,
author = {Dole{\v{z}}al, Tyler D and Tekoglu, Emre and Bae, Jong-Soo and Sim, Gi-Dong and Freitas, Rodrigo and Li, Ju},
doi = {10.1016/j.commatsci.2025.113858},
issn = {0927-0256},
journal = {Comput. Mater. Sci.},
pages = {113858},
title = {{Atomistic simulations of short-range ordering with light interstitials in Inconel superalloys}},
url = {https://www.sciencedirect.com/science/article/pii/S0927025625002010},
volume = {253},
year = {2025}
}

@article{Chen2021c,
  author = {Chen, Shuai and Aitken, Zachary H. and Pattamatta, Subrahmanyam and Wu, Zhaoxuan and Yu, Zhi Gen and
  Banerjee, Rajarshi and Srolovitz, David J. and Liaw, Peter K. and Zhang, Yong-Wei},
  doi = {10.1016/j.actamat.2021.116638},
  journal = {Acta Mater.},
  month = {mar},
  pages = {116638},
  publisher = {Elsevier BV},
  title = {{Chemical-affinity disparity and exclusivity drive atomic segregation, short-range ordering, and cluster formation in high-entropy alloys}},
  url = {https://doi.org/10.1016/j.actamat.2021.116638},
  volume = {206},
  year = {2021}
  }

@article{Xu2025a,
author = {Xu, Xinyu and Cai, Kehan and Shi, Yubai and Zhong, Peichen and Xie, Pinchen},
doi = {10.1103/fz5h-1t5b},
issn = {2469-9950},
journal = {Phys. Rev. B},
month = {jul},
number = {2},
pages = {024204},
title = {{Intrinsic structure of relaxor ferroelectrics from first principles}},
url = {https://link.aps.org/doi/10.1103/fz5h-1t5b},
volume = {114},
year = {2026}
}

@article{Liao2025,
author = {Liao, Junhong and Chen, Hao and Xie, Yaoshu and Li, Zihui and Tan, Shendong and Zhou, Shuyu and Jiang, Lu and Zhang, Xiang and Liu, Ming and He, Yan-Bing and Kang, Feiyu and Lun, Zhengyan and Zhao, Shixi and Hou, Tingzheng},
doi = {10.1002/aenm.202501857},
issn = {1614-6832},
journal = {Adv. Energy Mater.},
month = {aug},
number = {32},
pages = {2501857},
publisher = {John Wiley & Sons, Ltd},
title = {{Modeling short-range order in high-entropy cation-disordered rocksalt-type cathodes}},
url = {https://doi.org/10.1002/aenm.202501857},
volume = {15},
year = {2025}
}

@article{Wales1997,
author = {Wales, David J and Doye, Jonathan P K},
doi = {10.1021/jp970984n},
issn = {1089-5639},
journal = {J. Phys. Chem. A},
month = {jul},
number = {28},
pages = {5111--5116},
publisher = {American Chemical Society},
title = {{Global optimization by basin-hopping and the lowest energy structures of Lennard-Jones clusters containing up to 110 atoms}},
url = {https://doi.org/10.1021/jp970984n},
volume = {101},
year = {1997}
}

@article{Zhang2024a,
author = {Zhang, Yuxuan and Li, Zhenjie and Han, Zhong-Kang and Ouyang, Runhai},
doi = {10.1021/acs.jctc.4c00460},
issn = {1549-9618},
journal = {J. Chem. Theory Comput.},
month = {aug},
number = {15},
pages = {6971--6979},
publisher = {American Chemical Society},
title = {{Global optimization of cation ordering in perovskites by recommendation-based basin-hopping}},
url = {https://doi.org/10.1021/acs.jctc.4c00460},
volume = {20},
year = {2024}
}

@article{Goedecker2004,
author = {Goedecker, Stefan},
doi = {10.1063/1.1724816},
issn = {0021-9606},
journal = {J. Chem. Phys.},
month = {jun},
number = {21},
pages = {9911--9917},
title = {{Minima hopping: An efficient search method for the global minimum of the potential energy surface of complex molecular systems}},
url = {https://doi.org/10.1063/1.1724816},
volume = {120},
year = {2004}
}

@article{Talapatra2015,
author = {Talapatra, Anjana and Arr{\'{o}}yave, Raymundo and Entel, Peter and Valencia-Jaime, I and Romero, Aldo H},
doi = {10.1103/PhysRevB.92.054107},
journal = {Phys. Rev. B},
month = {aug},
number = {5},
pages = {054107},
publisher = {American Physical Society},
title = {{Stability analysis of the martensitic phase transformation in Co2NiGa Heusler alloy}},
url = {https://link.aps.org/doi/10.1103/PhysRevB.92.054107},
volume = {92},
year = {2015}
}

@article{Le2016,
author = {Le, Tu C and Winkler, David A},
doi = {10.1021/acs.chemrev.5b00691},
issn = {0009-2665},
journal = {Chem. Rev.},
month = {may},
number = {10},
pages = {6107--6132},
publisher = {American Chemical Society},
title = {{Discovery and optimization of materials using evolutionary approaches}},
url = {https://doi.org/10.1021/acs.chemrev.5b00691},
volume = {116},
year = {2016}
}

@article{Dean2020,
author = {Dean, James and Cowan, Michael J and Estes, Jonathan and Ramadan, Mahmoud and Mpourmpakis, Giannis},
doi = {10.1021/acsnano.0c01586},
issn = {1936-0851},
journal = {ACS Nano},
month = {jul},
number = {7},
pages = {8171--8180},
publisher = {American Chemical Society},
title = {{Rapid prediction of bimetallic mixing behavior at the nanoscale}},
url = {https://doi.org/10.1021/acsnano.0c01586},
volume = {14},
year = {2020}
}

@article{Mohn2009,
author = {Mohn, Chris E and Kob, Walter},
doi = {10.1016/j.commatsci.2008.03.046},
issn = {0927-0256},
journal = {Comput. Mater. Sci.},
number = {1},
pages = {111--117},
title = {{A genetic algorithm for the atomistic design and global optimisation of substitutionally disordered materials}},
url = {https://www.sciencedirect.com/science/article/pii/S0927025608003108},
volume = {45},
year = {2009}
}

@article{Anand2023,
author = {Anand, G},
doi = {10.1080/10426914.2023.2217909},
issn = {1042-6914},
journal = {Mater. Manuf. Process.},
month = {dec},
number = {16},
pages = {2044--2050},
publisher = {Taylor & Francis},
title = {{GAASP: Genetic algorithm-based atomistic sampling protocol for high-entropy materials}},
url = {https://doi.org/10.1080/10426914.2023.2217909},
volume = {38},
year = {2023}
}

@article{Guan2025,
author = {Guan, Pin-Wen and Spataru, Catalin D and Stavila, Vitalie and Jones, Reese and Sharma, Peter A and Witman, Matthew D},
doi = {10.1103/bsxd-qtph},
journal = {PRX Energy},
month = {sep},
number = {3},
pages = {033013},
publisher = {American Physical Society},
title = {{Thermodynamic modeling of complex solid solutions in the Lu-H-N system via graph neural network accelerated Monte Carlo simulations}},
url = {https://link.aps.org/doi/10.1103/bsxd-qtph},
volume = {4},
year = {2025}
}

@article{Fang2025,
author = {Fang, Zhenyao and Hsu, Ting-Wei and Yan, Qimin},
doi = {10.1021/acsnano.5c13080},
issn = {1936-0851},
journal = {ACS Nano},
month = {oct},
number = {42},
pages = {37353--37363},
publisher = {American Chemical Society},
title = {{A machine learning framework for modeling ensemble properties of atomically disordered materials}},
url = {https://doi.org/10.1021/acsnano.5c13080},
volume = {19},
year = {2025}
}

@article{Liu2025c,
author = {Liu, Xianglin and Yang, Kai and Zhou, Fanli and Xu, Pengxiang},
doi = {10.1021/acs.jctc.5c01614},
issn = {1549-9618},
journal = {J. Chem. Theory Comput.},
month = {dec},
number = {24},
pages = {12784--12795},
publisher = {American Chemical Society},
title = {{SMC-X: A Distributed, scalable Monte Carlo simulation method for chemically complex alloys}},
url = {https://doi.org/10.1021/acs.jctc.5c01614},
volume = {21},
year = {2025}
}

@article{Liu2025d,
author = {Liu, Xianglin and Yang, Kai and Liu, Yongxiang and Zhou, Fanli and Fan, Dengdong and Pei, Zongrui and Xu, Pengxiang and Tian, Yonghong},
doi = {10.1038/s41524-025-01762-8},
issn = {2057-3960},
journal = {npj Comput. Mater.},
number = {1},
pages = {267},
title = {{Revealing nanostructures in high-entropy alloys via machine-learning accelerated scalable Monte Carlo simulation}},
url = {https://doi.org/10.1038/s41524-025-01762-8},
volume = {11},
year = {2025}
}

@article{Niu2025,
author = {Niu, Caimei and Liu, Lifeng},
doi = {10.1016/j.commatsci.2025.113792},
issn = {0927-0256},
journal = {Comput. Mater. Sci.},
pages = {113792},
title = {{Short-range order based ultra fast large-scale modeling of high-entropy alloys}},
url = {https://www.sciencedirect.com/science/article/pii/S0927025625001351},
volume = {253},
year = {2025}
}

@article{Camino2025,
author = {Camino, Bruno and Buckeridge, John and Chancellor, Nicholas and Catlow, C Richard A and Ferrari, Anna Maria and Warburton, Paul A and Sokol, Alexey A and Woodley, Scott M},
doi = {10.1126/sciadv.adt7156},
journal = {Sci. Adv.},
month = {mar},
number = {23},
pages = {eadt7156},
publisher = {American Association for the Advancement of Science},
title = {{Exploring the thermodynamics of disordered materials with quantum computing}},
url = {https://doi.org/10.1126/sciadv.adt7156},
volume = {11},
year = {2025}
}

@article{Chen2023c,
author = {Chen, Wei and Hilhorst, Antoine and Bokas, Georgios and Gorsse, St{\'{e}}phane and Jacques, Pascal J and Hautier, Geoffroy},
doi = {10.1038/s41467-023-38423-7},
issn = {2041-1723},
journal = {Nat. Commun.},
number = {1},
pages = {2856},
title = {{A map of single-phase high-entropy alloys}},
url = {https://doi.org/10.1038/s41467-023-38423-7},
volume = {14},
year = {2023}
}

@book{Lukas2007,
address = {New York},
author = {Lukas, Hans L. and Fries, Suzana G. and Sundman, Bo},
publisher = {Cambridge University Press},
title = {{Computational Thermodynamics: The Calphad Method}},
year = {2007}
}

@article{Ansara1997,
author = {Ansara, Ibrahim and Dupin, Nathalie and Lukas, Hans Leo and Sundman, Bo},
doi = {10.1016/S0925-8388(96)02652-7},
issn = {0925-8388},
journal = {J. Alloys Compd.},
number = {1},
pages = {20--30},
title = {{Thermodynamic assessment of the Al-Ni system}},
url = {https://www.sciencedirect.com/science/article/pii/S0925838896026527},
volume = {247},
year = {1997}
}

@article{vandeWalle2017,
author = {van de Walle, Axel and Sun, Ruoshi and Hong, Qi-Jun and Kadkhodaei, Sara},
doi = {10.1016/j.calphad.2017.05.005},
issn = {0364-5916},
journal = {Calphad},
pages = {70--81},
title = {{Software tools for high-throughput CALPHAD from first-principles data}},
url = {https://www.sciencedirect.com/science/article/pii/S0364591617300305},
volume = {58},
year = {2017}
}

@article{Lederer2018,
author = {Lederer, Yoav and Toher, Cormac and Vecchio, Kenneth S and Curtarolo, Stefano},
doi = {10.1016/j.actamat.2018.07.042},
issn = {1359-6454},
journal = {Acta Mater.},
pages = {364--383},
title = {{The search for high entropy alloys: A high-throughput ab-initio approach}},
url = {https://www.sciencedirect.com/science/article/pii/S1359645418305706},
volume = {159},
year = {2018}
}

@article{Shen2025,
author = {Shen, Chen},
doi = {10.1016/j.commatsci.2025.113970},
issn = {0927-0256},
journal = {Comput. Mater. Sci.},
pages = {113970},
title = {{The synergy of machine learning and CALPHAD: Revitalizing traditional approaches}},
url = {https://www.sciencedirect.com/science/article/pii/S0927025625003131},
volume = {258},
year = {2025}
}

@article{Zhu2025a,
author = {Zhu, Siya and Sarıt{\"{u}}rk, Doğuhan and Arr{\'{o}}yave, Raymundo},
doi = {10.1038/s41524-025-01814-z},
issn = {2057-3960},
journal = {npj Comput. Mater.},
number = {1},
pages = {340},
title = {{Machine learning potentials for alloys: A detailed workflow to predict phase diagrams and benchmark accuracy}},
url = {https://doi.org/10.1038/s41524-025-01814-z},
volume = {11},
year = {2025}
}

@article{Zhu2025b,
author = {Zhu, Siya and Sarıt{\"{u}}rk, Doğuhan and Arr{\'{o}}yave, Raymundo},
doi = {10.1016/j.actamat.2025.120747},
issn = {1359-6454},
journal = {Acta Mater.},
pages = {120747},
title = {{Accelerating CALPHAD-based phase diagram predictions in complex alloys using universal machine learning potentials: Opportunities and challenges}},
url = {https://www.sciencedirect.com/science/article/pii/S1359645425000400},
volume = {286},
year = {2025}
}

@article{Kunselman2025,
author = {Kunselman, Courtney and Zhu, Siya and Sarıt{\"{u}}rk, Doğuhan and Arr{\'{o}}yave, Raymundo},
doi = {10.1007/s11669-025-01222-2},
issn = {1863-7345},
journal = {J. Phase Equilib. Diffus.},
pages = {10.1007/s11669-025-01222-2},
title = {{Construction and tuning of CALPHAD models using machine-learned interatomic potentials and experimental data: A case study of the Pt–W system}},
url = {https://doi.org/10.1007/s11669-025-01222-2},
year = {2025}
}

@article{Cao2025,
author = {Cao, Yifan and Sheriff, Killian and Freitas, Rodrigo},
doi = {10.1038/s41524-025-01722-2},
issn = {2057-3960},
journal = {npj Comput. Mater.},
number = {1},
pages = {268},
title = {{Capturing short-range order in high-entropy alloys with machine learning potentials}},
url = {https://doi.org/10.1038/s41524-025-01722-2},
volume = {11},
year = {2025}
}

@article{Friederich2021,
author = {Friederich, Pascal and H{\"{a}}se, Florian and Proppe, Jonny and Aspuru-Guzik, Al{\'{a}}n},
doi = {10.1038/s41563-020-0777-6},
issn = {1476-4660},
journal = {Nat. Mater.},
number = {6},
pages = {750--761},
title = {{Machine-learned potentials for next-generation matter simulations}},
url = {https://doi.org/10.1038/s41563-020-0777-6},
volume = {20},
year = {2021}
}

@article{Unke2021,
author = {Unke, Oliver T and Chmiela, Stefan and Sauceda, Huziel E and Gastegger, Michael and Poltavsky, Igor and Sch{\"{u}}tt, Kristof T and Tkatchenko, Alexandre and M{\"{u}}ller, Klaus-Robert},
doi = {10.1021/acs.chemrev.0c01111},
issn = {0009-2665},
journal = {Chem. Rev.},
month = {aug},
number = {16},
pages = {10142--10186},
publisher = {American Chemical Society},
title = {{Machine learning force fields}},
url = {https://doi.org/10.1021/acs.chemrev.0c01111},
volume = {121},
year = {2021}
}

@article{Ko2023,
author = {Ko, Tsz Wai and Ong, Shyue Ping},
doi = {10.1038/s43588-023-00561-9},
issn = {2662-8457},
journal = {Nat. Comput. Sci.},
number = {12},
pages = {998--1000},
title = {{Recent advances and outstanding challenges for machine learning interatomic potentials}},
url = {https://doi.org/10.1038/s43588-023-00561-9},
volume = {3},
year = {2023}
}

@article{Kalita2025,
author = {Kalita, Bhupalee and Gokcan, Hatice and Isayev, Olexandr},
doi = {10.1038/s43588-025-00930-6},
issn = {2662-8457},
journal = {Nat. Comput. Sci.},
number = {12},
pages = {1120--1132},
title = {{Machine learning interatomic potentials at the centennial crossroads of quantum mechanics}},
url = {https://doi.org/10.1038/s43588-025-00930-6},
volume = {5},
year = {2025}
}

@article{Liu2024b,
author = {Liu, Jiaheng and Wang, Pengbo and Luan, Jun and Chen, Junwei and Cai, Pengcheng and Chen, Jianhua and Lu, Xionggang and Fan, Yunying and Yu, Zhigang and Chou, Kuochih},
doi = {10.1021/acs.jctc.4c00340},
issn = {1549-9618},
journal = {J. Chem. Theory Comput.},
month = {dec},
number = {24},
pages = {11082--11092},
publisher = {American Chemical Society},
title = {{VASE: A high-entropy alloy short-range order structural descriptor for machine learning}},
url = {https://doi.org/10.1021/acs.jctc.4c00340},
volume = {20},
year = {2024}
}

@article{Ibrahim2026,
author = {Ibrahim, Shehu Adam and Yang, Jinxue and Shi, Tan and Zhang, Chen and Chen, Da and Li, Jing and Li, Yang and Mbazor, Jeremiah Chinonso and Zhang, Yizhuo and Su, Zhengxiong and Lu, Chenyang},
doi = {10.1021/acs.jpcc.5c07217},
issn = {1932-7447},
journal = {J. Phys. Chem. C},
month = {mar},
number = {12},
pages = {4538--4553},
publisher = {American Chemical Society},
title = {{Machine learning prediction of vacancy formation energies in CoNiCrFe high-entropy alloy: The role of atomic descriptors and local chemical order}},
url = {https://doi.org/10.1021/acs.jpcc.5c07217},
volume = {130},
year = {2026}
}

@article{Chen2021d,
author = {Chen, Chi and Zuo, Yunxing and Ye, Weike and Li, Xiangguo and Ong, Shyue Ping},
doi = {10.1038/s43588-020-00002-x},
issn = {2662-8457},
journal = {Nat. Comput. Sci.},
number = {1},
pages = {46--53},
title = {{Learning properties of ordered and disordered materials from multi-fidelity data}},
url = {https://doi.org/10.1038/s43588-020-00002-x},
volume = {1},
year = {2021}
}

@article{Deshmukh2024,
author = {Deshmukh, Gaurav and Wichrowski, Noah J and Evangelou, Nikolaos and Ghanekar, Pushkar G and Deshpande, Siddharth and Kevrekidis, Ioannis G and Greeley, Jeffrey},
doi = {10.1038/s41524-024-01256-z},
issn = {2057-3960},
journal = {npj Comput. Mater.},
number = {1},
pages = {116},
title = {{Active learning of ternary alloy structures and energies}},
url = {https://doi.org/10.1038/s41524-024-01256-z},
volume = {10},
year = {2024}
}

@article{Fang2024,
author = {Fang, Zhenyao and Yan, Qimin},
doi = {10.1038/s41524-024-01283-w},
issn = {2057-3960},
journal = {npj Comput. Mater.},
number = {1},
pages = {91},
title = {{Towards accurate prediction of configurational disorder properties in materials using graph neural networks}},
url = {https://doi.org/10.1038/s41524-024-01283-w},
volume = {10},
year = {2024}
}

@article{Zhang2025,
author = {Zhang, Hengrui and Huang, Ruishu and Chen, Jie and Rondinelli, James M and Chen, Wei},
doi = {10.1088/2632-2153/adc0e1},
issn = {2632-2153},
journal = {Mach. Learn.: Sci. Technol.},
number = {2},
pages = {025005},
publisher = {IOP Publishing},
title = {{Graph representation of local environments for learning high-entropy alloy properties}},
url = {https://doi.org/10.1088/2632-2153/adc0e1},
volume = {6},
year = {2025}
}

@article{Wang2025b,
author = {Wang, Lin and He, Tanjin and Ouyang, Bin},
doi = {10.1021/acsmaterialslett.5c00726},
journal = {ACS Mater. Lett.},
month = {aug},
number = {8},
pages = {2708--2715},
publisher = {American Chemical Society},
title = {{Impact of domain knowledge on the property prediction of specialized machine learning models}},
url = {https://doi.org/10.1021/acsmaterialslett.5c00726},
volume = {7},
year = {2025}
}

@article{Fang2026,
archivePrefix = {arXiv},
author = {Fang, Zhenyao and Yan, Qimin},
journal = {Preprint at arXiv},
pages = {10.48550/arXiv.2607.07456},
eprint = {2607.07456},
primaryClass = {cond-mat.mtrl-sci},
title = {{A multi-scale machine learning framework for coupled chemical, spin, and structural disorder in alloys}},
url = {https://arxiv.org/abs/2607.07456},
year = {2026}
}

@article{Behler2007,
author = {Behler, J{\"{o}}rg and Parrinello, Michele},
journal = {Phys. Rev. Lett.},
month = {apr},
number = {14},
pages = {146401},
publisher = {American Physical Society},
title = {{Generalized neural-network representation of high-dimensional potential-energy surfaces}},
volume = {98},
year = {2007}
}

@article{Bartok2010,
author = {Bart{\'{o}}k, Albert P and Payne, Mike C and Kondor, Risi and Cs{\'{a}}nyi, G{\'{a}}bor},
doi = {10.1103/PhysRevLett.104.136403},
journal = {Phys. Rev. Lett.},
month = {apr},
number = {13},
pages = {136403},
publisher = {American Physical Society},
title = {{Gaussian approximation potentials: The accuracy of quantum mechanics, without the electrons}},
url = {https://link.aps.org/doi/10.1103/PhysRevLett.104.136403},
volume = {104},
year = {2010}
}

@article{Bartok2013,
author = {Bart{\'{o}}k, Albert P and Kondor, Risi and Cs{\'{a}}nyi, G{\'{a}}bor},
doi = {10.1103/PhysRevB.87.184115},
journal = {Phys. Rev. B},
month = {may},
number = {18},
pages = {184115},
publisher = {American Physical Society},
title = {{On representing chemical environments}},
url = {https://link.aps.org/doi/10.1103/PhysRevB.87.184115},
volume = {87},
year = {2013}
}

@article{Xie2018,
author = {Xie, Tian and Grossman, Jeffrey C},
doi = {10.1103/PhysRevLett.120.145301},
journal = {Phys. Rev. Lett.},
month = {apr},
number = {14},
pages = {145301},
publisher = {American Physical Society},
title = {{Crystal graph convolutional neural networks for an accurate and interpretable prediction of material properties}},
volume = {120},
year = {2018}
}

@article{Chen2019,
author = {Chen, Chi and Ye, Weike and Zuo, Yunxing and Zheng, Chen and Ong, Shyue Ping},
doi = {10.1021/acs.chemmater.9b01294},
issn = {0897-4756},
journal = {Chem. Mater.},
month = {may},
number = {9},
pages = {3564--3572},
publisher = {American Chemical Society},
title = {{Graph networks as a universal machine learning framework for molecules and crystals}},
url = {https://doi.org/10.1021/acs.chemmater.9b01294},
volume = {31},
year = {2019}
}

@article{Damewood2023,
author = {Damewood, James and Karaguesian, Jessica and Lunger, Jaclyn R and Tan, Aik Rui and Xie, Mingrou and Peng, Jiayu and G{\'{o}}mez-Bombarelli, Rafael},
doi = {10.1146/annurev-matsci-080921-085947},
issn = {1531-7331},
journal = {Annu. Rev. Mater. Res.},
month = {jul},
number = {1},
pages = {399--426},
publisher = {Annual Reviews},
title = {{Representations of materials for machine learning}},
url = {https://doi.org/10.1146/annurev-matsci-080921-085947},
volume = {53},
year = {2023}
}

@article{Corso2024,
author = {Corso, Gabriele and Stark, Hannes and Jegelka, Stefanie and Jaakkola, Tommi and Barzilay, Regina},
doi = {10.1038/s43586-024-00294-7},
issn = {2662-8449},
journal = {Nat. Rev. Methods Primers},
number = {1},
pages = {17},
title = {{Graph neural networks}},
url = {https://doi.org/10.1038/s43586-024-00294-7},
volume = {4},
year = {2024}
}

@article{Lunger2024,
author = {Lunger, Jaclyn R and Karaguesian, Jessica and Chun, Hoje and Peng, Jiayu and Tseo, Yitong and Shan, Chung Hsuan and Han, Byungchan and Shao-Horn, Yang and G{\'{o}}mez-Bombarelli, Rafael},
doi = {10.1038/s41524-024-01273-y},
issn = {2057-3960},
journal = {npj Comput. Mater.},
number = {1},
pages = {80},
title = {{Towards atom-level understanding of metal oxide catalysts for the oxygen evolution reaction with machine learning}},
url = {https://doi.org/10.1038/s41524-024-01273-y},
volume = {10},
year = {2024}
}

@article{Schutt2018,
author = {Sch{\"{u}}tt, K T and Sauceda, H E and Kindermans, P.-J. and Tkatchenko, A and M{\"{u}}ller, K.-R.},
doi = {10.1063/1.5019779},
issn = {0021-9606},
journal = {J. Chem. Phys.},
month = {mar},
number = {24},
pages = {241722},
publisher = {American Institute of Physics},
title = {{SchNet – A deep learning architecture for molecules and materials}},
url = {https://doi.org/10.1063/1.5019779},
volume = {148},
year = {2018}
}

@inproceedings{Gasteiger2021,
author = {Gasteiger, Johannes and Yeshwanth, Chandan and G\"{u}nnemann, Stephan},
booktitle = {Advances in Neural Information Processing Systems},
editor = {M. Ranzato and A. Beygelzimer and Y. Dauphin and P.S. Liang and J. Wortman Vaughan},
pages = {15421--15433},
publisher = {Curran Associates, Inc.},
title = {Directional Message Passing on Molecular Graphs via Synthetic Coordinates},
url = {https://proceedings.neurips.cc/paper_files/paper/2021/file/82489c9737cc245530c7a6ebef3753ec-Paper.pdf},
volume = {34},
year = {2021}
}

@article{Park2021,
author = {Park, Cheol Woo and Kornbluth, Mordechai and Vandermause, Jonathan and Wolverton, Chris and Kozinsky, Boris and Mailoa, Jonathan P},
doi = {10.1038/s41524-021-00543-3},
issn = {2057-3960},
journal = {npj Comput. Mater.},
number = {1},
pages = {73},
title = {{Accurate and scalable graph neural network force field and molecular dynamics with direct force architecture}},
url = {https://doi.org/10.1038/s41524-021-00543-3},
volume = {7},
year = {2021}
}

@article{Smidt2021,
author = {Smidt, Tess E},
doi = {10.1016/j.trechm.2020.10.006},
issn = {2589-5974},
journal = {Trends Chem.},
number = {2},
pages = {82--85},
title = {{Euclidean symmetry and equivariance in machine learning}},
url = {https://www.sciencedirect.com/science/article/pii/S2589597420302641},
volume = {3},
year = {2021}
}

@article{Duval2023,
archivePrefix = {arXiv},
author = {Duval, Alexandre and Mathis, Simon V and Joshi, Chaitanya K and Schmidt, Victor and Miret, Santiago and Malliaros, Fragkiskos D and Cohen, Taco and Lio, Pietro and Bengio, Yoshua and Bronstein, Michael},
journal = {Preprint at arXiv},
pages = {10.48550/arXiv.2312.07511},
eprint = {2312.07511},
primaryClass = {cs.LG},
title = {{A hitchhiker's guide to geometric GNNs for 3D atomic systems}},
url = {https://arxiv.org/abs/2312.07511},
year = {2023}
}

@article{Batzner2022,
author = {Batzner, Simon and Musaelian, Albert and Sun, Lixin and Geiger, Mario and Mailoa, Jonathan P and Kornbluth, Mordechai and Molinari, Nicola and Smidt, Tess E and Kozinsky, Boris},
doi = {10.1038/s41467-022-29939-5},
issn = {2041-1723},
journal = {Nat. Commun.},
number = {1},
pages = {2453},
title = {{E(3)-equivariant graph neural networks for data-efficient and accurate interatomic potentials}},
url = {https://doi.org/10.1038/s41467-022-29939-5},
volume = {13},
year = {2022}
}

@inproceedings{Batatia2022,
author = {Batatia, Ilyes and Kovacs, David P and Simm, Gregor and Ortner, Christoph and Csanyi, Gabor},
booktitle = {Advances in Neural Information Processing Systems},
pages = {11423--11436},
publisher = {Curran Associates, Inc.},
title = {{MACE: Higher order equivariant message passing neural networks for fast and accurate force fields}},
year = {2022}
}

@article{Batatia2025a,
author = {Batatia, Ilyes and Batzner, Simon and Kov{\'{a}}cs, D{\'{a}}vid P{\'{e}}ter and Musaelian, Albert and Simm, Gregor N C and Drautz, Ralf and Ortner, Christoph and Kozinsky, Boris and Cs{\'{a}}nyi, G{\'{a}}bor},
doi = {10.1038/s42256-024-00956-x},
issn = {2522-5839},
journal = {Nat. Mach. Intell.},
number = {1},
pages = {56--67},
title = {{The design space of E(3)-equivariant atom-centred interatomic potentials}},
url = {https://doi.org/10.1038/s42256-024-00956-x},
volume = {7},
year = {2025}
}

@article{Wen2024,
author = {Wen, Mingjian and Horton, Matthew K and Munro, Jason M and Huck, Patrick and Persson, Kristin A},
doi = {10.1039/D3DD00233K},
journal = {Digit. Discov.},
number = {5},
pages = {869--882},
publisher = {RSC},
title = {{An equivariant graph neural network for the elasticity tensors of all seven crystal systems}},
volume = {3},
year = {2024}
}

@article{Yan2025,
author = {Yan, Wenjie and Lai, Xinming and Chen, Yicheng and Zhang, Wenhao and Wu, Jianming and Xu, Xin},
doi = {10.1021/jacs.5c12428},
issn = {0002-7863},
journal = {J. Am. Chem. Soc.},
month = {dec},
number = {51},
pages = {47044--47056},
publisher = {American Chemical Society},
title = {{General framework for geometric deep learning on tensorial properties of molecules and crystals}},
url = {https://doi.org/10.1021/jacs.5c12428},
volume = {147},
year = {2025}
}

@article{Li2024b,
author = {Li, Kangming and Choudhary, Kamal and DeCost, Brian and Greenwood, Michael and Hattrick-Simpers, Jason},
doi = {10.1039/D4TA00982G},
issn = {2050-7488},
journal = {J. Mater. Chem. A},
number = {21},
pages = {12412--12422},
publisher = {The Royal Society of Chemistry},
title = {{Efficient first principles based modeling via machine learning: from simple representations to high entropy materials}},
url = {http://dx.doi.org/10.1039/D4TA00982G},
volume = {12},
year = {2024}
}

@article{Sheriff2026,
author = {Sheriff, Killian and Xiao, Daniel Z and Cao, Yifan and Owen, Lewis R and Freitas, Rodrigo},
doi = {10.1126/sciadv.aea9951},
journal = {Sci. Adv.},
month = {aug},
number = {25},
pages = {eaea9951},
publisher = {American Association for the Advancement of Science},
title = {{Machine learning potentials for modeling alloys across compositions}},
url = {https://doi.org/10.1126/sciadv.aea9951},
volume = {12},
year = {2026}
}

@article{Chen2024a,
author = {Chen, Shunda and Jin, Xiaochen and Zhao, Wanyu and Li, Tianshu},
doi = {10.1103/PhysRevMaterials.8.043805},
journal = {Phys. Rev. Mater.},
month = {apr},
number = {4},
pages = {043805},
publisher = {American Physical Society},
title = {{Intricate short-range order in GeSn alloys revealed by atomistic simulations with highly accurate and efficient machine-learning potentials}},
url = {https://link.aps.org/doi/10.1103/PhysRevMaterials.8.043805},
volume = {8},
year = {2024}
}

@article{An2025,
author = {An, Liying and Ma, Huan and Liu, Jinjia and Guo, Wenping and Wen, Xiaodong},
doi = {10.1038/s41524-025-01694-3},
issn = {2057-3960},
journal = {npj Comput. Mater.},
number = {1},
pages = {226},
title = {{Accelerating structure relaxation in chemically disordered materials with a chemistry-driven model}},
url = {https://doi.org/10.1038/s41524-025-01694-3},
volume = {11},
year = {2025}
}

@article{Choyal2024,
author = {Choyal, Vijay and Sagar, Nidhish and {Sai Gautam}, Gopalakrishnan},
doi = {10.1021/acs.jctc.4c00039},
issn = {1549-9618},
journal = {J. Chem. Theory Comput.},
month = {jun},
number = {11},
pages = {4844--4856},
publisher = {American Chemical Society},
title = {{Constructing and evaluating machine-learned interatomic potentials for Li-based disordered rocksalts}},
url = {https://doi.org/10.1021/acs.jctc.4c00039},
volume = {20},
year = {2024}
}

@article{Riebesell2025,
author = {Riebesell, Janosh and Goodall, Rhys E A and Benner, Philipp and Chiang, Yuan and Deng, Bowen and Ceder, Gerbrand and Asta, Mark and Lee, Alpha A and Jain, Anubhav and Persson, Kristin A},
doi = {10.1038/s42256-025-01055-1},
issn = {2522-5839},
journal = {Nat. Mach. Intell.},
number = {6},
pages = {836--847},
title = {{A framework to evaluate machine learning crystal stability predictions}},
url = {https://doi.org/10.1038/s42256-025-01055-1},
volume = {7},
year = {2025}
}

@article{Liu2025e,
archivePrefix = {arXiv},
author = {Yan Liu and Jiantao Wang and Hongkun Deng and Yan Sun and Xing-Qiu Chen and Peitao Liu},
journal = {Preprint at arXiv},
pages = {10.48550/arXiv.2510.16697},
eprint = {2510.16697},
primaryClass = {cond-mat.mtrl-sci},
title = {{Efficient small-cell sampling for machine-learning potentials of multi-principal element alloys}},
url = {https://arxiv.org/abs/2510.16697},
year = {2025}
}

@article{Deng2023,
author = {Deng, Bowen and Zhong, Peichen and Jun, KyuJung and Riebesell, Janosh and Han, Kevin and Bartel, Christopher J and Ceder, Gerbrand},
doi = {10.1038/s42256-023-00716-3},
issn = {2522-5839},
journal = {Nat. Mach. Intell.},
pages = {1031--1041},
title = {{CHGNet as a pretrained universal neural network potential for charge-informed atomistic modelling}},
url = {https://doi.org/10.1038/s42256-023-00716-3},
volume = {5},
year = {2023}
}

@article{Merchant2023,
author = {Merchant, Amil and Batzner, Simon and Schoenholz, Samuel S and Aykol, Muratahan and Cheon, Gowoon and Cubuk, Ekin Dogus},
doi = {10.1038/s41586-023-06735-9},
issn = {1476-4687},
journal = {Nature},
number = {7990},
pages = {80--85},
title = {{Scaling deep learning for materials discovery}},
url = {https://doi.org/10.1038/s41586-023-06735-9},
volume = {624},
year = {2023}
}

@article{Schmidt2024,
author = {Schmidt, Jonathan and Cerqueira, Tiago F T and Romero, Aldo H and Loew, Antoine and J{\"{a}}ger, Fabian and Wang, Hai-Chen and Botti, Silvana and Marques, Miguel A L},
doi = {10.1016/j.mtphys.2024.101560},
issn = {2542-5293},
journal = {Mater. Today Phys.},
pages = {101560},
title = {{Improving machine-learning models in materials science through large datasets}},
url = {https://www.sciencedirect.com/science/article/pii/S2542529324002360},
volume = {48},
year = {2024}
}

@article{Kaplan2025,
archivePrefix = {arXiv},
author = {Aaron D. Kaplan and Runze Liu and Ji Qi and Tsz Wai Ko and Bowen Deng and Janosh Riebesell and Gerbrand Ceder and Kristin A. Persson and Shyue Ping Ong},
journal = {Preprint at arXiv},
pages = {10.48550/arXiv.2503.04070},
eprint = {2503.04070},
primaryClass = {cond-mat.mtrl-sci},
title = {{A foundational potential energy surface dataset for materials}},
url = {https://arxiv.org/abs/2503.04070},
year = {2025}
}

@article{Mazitov2025a,
author = {Mazitov, Arslan and Chorna, Sofiia and Fraux, Guillaume and Bercx, Marnik and Pizzi, Giovanni and De, Sandip and Ceriotti, Michele},
doi = {10.1038/s41597-025-06109-y},
issn = {2052-4463},
journal = {Sci. Data},
number = {1},
pages = {1857},
title = {{Massive Atomic Diversity: A compact universal dataset for atomistic machine learning}},
url = {https://doi.org/10.1038/s41597-025-06109-y},
volume = {12},
year = {2025}
}

@article{Levine2026,
archivePrefix = {arXiv},
author = {Daniel S. Levine and Muhammed Shuaibi and Evan Walter Clark Spotte-Smith and Michael G. Taylor and Muhammad R. Hasyim and Kyle Michel and Ilyes Batatia and G{\'{a}}bor Cs{\'{a}}nyi and Misko Dzamba and Peter Eastman and Nathan C. Frey and Xiang Fu and Vahe Gharakhanyan and Aditi S. Krishnapriyan and Joshua A. Rackers and Sanjeev Raja and Ammar Rizvi and Andrew S. Rosen and Zachary Ulissi and Santiago Vargas and C. Lawrence Zitnick and Samuel M. Blau and Brandon M. Wood},
journal = {Preprint at arXiv},
pages = {10.48550/arXiv.2505.08762},
eprint = {2505.08762},
primaryClass = {physics.chem-ph},
title = {{The Open Molecules 2025 (OMol25) dataset, evaluations, and models}},
url = {https://arxiv.org/abs/2505.08762},
year = {2026}
}

@article{Malosso2026,
archivePrefix = {arXiv},
author = {Cesare Malosso and Filippo Bigi and Paolo Pegolo and Joseph W. Abbott and Philip Loche and Mariana Rossi and Michele Ceriotti and Arslan Mazitov},
journal = {Preprint at arXiv},
pages = {10.48550/arXiv.2603.02089},
eprint = {2603.02089},
primaryClass = {physics.chem-ph},
title = {{High-quality, high-information datasets for universal atomistic machine learning}},
url = {https://arxiv.org/abs/2603.02089},
year = {2026}
}

@article{BarrosoLuque2026,
author = {Barroso-Luque, Luis and Shuaibi, Muhammed and Fu, Xiang and Wood, Brandon M and Dzamba, Misko and Gao, Meng and Rizvi, Ammar and Uyttendaele, Matt and Zitnick, C Lawrence and Ulissi, Zachary W},
doi = {10.1038/s43588-026-00996-w},
issn = {2662-8457},
journal = {Nat. Comput. Sci.},
number = {6},
pages = {642--652},
title = {{The Open Materials 2024 (OMat24) inorganic materials dataset and models}},
url = {https://doi.org/10.1038/s43588-026-00996-w},
volume = {6},
year = {2026}
}

@article{Peng2026,
author = {Peng, Jiayu},
doi = {10.1038/s43588-026-00994-y},
issn = {2662-8457},
journal = {Nat. Comput. Sci.},
number = {6},
pages = {552--553},
title = {{Elevating universal interatomic potentials with large-scale off-equilibrium data}},
url = {https://doi.org/10.1038/s43588-026-00994-y},
volume = {6},
year = {2026}
}

@article{Choi2025,
author = {Choi, Junyoung and Nam, Gunwook and Choi, Jaesik and Jung, Yousung},
doi = {10.1021/jacsau.4c01160},
journal = {JACS Au},
month = {apr},
number = {4},
pages = {1499--1518},
publisher = {American Chemical Society},
title = {{A perspective on foundation models in chemistry}},
url = {https://doi.org/10.1021/jacsau.4c01160},
volume = {5},
year = {2025}
}

@article{Chen2022,
author = {Chen, Chi and Ong, Shyue Ping},
doi = {10.1038/s43588-022-00349-3},
issn = {2662-8457},
journal = {Nat. Comput. Sci.},
number = {11},
pages = {718--728},
title = {{A universal graph deep learning interatomic potential for the periodic table}},
url = {https://doi.org/10.1038/s43588-022-00349-3},
volume = {2},
year = {2022}
}

@article{Zhang2024b,
author = {Zhang, Duo and Liu, Xinzijian and Zhang, Xiangyu and Zhang, Chengqian and Cai, Chun and Bi, Hangrui and Du, Yiming and Qin, Xuejian and Peng, Anyang and Huang, Jiameng and Li, Bowen and Shan, Yifan and Zeng, Jinzhe and Zhang, Yuzhi and Liu, Siyuan and Li, Yifan and Chang, Junhan and Wang, Xinyan and Zhou, Shuo and Liu, Jianchuan and Luo, Xiaoshan and Wang, Zhenyu and Jiang, Wanrun and Wu, Jing and Yang, Yudi and Yang, Jiyuan and Yang, Manyi and Gong, Fu-Qiang and Zhang, Linshuang and Shi, Mengchao and Dai, Fu-Zhi and York, Darrin M and Liu, Shi and Zhu, Tong and Zhong, Zhicheng and Lv, Jian and Cheng, Jun and Jia, Weile and Chen, Mohan and Ke, Guolin and E, Weinan and Zhang, Linfeng and Wang, Han},
doi = {10.1038/s41524-024-01493-2},
issn = {2057-3960},
journal = {npj Comput. Mater.},
number = {1},
pages = {293},
title = {{DPA-2: A large atomic model as a multi-task learner}},
url = {https://doi.org/10.1038/s41524-024-01493-2},
volume = {10},
year = {2024}
}

@article{Clausen2024,
author = {Clausen, Christian M and Rossmeisl, Jan and Ulissi, Zachary W},
doi = {10.1021/acs.jpcc.4c01704},
issn = {1932-7447},
journal = {J. Phys. Chem. C},
month = {jul},
number = {27},
pages = {11190--11195},
publisher = {American Chemical Society},
title = {{Adapting OC20-trained EquiformerV2 models for high-entropy materials}},
url = {https://doi.org/10.1021/acs.jpcc.4c01704},
volume = {128},
year = {2024}
}

@article{Batatia2025b,
author = {Batatia, Ilyes and Benner, Philipp and Chiang, Yuan and Elena, Alin M and Kov{\'{a}}cs, D{\'{a}}vid P and Riebesell, Janosh and Advincula, Xavier R and Asta, Mark and Avaylon, Matthew and Baldwin, William J and Berger, Fabian and Bernstein, Noam and Bhowmik, Arghya and Bigi, Filippo and Blau, Samuel M and Cărare, Vlad and Ceriotti, Michele and Chong, Sanggyu and Darby, James P and De, Sandip and {Della Pia}, Flaviano and Deringer, Volker L and Elijo{\v{s}}ius, Rokas and El-Machachi, Zakariya and Fako, Edvin and Falcioni, Fabio and Ferrari, Andrea C and Gardner, John L A and Gawkowski, Miko{\l}aj J and Genreith-Schriever, Annalena and George, Janine and Goodall, Rhys E A and Grandel, Jonas and Grey, Clare P and Grigorev, Petr and Han, Shuang and Handley, Will and Heenen, Hendrik H and Hermansson, Kersti and Ho, Cheuk Hin and Hofmann, Stephan and Holm, Christian and Jaafar, Jad and Jakob, Konstantin S and Jung, Hyunwook and Kapil, Venkat and Kaplan, Aaron D and Karimitari, Nima and Kermode, James R and Kourtis, Panagiotis and Kroupa, Namu and Kullgren, Jolla and Kuner, Matthew C and Kuryla, Domantas and Liepuoniute, Guoda and Lin, Chen and Margraf, Johannes T and Magdău, Ioan-Bogdan and Michaelides, Angelos and Moore, J Harry and Naik, Aakash A and Niblett, Samuel P and Norwood, Sam Walton and O'Neill, Niamh and Ortner, Christoph and Persson, Kristin A and Reuter, Karsten and Rosen, Andrew S and Rosset, Louise A M and Schaaf, Lars L and Schran, Christoph and Shi, Benjamin X and Sivonxay, Eric and Stenczel, Tam{\'{a}}s K and Sutton, Christopher and Svahn, Viktor and Swinburne, Thomas D and Tilly, Jules and van der Oord, Cas and Vargas, Santiago and Varga-Umbrich, Eszter and Vegge, Tejs and Vondr{\'{a}}k, Martin and Wang, Yangshuai and Witt, William C and Wolf, Thomas and Zills, Fabian and Cs{\'{a}}nyi, G{\'{a}}bor},
doi = {10.1063/5.0297006},
issn = {0021-9606},
journal = {J. Chem. Phys.},
month = {nov},
number = {18},
pages = {184110},
title = {{A foundation model for atomistic materials chemistry}},
url = {https://doi.org/10.1063/5.0297006},
volume = {163},
year = {2025}
}

@article{Mazitov2025b,
author = {Mazitov, Arslan and Bigi, Filippo and Kellner, Matthias and Pegolo, Paolo and Tisi, Davide and Fraux, Guillaume and Pozdnyakov, Sergey and Loche, Philip and Ceriotti, Michele},
doi = {10.1038/s41467-025-65662-7},
issn = {2041-1723},
journal = {Nat. Commun.},
number = {1},
pages = {10653},
title = {{PET-MAD as a lightweight universal interatomic potential for advanced materials modeling}},
url = {https://doi.org/10.1038/s41467-025-65662-7},
volume = {16},
year = {2025}
}

@inproceedings{Wood2025,
author = {Wood, Brandon and Dzamba, Misko and Fu, Xiang and Gao, Meng and Shuaibi, Muhammed and Barroso-Luque, Luis and Abdelmaqsoud, Kareem and Gharakhanyan, Vahe and Kitchin, John and Levine, Daniel and Michel, Kyle and Sriram, Anuroop and Cohen, Taco and Das, Abhishek and Sahoo, Sushree and Rizvi, Ammar and Ulissi, Zachary and Zitnick, Larry},
booktitle = {Advances in Neural Information Processing Systems},
editor = {Belgrave, D and Zhang, C and Lin, H and Pascanu, R and Koniusz, P and Ghassemi, M and Chen, N},
pages = {129391--129427},
publisher = {Curran Associates, Inc.},
title = {{UMA: A family of universal models for atoms}},
url = {https://proceedings.neurips.cc/paper_files/paper/2025/file/bbf23e81b0ad7637fe9a731d0b676ca6-Paper-Conference.pdf},
volume = {38},
year = {2025}
}

@article{Biswas2026,
author = {Biswas, Maitreyo and Desai, Rushik and Bidna, Gavin and Mannodi-Kanakkithodi, Arun},
doi = {10.1021/acs.jcim.5c01611},
issn = {1549-9596},
journal = {J. Chem. Inf. Model.},
month = {feb},
number = {3},
pages = {1353--1370},
publisher = {American Chemical Society},
title = {{Unified graph-based interatomic potential for perovskite structure optimization}},
url = {https://doi.org/10.1021/acs.jcim.5c01611},
volume = {66},
year = {2026}
}

@article{Ullberg2026,
author = {Ullberg, R Seaton and Langhout, John D and Butala, Megan M and Phillpot, Simon R},
doi = {10.1021/acsaem.5c02676},
journal = {ACS Appl. Energy Mater.},
month = {apr},
number = {7},
pages = {3703--3715},
publisher = {American Chemical Society},
title = {{Predicting short-range order in disordered rocksalt Li-oxide cathode materials with density functional theory and crystal graph neural networks}},
url = {https://doi.org/10.1021/acsaem.5c02676},
volume = {9},
year = {2026}
}

\clearpage

\end{document}